\documentclass{jfm}

\begin{document}

\newtheorem{lemma}{Lemma}
\newtheorem{corollary}{Corollary}

\shorttitle{Nanoparticle diffusion in blood flow} 
\shortauthor{Z. Liu, J. R. Clausen, R. R. Rao and C. K. Aidun} 

\title{Nanoparticle diffusion in sheared cellular blood flow}
\author
{
  Zixiang Liu\aff{1},
  Jonathan R. Clausen\aff{3},
  Rekha R. Rao\aff{3}
  \and Cyrus K. Aidun\aff{1,2}
  \corresp{\email{cyrus.aidun@me.gatech.edu}}
}
\affiliation
{
  \aff{1}George W. Woodruff School of Mechanical Engineering, Georgia Institute of Technology, Atlanta, GA 30332 USA
  \aff{2}Parker H. Petit Institute for Bioengineering and Bioscience, Georgia Institute of Technology, Atlanta, GA 30332 US
  \aff{3}Sandia National Laboratories, Albuquerque, NM 87185, USA
}
\maketitle

\begin{abstract}
Using a multiscale blood flow solver, the complete diffusion tensor of nanoparticle (NP) in sheared cellular blood flow is calculated over a wide range of shear rate and haematocrit. In the short-time regime, NPs exhibit anomalous dispersive behaviors under high shear and high haematocrit due to the transient elongation and alignment of the red blood cells (RBCs). In the long-time regime, the NP diffusion tensor features high anisotropy. Particularly, there exists a critical shear rate ($\sim$100 $s^{-1}$) around which the shear-rate dependence of the diffusivity tensor changes from linear to nonlinear scale. Above the critical shear rate, the cross-stream diffusivity terms vary sublinearly with shear rate, while the longitudinal term varies superlinearly. The dependence on haematocrit is linear in general except at high shear rates, where a sublinear scale is found for the vorticity term and a quadratic scale for the longitudinal term. Through analysis of the suspension microstructure and numerical experiments, the nonlinear hemorheological dependence of the NP diffusion tensor is attributed to the streamwise elongation and cross-stream contraction of RBCs under high shear, quantified by a Capillary number. The RBC size is shown to be the characteristic length scale affecting the RBC-enhanced shear-induced diffusion (RESID), while the NP size at submicron exhibits negligible influence on the RESID. Based on the observed scaling behaviors, empirical correlations are proposed to bridge the NP diffusion tensor to specific shear rate and haematocrit. The characterized NP diffusion tensor provides a constitutive relation that can lead to more effective continuum models to tackle large-scale NP biotransport applications.
\end{abstract}



\section{Introduction}
The rapid advancement of nanotechnology and nanomaterial manufacturing has led to emerging exposure of human biological systems, such as cardiovascular systems, to nanosized particulate matters \citep{Albanese2012,Malysheva2015}, ranging from engineered nanocarriers functioning as medicine/imaging agents \citep{Yoo2011,Blanco2015,Griffin2018} to aerosol pollutant particulates potentially causing fatal cardiovascular disease \citep{Brook2010,Newby2015,Miller2017}. Both the design of optimal nanocarrier systems and the prevention and control of nanoparticle (NP) toxicity rely heavily on the knowledge of NP distribution in vascular systems \citep{Albanese2012}. Nevertheless, the biodistribution of NP remains elusive to date and limits the bio-availability of NP systems to the human bio-environment. One of the primary bottlenecks is the lack of understanding on the NP dispersive mechanisms in blood flow with respect to specific hemorheological conditions.

Multi-scale computational models \citep{Lee2013,Muller2014,Liu2018a} have been developed and applied to predict the NP distribution in micro-vessels by directly simulating NPs and red blood cells (RBCs) suspended in blood plasma. Although these models, as particle-based approaches, provide a straightforward means to predict the NP distribution in realistic, micro-scale vessels, they are still computationally intractable when it comes to organ- and circulation-level applications. Alternatively, continuum models \citep{Eckstein1991,Decuzzi2010,Hossain2013,MM2015} have the ability to predict NP distribution in large-scale vascular systems by solving the three-dimensional (3D) convection-diffusion equation
\begin{equation}
  \frac{\partial c}{\partial t} + \boldsymbol{u \cdot \nabla} c + \boldsymbol{\nabla \cdot J} = 0,
  \label{eqn:CDE}
\end{equation}
where $t$ is time, $\mathit{c}$ is the NP concentration, $\boldsymbol{u}$ is the local fluid velocity and $\boldsymbol{J}$ is the flux of NP concentration. Here, $\boldsymbol{J}$ is often estimated by the Fick's law as $\boldsymbol{J}=-\mathsfbi{D}^{\infty} \boldsymbol{\cdot \nabla}c$, where the NP diffusion tensor, $\mathsfbi{D}^{\infty}$, is so far treated as isotropic, Brownian diffusivity \citep{Hossain2013} or solute diffusivity measured in a single principal direction \citep{Zydney1988}. Since the particle diffusion tensor is anisotropic in nature even for monodisperse rigid sphere suspensions \citep{FossJFM1999,FossJFM2000}, an improved constitutive relation capturing the anisotropy of $\mathsfbi{D}^{\infty}$ subject to local hemorheological properties is therefore necessary to form a better closure of this convection-diffusion problem.

Using Couette-type flow devices, experiments have been conducted to characterize the particle self-diffusivity \citep{Eckstein1977,Breedveld1998} and effective solute diffusivity \citep{Wang1985,Zydney1988,Breedveld1998} in non-colloidal particle suspensions under linear shear flow for various shear rates and particle volume fractions. However, due to the difficulties in particle tracking in the presence of the affine flow effect, only the particle diffusivity in the cross-stream directions are reported.  Apart from the experimental efforts, particle-scale simulations have become an important tool for characterizing the anisotropic particle diffusivity tensor in both colloidal \citep{FossJFM1999,FossJFM2000,FossJR2000} and noncolloidal \citep{Sierou2004,Yeo2010,ClausenJFM2011} suspensions under shear flow. Owing to the success of those particle-scale simulation techniques, substantial progress have been made in understanding the rheological and hydrodynamic response of the particle diffusion tensor in sheared monodisperse suspensions. Nonetheless, the bidisperse RBC-NP suspension system remains largely unexplored and entails unique transport phenomenology that is unavailable to the conventional monodisperse particle suspensions. 

First, there is a large length-scale discrepancy between NPs $\sim$$\textit{O}(10\ nm)$ and RBCs $\sim$$\textit{O}(10\ \mu m)$. Consequently, NPs are subject to both molecular level thermal fluctuations (Brownian motion) and cellular level interactions with RBCs. On macroscopic scales, the two effects synergistically give rise to an apparent diffusivity contributed by both Brownian diffusivity (BD) and the so-called RBC-enhanced shear-induced diffusivity (RESID) \citep{MM2015,MM2016,Liu2018a}. Second, the NP phase shows infinite dilution while the RBC phase exhibits a range of physiological concentrations from $\sim$$10\%$ to $\sim$$40\%$. Consequently, the BD shows insignificant dependence on shear rate and haematocrit, while the RESID is highly dependent on the hemorheological conditions \citep{MM2016}. Third, RBCs deform considerably under shear, from biconcave shape in equilibrium to large elongation and tank treading of the membrane under high shear \citep{Gross2014}. Such geometric asymmetry and morphological changes of RBCs could alter RESID substantially.

Therefore, the objective of the present work is to characterize the bulk diffusivity tensor of NP in sheared blood flow and interrogate the NP dispersive mechanism specific to a broad range of haematocrit and shear rate. Given the large length-scale discrepancy (3$\sim$4 orders of magnitude) between NP and RBC, resolving both particle phases using direct numerical simulation (DNS) is computationally prohibitive. Therefore, a multiscale complex blood flow solver \citep{Aidun2010,Reasor2012,ReasorJFM2013,Liu2018a,Liu2018b} is employed to treat NP as effective Brownian particles while directly resolving the RBC phase. Such multiscale treatment can substantially reduce the computational expense but still preserve the critical suspension physics at distinct scales. As will be shown in the following section (\S $\ref{sec:sr}$), good comparison between the simulation and experimental results can be obtained using this multiscale approach. Since confinement (wall) effects in general lead to spatial heterogeneity of the blood flow \citep{Kumar2012} that forbids the calculation of NP diffusivity related to specific haematocrit and shear rate, the Lees-Edwards boundary condition (LEbc) \citep{LEbc1972} is implemented \citep{MacJFM2009} to impose unconfined simple shear flow to obtain the NP bulk diffusive properties. 

Note that blood flow typically occurs under confinement (e.g., blood flow in arteries) involving heterogeneous flow structures that could induce cell segregation \citep{Kumar2012,Ahmed2018} and margination \citep{Zhao2011,Zhao2012,ReasorABE2013,MM2015,MM2016}. Such phenomena are found to be a synergistic outcome of the RBC-induced diffusion in the RBC-laden region and the formation of cell free layer (CFL) near the wall \citep{MM2015,MM2016}. For NPs with negligible inertia and much smaller length scale compared to the CFL thickness \citep{Zhao2012,MM2016}, it is expected that the presence of wall has insignificant direct influence on the NP diffusion in the RBC-laden region. Besides, large-scale problems suitable for continuum modeling (such as blood flow through coronary arteries) typically feature much larger length and time scales compared to those considered in the current cellular-scale studies. It is therefore plausible to hypothesize the long-time NP diffusion tensor evaluated in an unbounded simple shear flow should closely capture the NP diffusive behavior subject to the same local hemorheological condition in a macroscale heterogeneous blood flow environment.

One unique contribution of this work is the development of a multiscale-simulation-informed empirical expression that links the anisotropic NP diffusion tensor to the local hemorheological conditions. Such 3D NP diffusive information in sheared blood is intractable to measure through either experiment or DNS simulation. On the application side, the developed NP diffusion tensor provides a constitutive relation that can lead to more effective continuum-level models to tackle large-scale NP biotransport problems. On the suspension rheology side, the diffusive phenomenology observed in such a biophysical, bidisperse RBC-NP suspension system could entail novel suspension physics that is unavailable to conventional suspension flows.

The remainder of this article is organized as follows.  In \S \ref{sec:method}, the details of the computational methodology are presented.  In \S \ref{sec:results}, we present results and perform numerical experiments. In \S \ref{sec:prob}, the numerical problem is formulated with careful consideration of numerical resolution as well as physiological significance. In \S \ref{sec:msd}, the transient dispersive behaviors of NP are presented. As follows in \S \ref{sec:hemo}, the long-time diffusive behaviors of NP subject to a wide range of shear rate and haematocrit are validated with available experimental data and interrogated with various hemorheological scaling behaviors. In \S \ref{sec:mst}, the NP-RBC suspension microstructure is analyzed to give mechanistic insights to the hemorheological scaling observations. In \S \ref{sec:RBCm} and \S \ref{sec:NPsize}, numerical experiments are conducted to shed light on the physical mechanisms governing the nonlinear shear-rate dependence of the NP diffusion tensor. In \S \ref{sec:corr}, empirical correlations are proposed based on the hemorheological scalings observed in previous sections. In \S \ref{sec:discon}, we conclude this work with some remarks.


\section{Methodology}\label{sec:method}
\begin{figure}
  \centerline{\includegraphics[width=0.95\columnwidth]{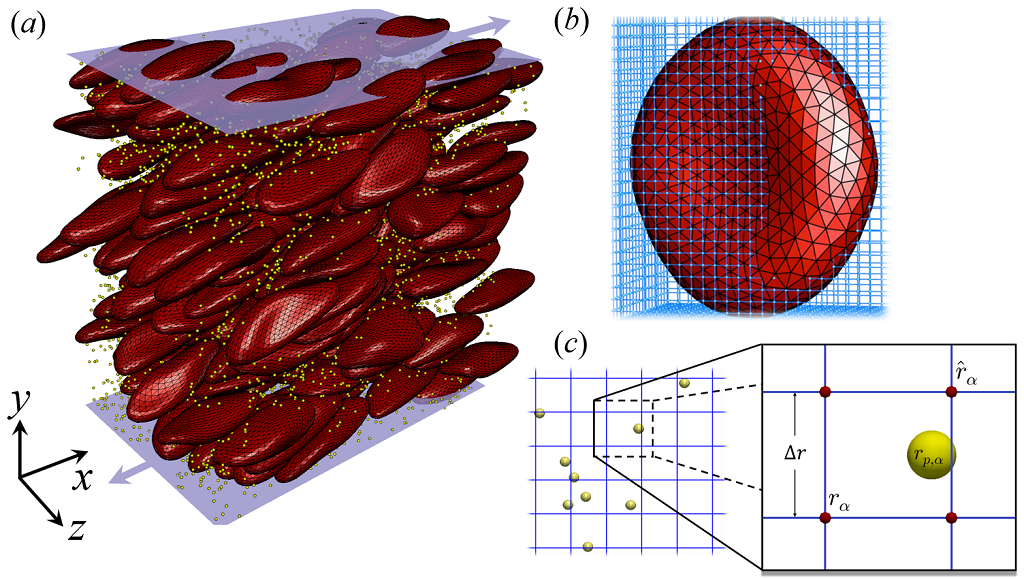}}
  \caption{(a) The RBC-NP bidisperse suspension in a triply periodic domain at shear rate $\dot{\gamma}=1\ 000\ s^{-1}$ and haematocrit, $\phi=40\%$. (b) The RBC membrane immersed in the LB lattice domain. The RBC membrane dynamics and deformation is captured by the course-grained spectrin-link method \citep{Pivkin2008,FedosovBJ2010,Reasor2012}. (c) The NP suspension dynamics is captured by the LB-LD method \citep{Liu2018a,Liu2018b}, which captures the NP Brownian motion and long-ranged HI in an off-lattice fashion.}
\label{fig:method}
\end{figure}

The numerical method for this study is through a 3D lattice-Boltzmann based multiscale complex blood flow solver that efficiently resolves both the dynamics and interactions of nanoscale particles and microscale capsules \citep{Aidun1998,Aidun1998jsp,Aidun2010,Reasor2012,Liu2018a,Liu2018b}, as demonstrated in figure \ref{fig:method}. The LB method is a well-established numerical model for hydrodynamics and proves to be a highly scalable method for direct numerical simulation (DNS) of dense particulate suspensions \citep{ClausenCPC2010,Aidun2010}. Modeling of the RBC dynamics and deformation is via a coarse-grained spectrin-link membrane method \citep{Pivkin2008,FedosovBJ2010} coupled to the LB method\citep{Reasor2012}, which has been validated against experimental results\citep{Reasor2012,ReasorJFM2013}. The NP suspension dynamics are resolved via a two-way coupled lattice-Boltzmann Langevin-dynamics (LB-LD) method with both particle Brownian motion and long-range hydrodynamic interactions (HI) directly resolved and validated \citep{Liu2018a,Liu2018b}. The solver has been successfully applied to several studies of particle and biopolymer transport in cellular blood flow \citep{ReasorJFM2013,ReasorABE2013,MM2016,Ahmed2018,Liu2018a,Griffin2018}. 

\subsection{Lattice-Boltzmann method}\label{sec:lb}
\noindent Simulation of the suspending fluid is based on the LB method developed by \cite{Aidun1995,Aidun1998,Aidun2010}. The LB method solves the discretized Boltzmann transport equation in velocity space through the streaming-collision process. In streaming, the fictitious fluid particles propagate along discrete velocity vectors forming a lattice space. In collision, the fluid particles at each lattice site collide with each other, causing the relaxation of the particle distribution function (PDF) towards a local `Maxwellian' equilibrium PDF. The collision term is linearized based on the single-relaxation-time Bhatnagar, Gross, and Krook (BGK) operator \citep{Bhatnagarp}. The temporal evolution of the particle distribution function is given as
\begin{equation}
  f_i(\boldsymbol{r}+\Delta t \boldsymbol{e}_i, t+\Delta t) = f_i(\boldsymbol{r}, t) - \frac{\Delta t}{\tau}[f_i(\boldsymbol{r}, t) - f_i^{(0)}(\boldsymbol{r}, t)] + f_i^S(\boldsymbol{r}, t),
  \label{eqn:lb1}
\end{equation}
where $f_i$ is the fluid PDF, $f_i^{(0)}$ is the equilibrium PDF, $r$ is the lattice site, $e_i$ is the discrete lattice velocity, $t$ is time, $\tau$ is the single relaxation time and $f_i^S$ is a forcing source term introduced to account for the discrete external force effect (He et al 1997). This method has a pseudo speed of sound, $c_s=\Delta r/(\sqrt{3} \Delta t)$, and a fluid kinematic viscosity, $\nu$=$(\tau - \Delta t/2) c_s^2$, where $\Delta t$ is the time step and $\Delta r$ is the unit lattice distance. The positivity of $\nu$ requires $\tau$$>$$\Delta t/2$. In the LB method, time and space are typically normalized by $\Delta t$ and $\Delta r$, respectively, such that $\Delta t_{LB}$=$\Delta r_{LB}$=1 are employed to advance equation \ref{eqn:lb1}. In the near incompressible limit (i.e., the Mach number, $Ma$=$u/c_s$$\ll$1), the LB equation recovers the Navier-Stokes equation \citep{Junk2003} with the equilibrium PDF given in terms of local macroscopic variables as
\begin{equation}
  f_i^{(0)}(\boldsymbol{r}, t) = \omega_i \rho [1 +  \frac{1}{c_s^2}(\boldsymbol{e}_i \cdot \boldsymbol{u}) + \frac{1}{2c_s^4}(\boldsymbol{e}_i \cdot \boldsymbol{u})^2 - \frac{1}{2c_s^2}(\boldsymbol{u} \cdot \boldsymbol{u})],
  \label{eqn:lb2}
\end{equation}
where $\omega_i$ denotes the set of lattice weights defined by the LB stencil in use. The macroscopic properties such as the fluid density, $\rho$, velocity, $\boldsymbol{u}$, and pressure, $p$, are obtained via moments of the equilibrium distribution functions as
\begin{equation}
  \rho=\sum_{i=1}^Q f_i^{(0)}(\boldsymbol{r},t),\ \ \boldsymbol{u}=\frac{1}{\rho}\sum_{i=1}^Q f_i^{(0)}(\boldsymbol{r},t) \boldsymbol{e}_i, \ \ p\mathsfbi{I}=\sum_{i=1}^Q f_i^{(0)}(\boldsymbol{r},t) \boldsymbol{e}_i\boldsymbol{e}_i-\rho\boldsymbol{u}\boldsymbol{u},
  \label{eqn:lb3}
\end{equation}
respectively. Here, $\mathsfbi{I}$ is the identity tensor and pressure can be related to density and the speed of sound through $p$=$\rho c_s^2$. For the D3Q19 stencil adopted in the current study, $Q$ is equal to 19. Along the rest, non-diagonal, and diagonal lattice directions, $\omega_i$ is equal to 1/3, 1/18, and 1/36, and $|\boldsymbol{e}_i|$ is equal to 0, $\Delta r/\Delta t$, and $\sqrt{2}(\Delta r/\Delta t)$, correspondingly. 

\subsection{Spectrin-link method}\label{sec:sl}
\noindent Modelling of the RBC membrane is through a coarse-grained spectrin-link (SL) method \citep{FedosovBJ2010,Pivkin2008} coupled with the LB method, which has been extensively validated with experimental results and proved to be a successful tool to capture both single RBC deformation and dynamics \citep{Reasor2012} and rheology of RBC suspensions at physiological haematocrit \citep{ReasorJFM2013}. 
In the SL model, the RBC membrane is modeled as a triangulated network with a collection of vertices mimicking actin vertex coordinates, denoted by $\{\boldsymbol{x}_n,\ n$$\in$$1,...,N\}$. The Helmholtz free energy of the network system, $E(\boldsymbol{x}_n)$, including in-plane, bending, volume and surface area energy components \citep{Dao2006}, is given by 
\begin{equation}
  E(\boldsymbol{x}_n) = 
  E_{IP} + E_{B} + E_{\Omega} + E_{A}.
  \label{eqn:sl1}
\end{equation}
Here, the in-plane energy, $E_{IP}$, characterizes the membrane shear modulus through a worm-like chain (WLC) potential \citep{Bustamante2003} coupled with a hydrostatic component \citep{FedosovBJ2010}. The bending energy, $E_B$, specifies the membrane bending stiffness, which is essential in characterizing the equilibrium RBC biconcave morphology \citep{Dao2006,FedosovBJ2010}. The volumetric contraint energy, $E_{\Omega}$, and the area constraint energy, $E_A$, preserve the RBC volume and area conservation, respectively, when subject to external forces. 

The dynamics of each vertice advance according to the Newton$\text{'}$s equations of motion,
\begin{equation}
  \frac{d\boldsymbol{x}_n}{dt}=\boldsymbol{v}_n,\ \ M\frac{d\boldsymbol{v}_n}{dt}=\mathbf{f}_n^{SL}+\mathbf{f}_n^{LB}+\mathbf{f}_n^{CC}
  \label{eqn:sl2}
\end{equation}
where $\boldsymbol{v}_n$ is the velocity of the vertice, $n$, and $M$ is the fictitious mass equal to the total mass of the cell divided by the number of vertices, $N$. The number of vertices used to discretize the RBC membrane is $N$=613, which has shown to yield adequate resolution to resolve hydrodynamic forces \citep{MacJFM2009} and capture single RBC dynamics \citep{Reasor2012} and concentrated RBC suspension rheology \citep{ReasorJFM2013} when coupled with the LB method. $\mathbf{f}_n^{LB}$ specifies the forces on the vertex due to the fluid-solid coupling. $\mathbf{f}_n^{CC}$ are the forces due to cell-cell interactions. The forces due to the Helmholtz free energy based on the SL model is determined by
\begin{equation}
  \mathbf{f}_n^{SL} = -\frac{\partial E(\boldsymbol{x}_n)}{\partial \boldsymbol{x}_n}.
  \label{eqn:sl3}
\end{equation}
The SL method is solved by integrating equations \ref{eqn:sl2} at each LB time step using a first-order-accurate forward Euler scheme in consistency with the LB evolution equation to avoid excessive computational expense \citep{Reasor2012,Liu2018a}.

\subsection{Langevin-dynamics approach}\label{sec:ld}
\noindent The nanoscale particle suspensions are resolved through a two-way coupled LB-LD method \citep{Liu2018a,Liu2018b}, which has been shown to correctly capture the theoretical Brownian diffusivity and long-range many-body HI. This approach treats suspended particles in Stokesian regimes as point particles, while the particles with Brownian effect are coupled to the non-fluctuating LB fluid in a two-way fashion through spatial extra/inter-polation schemes \citep{Ahlrichs1999,Peskin2002,Mynam2011}. The dynamics of the LD particles is described by the Langevin equation (LE),
\begin{equation}
  m \frac{d \boldsymbol{u}_p}{dt} = \boldsymbol{C}_p + \boldsymbol{F}_p + \boldsymbol{S}_p,
  \label{eqn:ld1}
\end{equation}
where $m$ is the mass of a single particle. The conservative force, $\boldsymbol{C}_p$, specifying the interparticle and particle-surface interaction forces, is determined by calculating the directional derivatives of the total potential energy $U_{total}$ as
\begin{equation}
  \boldsymbol{C}_p = - \frac{dU_{total}}{d\boldsymbol{r}_p},
  \label{eqn:ld2}
\end{equation}
where the details of $U_{total}$ are discussed in $\S \ref{sec:pp}$. The frictional force, $\boldsymbol{F}_p$, is assumed to be proportional to the relative velocity of the particle with respect to the local viscous fluid velocity \citep{Ahlrichs1998,Ahlrichs1999},
\begin{equation}
  \boldsymbol{F}_p = -\zeta[\boldsymbol{u}_p(t)-\boldsymbol{u}(\boldsymbol{r}_p,t)],
  \label{eqn:ld3}
\end{equation}
where $\boldsymbol{u}_p$ denotes the particle velocity, and $\boldsymbol{u}(\boldsymbol{r}_p,t)$ is the interpolated LB fluid velocity at the center of the particle. The friction coefficient, $\zeta$, is determined by the Stokes’ drag law, $\zeta=3\pi\mu d_p$, where $\mu$ is the dynamic viscosity of the suspending fluid. The stochastic force, $\boldsymbol{S}_p$, explicitly gives rise to the Brownian motion of the particle and satisfies the fluctuation-dissipation theorem (FDT) \citep{Kubo1966} by
\begin{equation}
  \langle S_{p,i}^{\alpha} (t)\rangle = 0, \ \ 
  \langle S_{p,i}^{\alpha} (t)S_{p,j}^{\beta} (t)\rangle = 2k_BT\zeta \delta_{ij}\delta_{\alpha\beta}\delta(t-t'),
  \label{eqn:ld5}
\end{equation}
where $i,j\in \{ x,y,z\}$, $\alpha$ and $\beta$ run through all the particle indices, $\delta_{ij}$ and $\delta_{\alpha\beta}$ are Kronecker deltas, $\delta(t-t')$ is the Dirac-delta function, $k_B$ is the Boltzmann constant and $T$ is the absolute temperature of the suspending fluid. The angle brackets denote the ensemble average over all the realizations of the random variables. Since only the time scales equal to and greater than the Brownian diffusion time scale is of interest, this study solves the over-damped discretized LE as suggested in \cite{Liu2018a,Liu2018b}.

\subsection{Fluid-solid coupling}\label{sec:fsi}
\noindent The coupling between fluid and RBC is accomplished through the Aidun, Lu, and Ding (ALD) fluid-solid interaction approach, of which the details are well documented in previous publications \citep{Aidun1998,Reasor2012,Aidun2010}. 
To resolve the NP dynamics subjected to the hydrodynamics and the long-ranged HI among NPs, the LD particle and the non-fluctuating LB fluid phase are coupled in a two-way fashion, as discussed and verified in previous studies \citep{Mynam2011,Liu2018a,Liu2018b}.
Specifically, the hydrodynamic force exerted on NP, $\boldsymbol{F}_p^H$, is systematically decomposed into frictional and stochastic components as
\begin{equation}
  \boldsymbol{F}_p^H = \boldsymbol{F}_p+\boldsymbol{S}_p=-\zeta[\boldsymbol{u}_p(t)-\boldsymbol{u}(\boldsymbol{r}_p,t)]+\boldsymbol{S}_p,
  \label{eqn:fsi3}
\end{equation}
where the fluid velocity at the particle site, $\boldsymbol{u}(\boldsymbol{r}_p,t)$, is interpolated based on surrounding LB velocities and applied to update the LD particle dynamics through equation \ref{eqn:ld1}. The weighting functions, $w(\boldsymbol{r},\boldsymbol{r}_p)$, for interpolation is constructed using a trilinear scheme \citep{Ahlrichs1998,Liu2018a}. Since $\boldsymbol{F}_p$ and $\boldsymbol{S}_p$ are both originated from the `collision' between NP and liquid molecules, $\boldsymbol{F}_p^H$ (instead of $\boldsymbol{F}_p$) is assigned back to the fluid phase to satisfy momentum conservation. The same weighting function is then applied to constructing the local forcing source term as
\begin{equation}
  f_{i}^S(\boldsymbol{r},t) = - \frac{w(\boldsymbol{r},\boldsymbol{r}_p) \omega_i \boldsymbol{F}_p^H \cdot \boldsymbol{e}_i}{c^2_s\Delta r^3},
  \label{eqn:fsi4}
\end{equation}
which is adopted by equation \ref{eqn:lb1} to update the local hydrodynamics. The coupled LB-LD method, similar to the external boundary force (EBF) method \citep{Wu2008}, modifies the conventional LB evolution equation into equation \ref{eqn:lb1} by adding the forcing distribution function $f_i^S(\boldsymbol{r},t)$, which is shown to approximate the Navier-Stokes equation in the macroscopic scale (Guo et al 2002).

\subsection{Contact modeling}\label{sec:pp}
\noindent 
The contact model for RBCs that specifies $\mathbf{f}_n^{CC}$ is based on the subgrid contact functions originally formulated in \citet{DingAidun2003} and later improved by \citet{MacJFM2009} and \citet{ClausenJFM2011}. It prevents the RBC from overlapping when cell-cell membrane separation is below one LB lattice spacing. In this model, the lubrication term is replaced with an exponential contact function to avoid numerical instability driven by the singular nature of the lubrication hydrodynamics and the discrete nature of the interparticle seperation calculation, as explained in detail in \cite{MacJFM2009,Clausen2010pof}.
The rheological insensitivities of the contact model to model parameters have been discussed in detail by \citet{ClausenJFM2011}. This contact model has also been previously applied in the characterization of rheological properties of concentrated deformable capsule \citep{ClausenJFM2011} and RBC \citep{ReasorJFM2013} suspensions with satisfactory agreement with experimental measurements. 
Although lubrication force is shown to play some role in particle self-diffusion in sheared monodisperse rigid particle suspensions \citep{FossJR2000,FossJFM2000}, we anticipate such effect to be insignificant in the current bidisperse suspension system given the NP diffusion is dominant by the RBC-NP interactions.

The NP-RBC contact model that provides $U_{total}$ is based on the Morse potential documented in \cite{Liu2004}, which is an empirical model for particle-particle interactions that can be calibrated to match experimental measurements \citep{Neu2002,Liu2004}. Due to the variety and somewhat lack of statistics for actual NP-RBC short-distance interaction, this study employs the measured cell-cell interaction potential \citep{Neu2002} for the NP-RBC interactions. The potential parameters are specified according to \cite{Liu2004} but with a cut-off distance selected to only preserve the repulsive effect. 
The detailed formulation of the Morse potential has been discussed in \S\ Appendix \ref{appA} with sensitivity analysis performed showing that the computation of NP diffusivity is insensitive to the change of the model parameters up to 60\%. This suggests that the NP diffusion is largely driven by the hydrodynamic interaction rather than the direct contact between NP and RBC membrane.
The short-distance NP-NP interaction is neglected due to the extreme dilution of the NP concentration ($\ll$1$\%$) considered.

This work does not attempt to model any adhesive forces between RBCs since above the shear rate of $5\ s^{-1}$ the aggregation of RBCs is not significant \citep{FedosovPNAS2011}. Although the adhesion or uptake of NPs to cells may be influential to the NP dispersive behavior \citep{Shang2014}, it is however not within the scope of this study.

\subsection{Lees-Edwards boundary condition}\label{sec:lebc}
\noindent Since the primary focus of this study is on the particle bulk diffusive behavior subject to no wall effect, simulations are performed in an unbounded, triply periodic cubic domain where a constant shear rate is imposed through the Lees-Edwards boundary condition (LEbc) \citep{LEbc1972}. This method, originally developed for molecular dynamics simulations, was extended to the LB method by \cite{Wagner2002} and later applied to deformable suspensions on parallel computing architectures by \cite{ClausenJFM2011,ReasorJFM2013}. 
In addition to the operations associated with regular periodic boundary conditions, both the particle (NP and RBC) phase and the fluid phase undergo a shift in position and velocity according to the LEbc scheme as they cross the top ($+y$) or bottom ($-y$) boundary.

\subsection{Characterization of the particle diffusion tensor}\label{sec:pdt}
\noindent The presence of shearing flow imposes a convective effect on the particle suspension and complicates the characterization of particle diffusion tensor. The major difficulty lies in determination of the longitudinal diffusivity ($\mathsfi{D}_{xx}^{\infty}$) and off-diagonal diffusivity ($\mathsfi{D}_{xy}^{\infty}$), which require careful subtraction of the affine particle displacement. The diffusion tensor in sheared monodisperse colloidal suspensions have been successfully quantified by sampling the non-affine particle mean square displacements (MSDs) \citep{Morris1996,FossJFM1999,FossJFM2000,Zia2010}. Therefore, this study calculates the long-time NP diffusion tensor, $\mathsfbi{D}^{\infty}$, in the form of
\begin{equation}
  \mathsfbi{D}^{\infty} = 
\begin{pmatrix}
\mathit{\mathsfi{D}}_{xx}^{\infty} & \mathit{\mathsfi{D}}_{xy}^{\infty} &                        0  \\ 
\mathit{\mathsfi{D}}_{yx}^{\infty} & \mathit{\mathsfi{D}}_{yy}^{\infty} &                        0  \\ 
                       0 &                        0 & \mathit{\mathsfi{D}}_{zz}^{\infty}
\end{pmatrix}
  \label{eqn:Dtensor1}
\end{equation}
where each non-zero diffusion component is calculated by
\begin{subequations}
\begin{equation}
  \mathit{\mathsfi{D}}_{xx}^{\infty} = \frac{1}{2} \frac{\mathit{d}}{\mathit{dt}} 
  [ \langle \mathit{x}^{na}(t) \mathit{x}^{na}(t) \rangle -
    \langle \mathit{x}^{na}(t) \rangle \langle \mathit{x}^{na}(t) \rangle ], 
  \label{eqn:Dtensor2}
\end{equation}
\begin{equation}
  \mathit{\mathsfi{D}}_{xy}^{\infty} = \frac{1}{2} \frac{\mathit{d}}{\mathit{dt}} 
  [ \langle \mathit{x}^{na}(t) \mathit{y}(t) \rangle -
    \langle \mathit{x}^{na}(t) \rangle \langle \mathit{y}(t) \rangle ], 
  \label{eqn:Dtensor3}
\end{equation}
\begin{equation}
  \mathit{\mathsfi{D}}_{yy}^{\infty} = \frac{1}{2} \frac{\mathit{d}}{\mathit{dt}} 
  [ \langle \mathit{y}(t) \mathit{y}(t) \rangle -
    \langle \mathit{y}(t) \rangle \langle \mathit{y}(t) \rangle ], 
  \label{eqn:Dtensor4}
\end{equation}
\begin{equation}
  \mathit{\mathsfi{D}}_{zz}^{\infty} = \frac{1}{2} \frac{\mathit{d}}{\mathit{dt}} 
  [ \langle \mathit{z}(t) \mathit{z}(t) \rangle -
    \langle \mathit{z}(t) \rangle \langle \mathit{z}(t) \rangle ], 
  \label{eqn:Dtensor5}
\end{equation}
\end{subequations}
as $\mathit{t}\rightarrow\infty$. Here, and hereinafter, the angle brackets denote an ensemble average over all NPs in the system; $x$, $y$ and $z$ denote the absolute displacement of NP in three principal flow directions, i.e., longitudinal, velocity-gradient, and vorticity direction, respectively. The diffusivity tensor is symmetric, thus $\mathsfi{D}_{xy}^{\infty}$ and $\mathsfi{D}_{yx}^{\infty}$ are equal; $\mathsfi{D}_{xz}^{\infty}$ and $\mathsfi{D}_{yz}^{\infty}$ (and the transpose) are insignificant \citep{BradyMorris1997}, which is also confirmed in our simulation. The non-affine displacement, $x^{na}$, is calculated by subtracting the absolute displacement with its affine component, $x^a$, i.e. $x^{na}(t)=x(t)-x^a(t)$, where $x^a(t)=\int_{0}^{t}\dot{\gamma}y(\tau)d\tau$ and $\dot{\gamma}$ is the imposed shear rate through LEbc. When calculating the absolute displacement of NPs that undergo a shift of position due to the LEbc, the particle reference position is shifted accordingly to subtract the shift effect. Particle displacements are followed every $\sim$$0.06\ \dot{\gamma}t$ to ensure the growth of MSDs captured with adequate accuracy. For clarity, the superscript of the affine displacement and the expectation terms are both dropped in the MSD notations as follows.


\section{Results}\label{sec:results}
In this section, we first formulate the simulations based on both physiological and numerical rationales.  As follows, the transient growth of the NP mobility is discussed to understand the short-time response of the NP dispersive behavior at various haematocrit and shear rate. Then, focus will be shifted to understanding the NP long-time diffusive behavior under different hemorheological conditions with appropriate scaling, where the simulation results are also compared with available experimental data. To gain insight into the mechanisms governing the nonlinear haematocrit and shear-rate dependence of the NP diffusion tensor, we visualize the NP-RBC microstructure and carry out numerical experiments. Eventually, we construct empirical correlations for the complete NP diffusion tensor. 

\subsection{Problem formulation}\label{sec:prob}
\begin{table}
 \begin{center}
  \begin{tabular}{ccccccccccc}
    $\dot{\gamma}\ [s^{-1}]$ & $\phi$ & $a_1/a_2$ & $N^{RBC}$ & $N^{NP}$ & $\Pen$ & $Ca_G$ & $\mathsfi{D}_{xx}^{\infty}/\mathsfi{D}^B$ & $D_{yy}^{\infty}/D^B$ & $\mathsfi{D}_{zz}^{\infty}/\mathsfi{D}^B$ & $\mathsfi{D}_{xy}^{\infty}/\mathsfi{D}^B$ \\[4pt]
       10 & 0.0 & 0.017 &   0 & 5\ 000 & 0.0066 & 0.0055 & 1.0 & 1.0 & 1.0 &  0.0\\[1pt]
       10 & 0.1 & 0.017 &  52 & 5\ 000 & 0.0066 & 0.0055 & 1.4 & 1.1 & 1.1 &  0.05\\[1pt]
       10 & 0.2 & 0.017 & 104 & 5\ 000 & 0.0066 & 0.0055 & 2.5 & 1.1 & 1.2 &  0.03\\[1pt]
       10 & 0.3 & 0.017 & 156 & 5\ 000 & 0.0066 & 0.0055 & 2.6 & 1.1 & 1.3 &  0.05\\[1pt]
       10 & 0.4 & 0.017 & 208 & 5\ 000 & 0.0066 & 0.0055 & 3.2 & 1.2 & 1.4 &  0.03\\[1pt]
       
       30 & 0.0 & 0.017 &   0 & 5\ 000 & 0.020 & 0.017 & 1.0 & 1.0 & 1.0 &  0.0\\[1.pt]
       30 & 0.1 & 0.017 &  52 & 5\ 000 & 0.020 & 0.017 & 2.6 & 1.3 & 1.2 &  -0.1\\[1.pt]
       30 & 0.2 & 0.017 & 104 & 5\ 000 & 0.020 & 0.017 & 4.8 & 1.6 & 1.3 & -0.2\\[1.pt]
       30 & 0.3 & 0.017 & 156 & 5\ 000 & 0.020 & 0.017 & 6.7 & 1.9 & 1.5 & -0.2\\[1.pt]
       30 & 0.4 & 0.017 & 208 & 5\ 000 & 0.020 & 0.017 & 7.2 & 2.0 & 1.8 & -0.4\\[1.pt]
       
       100 & 0.0 & 0.017 &   0 & 5\ 000 & 0.066 & 0.055 & 1.1 & 1.0 & 1.1 &  0.0\\[1.pt]
       100 & 0.1 & 0.017 &  52 & 5\ 000 & 0.066 & 0.055 & 7.3 & 1.9 & 1.6 &  -0.6\\[1.pt]
       100 & 0.2 & 0.017 & 104 & 5\ 000 & 0.066 & 0.055 & 14.5 & 2.9 & 2.0 & -1.0\\[1.pt]
       100 & 0.3 & 0.017 & 156 & 5\ 000 & 0.066 & 0.055 & 18.9 & 3.7 & 3.1 & -1.3\\[1.pt]
       100 & 0.4 & 0.017 & 208 & 5\ 000 & 0.066 & 0.055 & 29.2 & 4.1 & 3.8 & -2.0\\[1.pt]
       
       300 & 0.0 & 0.017 &   0 & 5\ 000 & 0.198 & 0.165 & 1.0 & 1.0 & 1.0 &  0.0\\[1.pt]
       300 & 0.1 & 0.017 &  52 & 5\ 000 & 0.198 & 0.165 & 28.0 & 3.3 & 2.9 &  -2.2\\[1.pt]
       300 & 0.2 & 0.017 & 104 & 5\ 000 & 0.198 & 0.165 & 67.9 & 5.7 & 3.7 & -4.3\\[1.pt]
       300 & 0.3 & 0.017 & 156 & 5\ 000 & 0.198 & 0.165 & 133.0 & 8.8 & 6.3 & -7.6\\[1.pt]
       300 & 0.4 & 0.017 & 208 & 5\ 000 & 0.198 & 0.165 & 216.4 & 10.8 & 8.0 & -9.1\\[1.pt]
       
       1\ 000 & 0.0 & 0.017 &   0 & 5\ 000 & 0.66 & 0.55 &    1.0 &  1.0 &  1.0 &   0.0\\[1.pt]
       1\ 000 & 0.1 & 0.017 &  52 & 5\ 000 & 0.66 & 0.55 &   73.8 &  7.0 &  8.1 & -10.1\\[1.pt]
       1\ 000 & 0.2 & 0.017 & 104 & 5\ 000 & 0.66 & 0.55 &  211.1 & 12.5 & 10.5 & -14.9\\[1.pt]
       1\ 000 & 0.3 & 0.017 & 156 & 5\ 000 & 0.66 & 0.55 &  674.2 & 19.1 & 16.1 & -26.9\\[1.pt]
       1\ 000 & 0.4 & 0.017 & 208 & 5\ 000 & 0.66 & 0.55 & 2095.9 & 23.2 & 18.8 & -38.1\\[1.pt]
       
       2\ 000 & 0.0 & 0.017 &   0 & 5\ 000 & 1.32 & 1.10 &     0.9 &  1.0 &  1.0 &   0.0\\[1.pt]
       2\ 000 & 0.1 & 0.017 &  52 & 5\ 000 & 1.32 & 1.10 &   122.8 & 12.5 & 15.6 & -19.9\\[1.pt]
       2\ 000 & 0.2 & 0.017 & 104 & 5\ 000 & 1.32 & 1.10 &   350.0 & 20.7 & 21.5 & -25.7\\[1.pt]
       2\ 000 & 0.3 & 0.017 & 156 & 5\ 000 & 1.32 & 1.10 &  1379.3 & 29.6 & 26.3 & -50.0\\[1.pt]
       2\ 000 & 0.4 & 0.017 & 208 & 5\ 000 & 1.32 & 1.10 &  4150.3 & 36.6 & 29.4 & -82.8\\[1.pt]
       
       10\ 000 & 0.0 & 0.017 &   0 & 5\ 000 & 6.60 & 5.52 &     1.1 &  1.0 &  1.0 &   0.0\\[1.pt]
       10\ 000 & 0.1 & 0.017 &  52 & 5\ 000 & 6.60 & 5.52 &   606.7 & 40.9 & 77.3 & -73.1\\[1.pt]
       10\ 000 & 0.2 & 0.017 & 104 & 5\ 000 & 6.60 & 5.52 &  1363.7 & 44.9 & 86.4 & -81.8\\[1.pt]
       10\ 000 & 0.3 & 0.017 & 156 & 5\ 000 & 6.60 & 5.52 &  2788.5 & 64.0 & 110.8 & -138.0\\[1.pt]
       10\ 000 & 0.4 & 0.017 & 208 & 5\ 000 & 6.60 & 5.52 &  7575.6 & 80.4 & 120.3 & -220.9\\[1.pt]
  \end{tabular}
  \caption{Simulation data. Each case is simulated independently in a LB domain of size 128$\times$128$\times$80 at specified $\Pen$, $Ca_G$ and $\phi$. NP long-time diffusivities, $\mathsfi{D}_{xx}^{\infty}$, $\mathsfi{D}_{yy}^{\infty}$, $\mathsfi{D}_{zz}^{\infty}$ and $\mathsfi{D}_{xy}^{\infty}$, normalized by the theoretical Brownian diffusivity, $\mathsfi{D}^B$, are listed for all cases. The unlisted off-diagonal diffusivities, $\mathsfi{D}_{xz}^{\infty}$ and $\mathsfi{D}_{yz}^{\infty}$, are found to be negligible. The measured diffusivities have a standard deviation less than 5$\%$. RBC has an effective radius of $a_2=2.9\ \mu m$. Brownian diffusivity is calculated by $\mathsfi{D}^B=k_BT/6\mu \pi a_1$ at temperature $T=310\ K$. }{\label{tab:srht}}
 \end{center}
\end{table}
\noindent The apparent diffusivity of NPs in unbounded blood flow under simple shear is determined by NP radius, $a_1$, shear rate, $\dot{\gamma}$, and haematocrit, $\phi$, with NP concentration in the dilute regime. The relevant dimensionless parameters (besides $\phi$) primarily include the NP P\'eclet number,  
\begin{equation}
  \Pen=\frac{\dot{\gamma}a_1^2}{\mathsfi{D}^B}, 
  \label{eqn:Pe}
\end{equation}
expressing the ratio of shear-induced diffusion to Brownian diffusion, and the RBC capillary number,
\begin{equation}
  Ca_G=\frac{\mu\dot{\gamma}a_2}{G}, 
  \label{eqn:Ca}
\end{equation}
quantifying the competition between the fluid viscous stress and the membrane elastic stress. Here, $\mu$ is the dynamic viscosity of suspending plasma, $a_2$ is the effective radius of RBC and $G$ is the elastic shear modulus of the RBC membrane; $\mathsfi{D}^B$ is the Brownian diffusivity, which is determined by the Stokes-Einstein relation, $\mathsfi{D}^B$=$k_B T/6\pi \mu a_1$, where $k_B$ is Boltzmann's constant and $T$ is the absolute temperature. The NP P\'eclet number quantifies the severity of the NP Brownian effect, while the RBC capillary number determines the deformability of the RBC capsule.

To obtain appropriate scaling relations, we performed a large number of independent 3D simulations. Table \ref{tab:srht} lists all the simulation parameters and the measured NP diffusivities. A wide range of shear rate ($10$$\leq$$\dot{\gamma}$$\leq $$10\ 000\ s^{-1}$) and haematocrit ($0$$\leq$$\phi$$\leq$$0.4$) with physiological relevance \citep{Lipo2005,Popel2005} is covered. Cases with $\dot{\gamma}$=$10\ 000\ s^{-1}$ are to match certain vascular pathological conditions, e.g., the high shear induced thrombosis \citep{Casa2017}. For discussions in this section, NP size is set to $2a_1$=$100\ nm$. RBCs are assumed to be in healthy state with an effective radius $a_2$=$2.9\ \mu m$ and a membrane shear modulus $G$=$0.0063\ dynes/cm$. The absolute temperature is set to $T$=$\SI{37}{\celsius}$, at which the plasma has a viscosity $\mu$=$1.2\ cP$ and a density $\rho$=$1.0\ g/cm^3$. The viscosity ratio of RBC cytoplasm to plasma is set to the physiological value $\lambda$=$5.0$. The density of cytoplasm is set to that of the plasma. The corresponding $\Pen$ and $Ca_G$ lie in the range of $0.0066\leq$$\Pen$$\leq6.60$ and $0.0055\leq$$Ca_G$$\leq5.52$, respectively. All simulations are formulated by matching the dimensionless group, i.e., $\phi$, $\Pen$ and $Ca_G$.

\begin{figure}
  \centerline{\includegraphics[width=0.98\columnwidth]{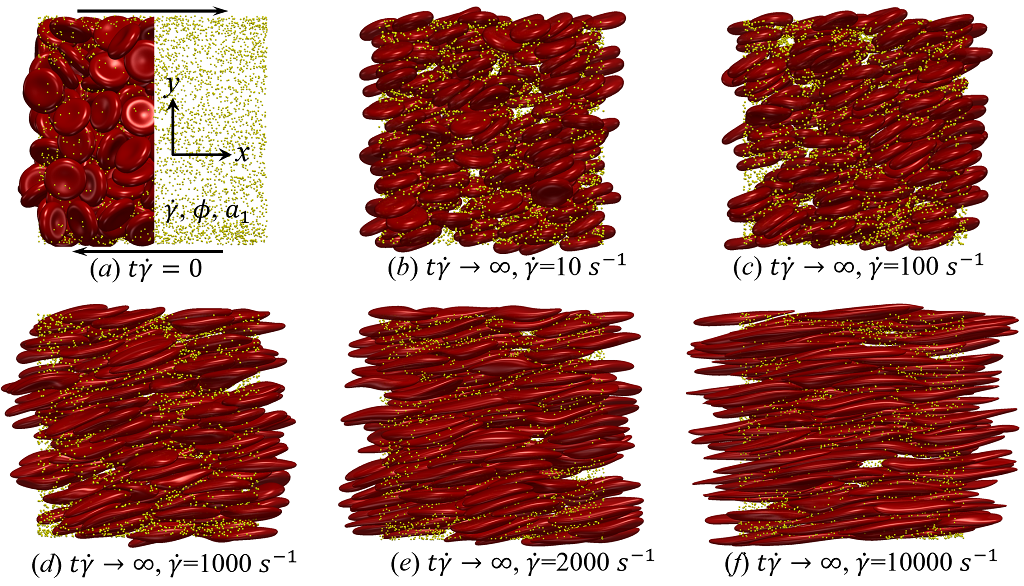}}
  \caption{(a) The RBC-NP complex configuration at $t\dot{\gamma}$=$0$. NPs and RBCs are randomly and uniformly mixed in an unbounded shear flow. RBCs are shown partially in the domain to visualize the NP phase. The athermal system is determined by shear rate, $\dot{\gamma}$, haematocrit, $\phi$ and NP radius, $a_1$. (b-f) The RBC-NP complex configuration in the long-time scale regime under a shear rate of $\dot{\gamma}=$ (b) 10, (c) 100, (d) $1\ 000$, (e) $2\ 000$ and (f) $10\ 000\ s^{-1}$ with $\phi=0.4$.}
\label{fig:RBCNP1}
\end{figure}

Simulations are initiated by imposing steady shear flow on the uniformly, randomly mixed NPs and RBCs at specific shear rate in a LEbc computational domain, as shown in figure \ref{fig:RBCNP1} (a). The domain has a dimension of $128\times128\times80$ ($42.7\times42.7\times26.7\ \mu m^3$) in longitudinal ($x$), velocity-gradient ($y$) and vorticity ($z$) directions, respectively. This LB domain size matches the highest resolution applied for the rheological characterization of cellular blood flow under shear by \citet{ReasorJFM2013}. The selected LB grid resolution ($300\ nm$ per lattice unit) and the equilibrium RBC mesh size (1.5 lattice units per link length), has previously proven to be fine enough to capture both the single RBC dynamics \citep{MacJFM2009,Reasor2012} and the rheological properties of concentrated cellular blood flow \citep{MacJFM2009,ReasorJFM2013}.

To obtain converged long-time diffusivity, sufficient strains ($t\dot{\gamma}$$\sim$$1\ 000$) and a large number of particles (5\ 000 NPs and up to 208 RBCs) are employed for each simulation. The resolution of these simulations in terms of strain units and number of particles is on the high end compared to other numerical studies on particle diffusion in colloidal/non-colloidal suspensions \citep{FossJFM1999,FossJFM2000,Sierou2004,Yeo2010,ClausenJFM2011,Gross2014,Mountrakis2016}. As will be discussed in \S \ref{sec:sr}, the selected resolution produces good agreement between the simulation results and the available experimental data. 

Simulations are performed on the Intel Xeon Skylake nodes of the TACC (Texas Advanced Computing Center) Stampede-2 system where each node features 48 cores and a 2.1 GHz clock rate. For the case at $\phi=0.4$ with 208 RBCs and 5\ 000 NPs, each run takes $\sim$$168$ hours on 32 cores ($\sim$$5\ 376$ core hours) to accomplish 1\ 000 strains ($t\dot{\gamma}$). The total computational cost for the 35 independent cases listed in table \ref{tab:srht} is approximately 140\ 000 core hours.

\subsection{Temporal growth of NP transient mobility}\label{sec:msd}
\noindent 
To understand the short-time dispersive characteristics of the NP phase, we examine the transient behavior of NP mobility by tracking the temporal growth of NP MSDs. Cross-stream MSDs, $\langle y(t)y(t)\rangle /a_1^2$ and $\langle z(t)z(t)\rangle /a_1^2$, are sampled from the initial configuration ($t\dot{\gamma}$=$0$) where NPs and RBCs are randomly and uniformly mixed, as shown in figure \ref{fig:RBCNP1} (a). Longitudinal and off-diagonal MSDs, $\langle x(t)x(t)\rangle /a_1^2$ and $\langle x(t)y(t)\rangle /a_1^2$, are sampled starting from a long-time configuration ($t\dot{\gamma}$$=$$400$) to avoid extra convective effect caused by the transient elongation and reorientation of RBCs. Such transient effects are found to introduce extra affine displacement leading to $\sim$$t^3$ growth of MSDs and hence jeopardize the measurement of the long-time diffusivity in the $x$-relevant directions. The necessity of eliminating affine effect for calculating streamwise diffusivities is elaborated by \citet{FossJFM2000} in the context of sheared colloidal suspensions.

\begin{figure}
  \centering
  \includegraphics[width=0.24\columnwidth]{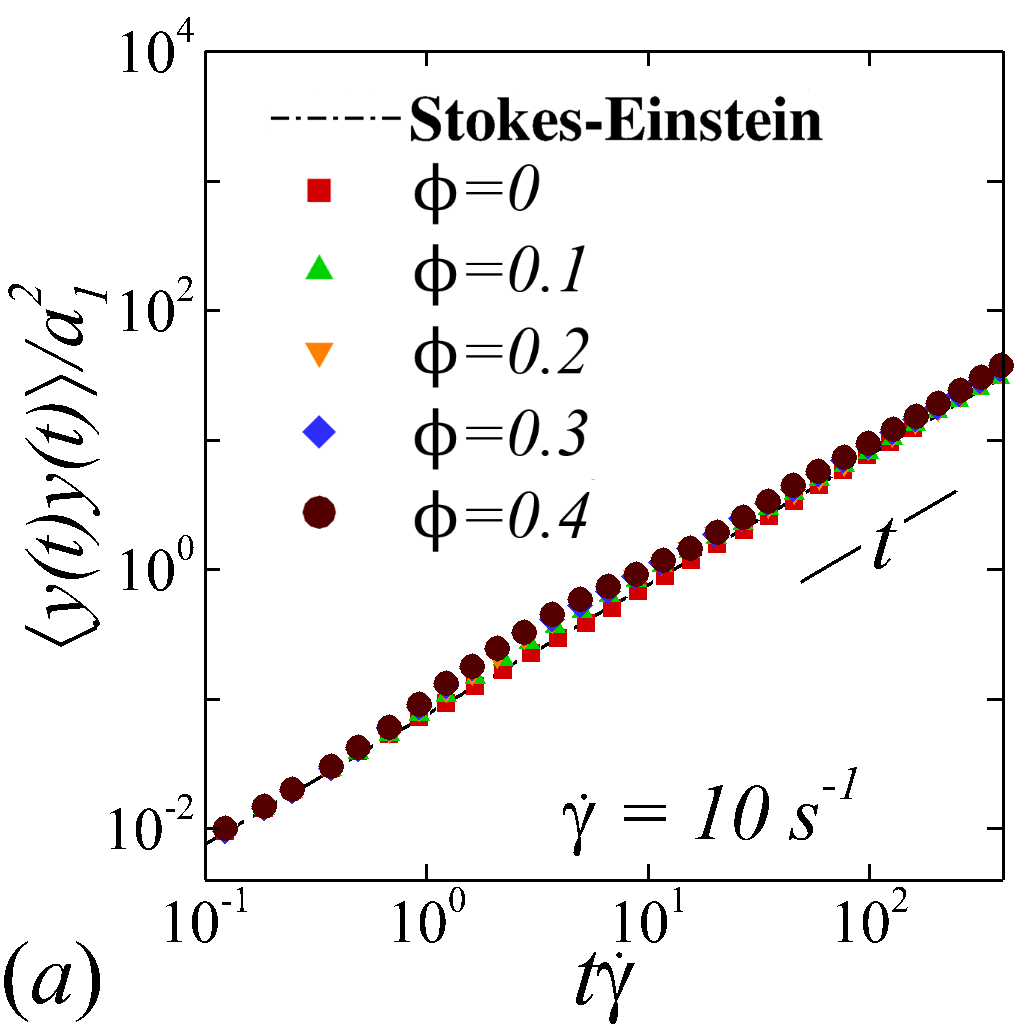}
  \includegraphics[width=0.24\columnwidth]{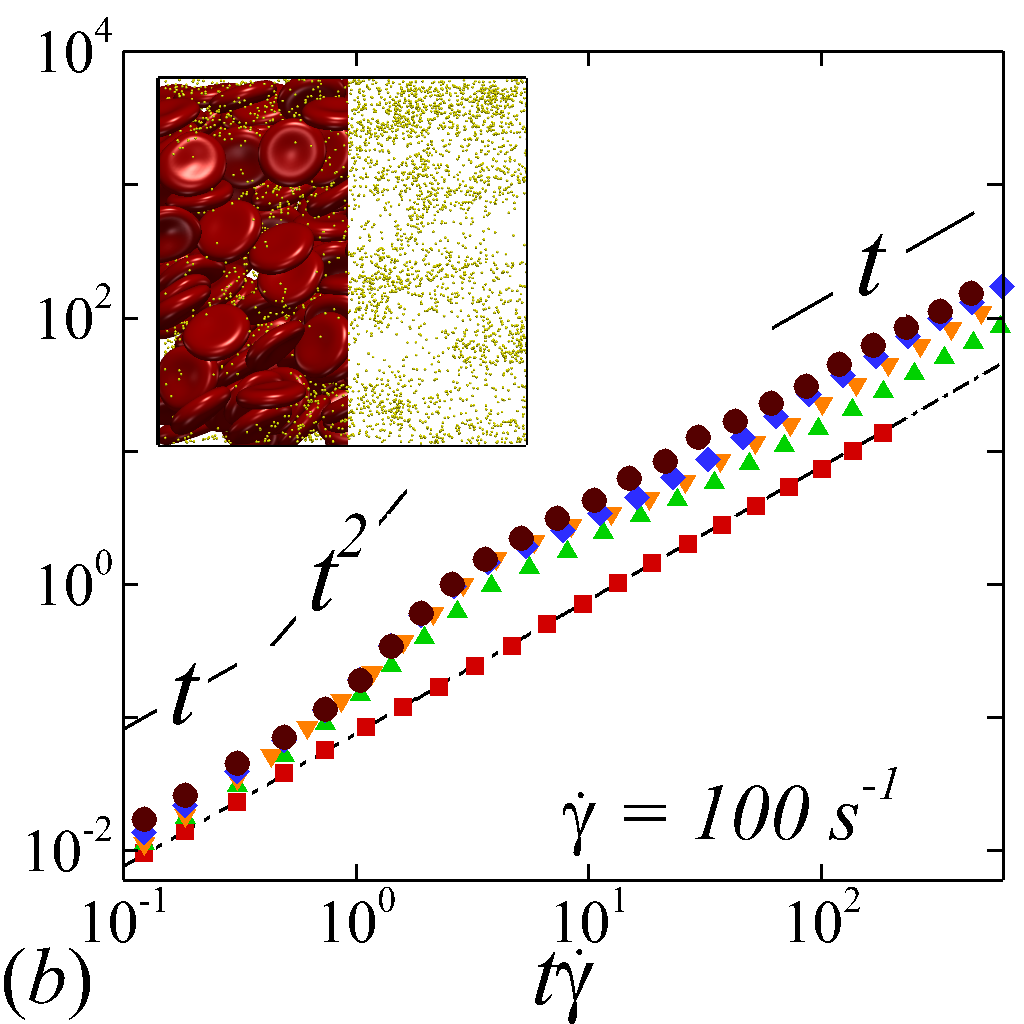}
  \includegraphics[width=0.24\columnwidth]{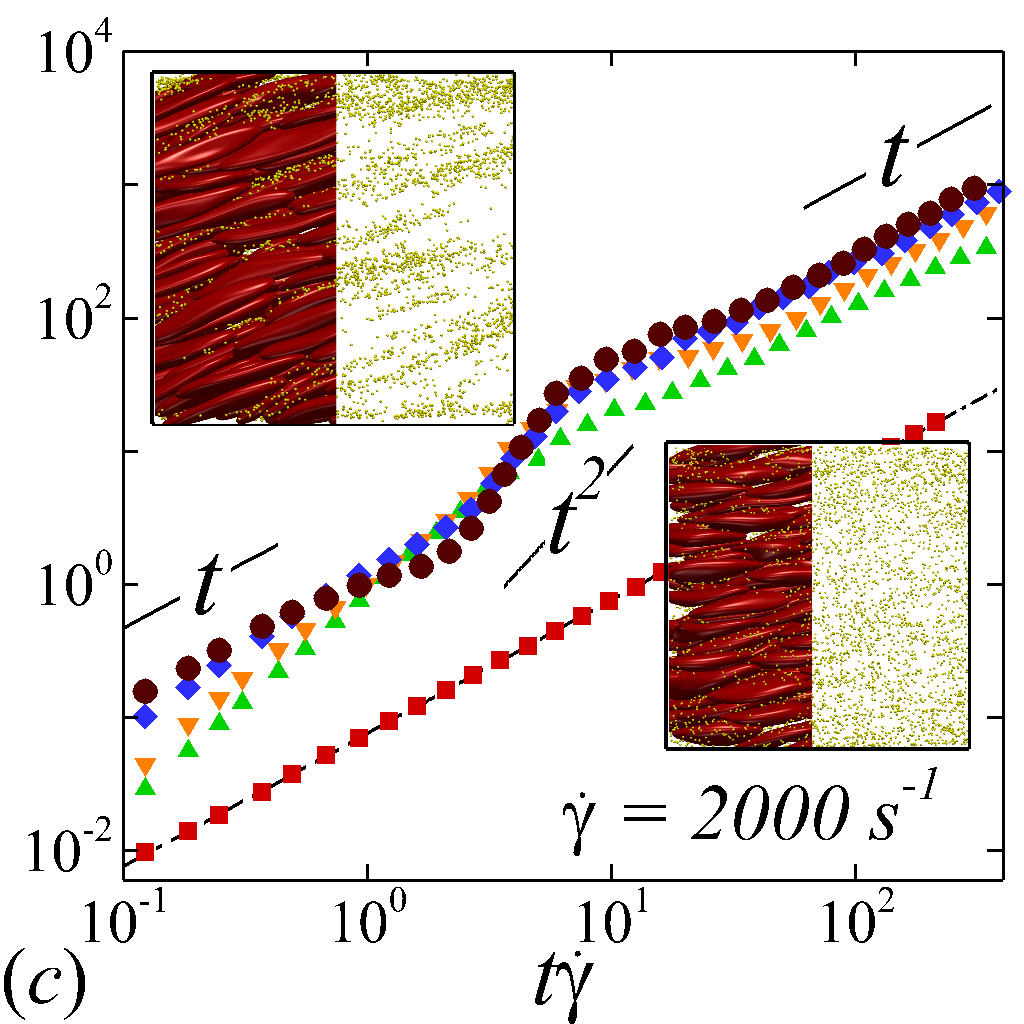}
  \includegraphics[width=0.24\columnwidth]{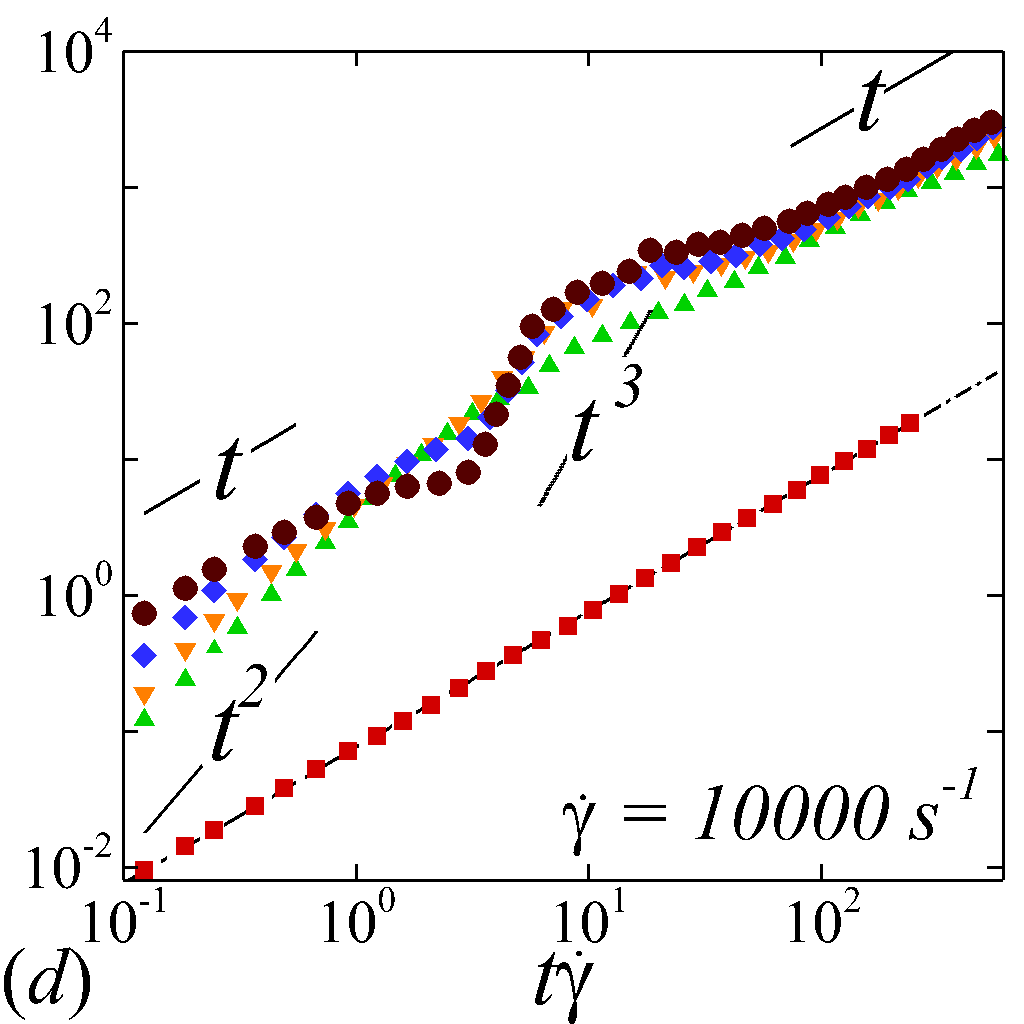}
  \includegraphics[width=0.24\columnwidth]{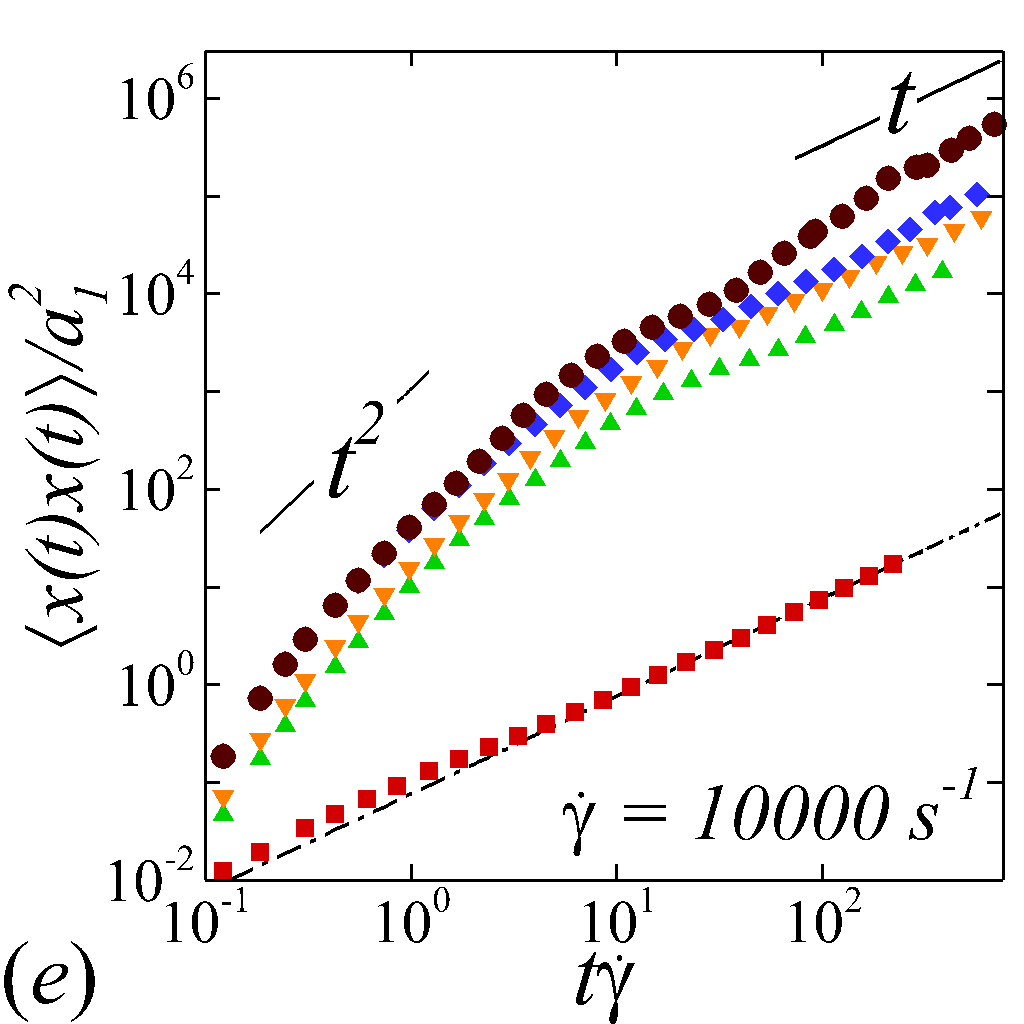}
  \includegraphics[width=0.24\columnwidth]{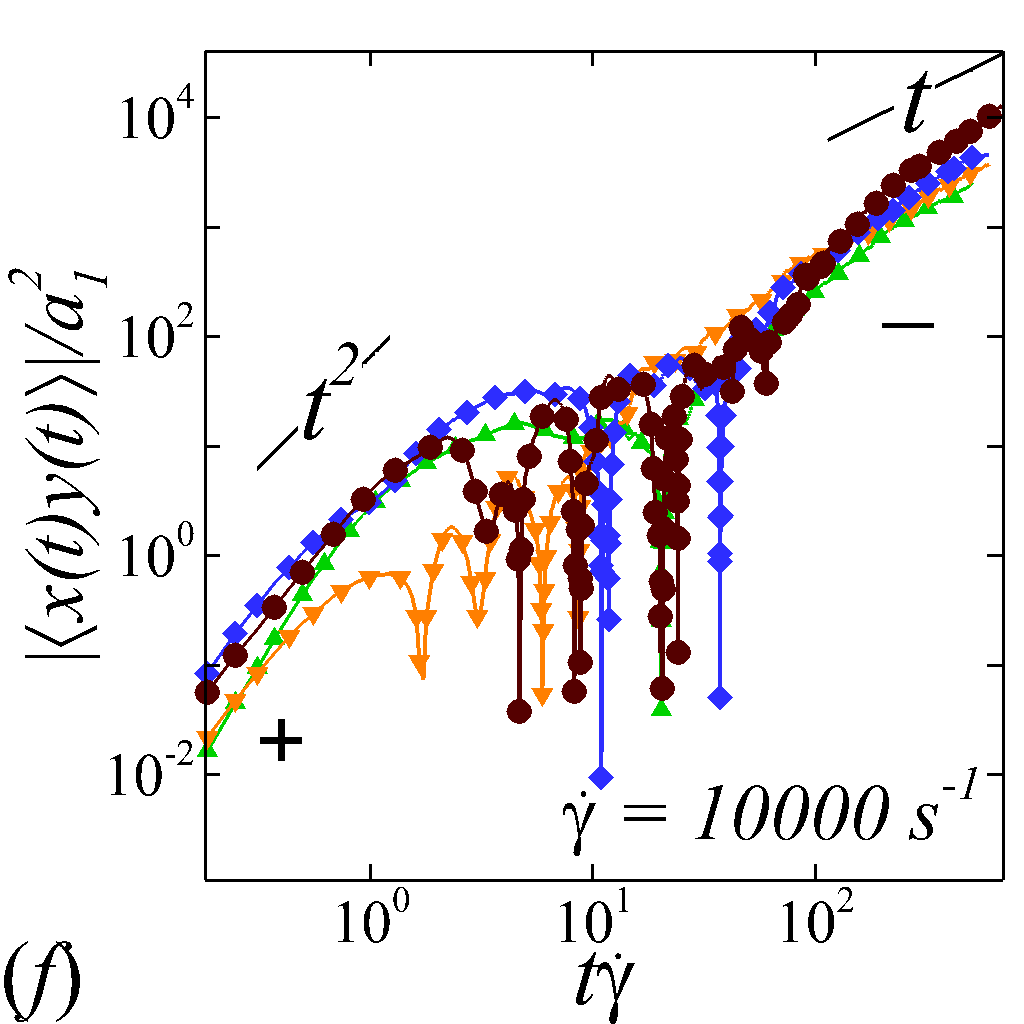}
  \includegraphics[width=0.24\columnwidth]{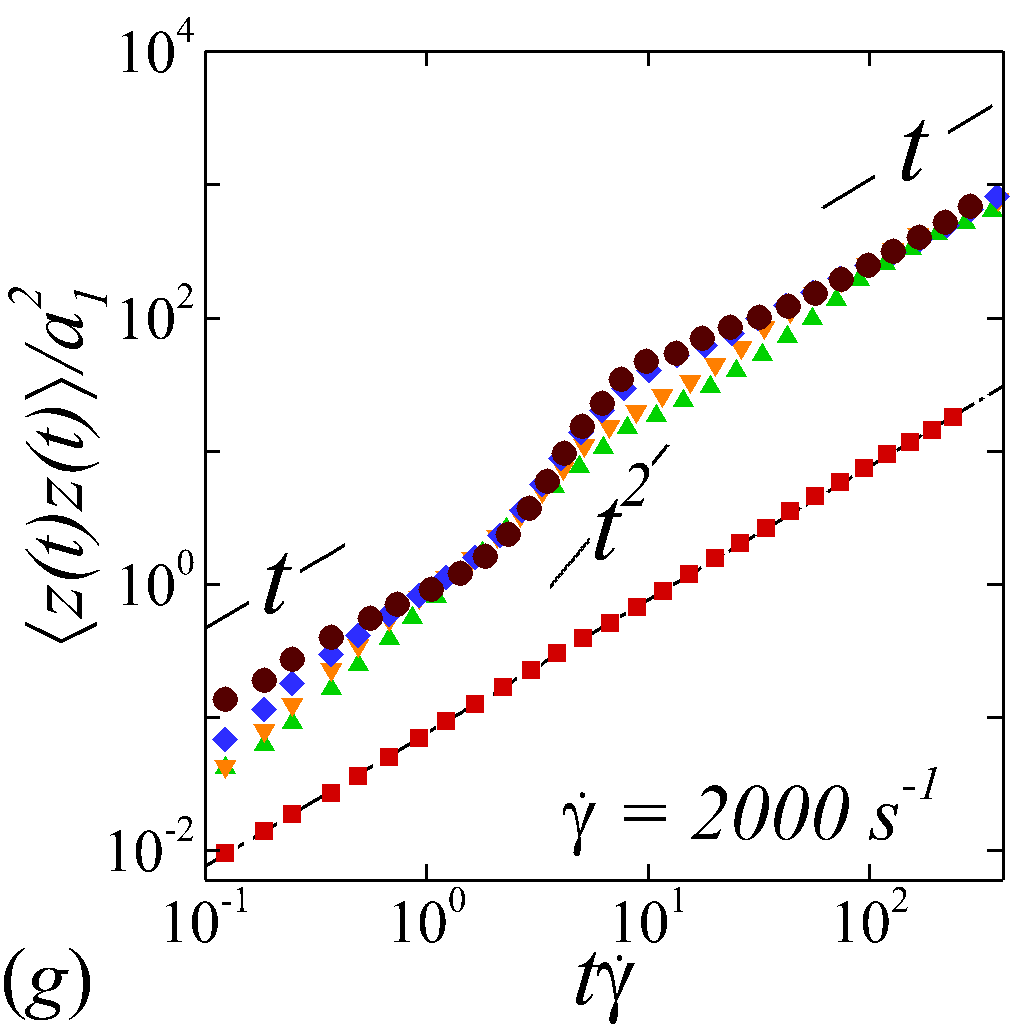}
  \includegraphics[width=0.24\columnwidth]{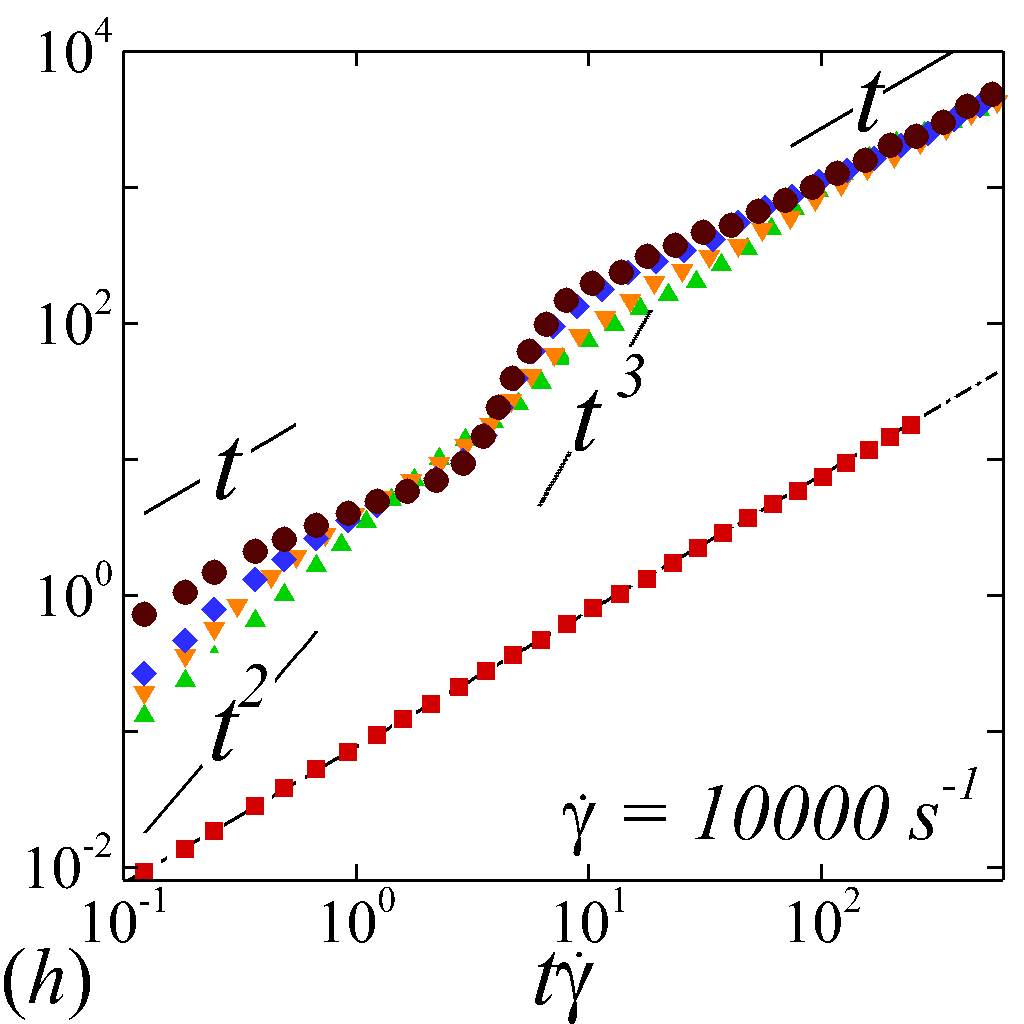}  
  \caption{Normalized MSDs plotted against relative shear strains in logarithm scale at various haematocrits for different shear rate. (a), (b), (c) and (d) are concerned with evolutions of $\langle y(t)y(t)\rangle /a_1^2$; (e) and (f) depict evolutions of $\langle x(t)x(t)\rangle /a_1^2$ and $|\langle x(t)y(t)\rangle|/a_1^2$, respectively; (g) and (h) depict evolutions of $\langle z(t)z(t)\rangle /a_1^2$. The imposed shear rate is denoted on each plot. Inset of (b) shows the RBC-NP distribution at $t\dot{\gamma}$$\sim$1 under $\dot{\gamma}=100 s^{-1}$. Insets of (c) shows the RBC-NP distribution at $t\dot{\gamma}$$\sim$1 (upper) and $t\dot{\gamma}$$\sim$100 (lower) under $\dot{\gamma}=2000 s^{-1}$ and $\phi=0.4$. A NP radius of $a_1$=50 $nm$ is used for normalization. Cross-stream MSDs (a-d) and (g,h) are sampled starting from $t\dot{\gamma}$$=$0 to capture the transient NP dispersive behavior; stream-wise and off-diagonal MSDs are sampled starting from $t\dot{\gamma}$=400 to avoid the initial extra affine displacement induced by RBC elongation and orientation.} 
\label{fig:MSD}
\end{figure}

Figure \ref{fig:MSD} plots the particle normalized MSDs growth with respect to the relative strains sampled.
At $\phi$=$0$, all diagonal MSDs grow according to the Stokes-Einstein (SE) prediction with no dependence on shear rate, while off-diagonal MSD produces zero value (not shown). This is consistent with the isotropic nature of NP Brownian motion in a dilute and unbounded solution, and also serves as a verification of the MSD calculations. At $\phi$$>$$0$, a deviation from SE relation occurs in all diagonal MSDs as a result of NP-RBC interactions. At the $\dot{\gamma}$$=$$10\ s^{-1}$, only slight deviation from SE relation is observed given the dominance of Brownian diffusion; see figure \ref{fig:MSD} (a).

The transient deviation of NP mobility from the SE relation can be further interrogated through the evolution of cross-stream MSDs, as depicted in figure \ref{fig:MSD} (a-d, g-h). 
In the short-time regime ($t\dot{\gamma}$$\ll$$1$), a linear growth of MSD ($\sim$$t$) is observed particularly under low shear rate (e.g. $\dot{\gamma}$$\leq$100 $s^{-1}$) or high shear rate and high hematocrit (e.g. $\dot{\gamma}$$\ge$2000 $s^{-1}$ and $\phi$=0.4) condition, as shown in figure \ref{fig:MSD} (b) and (c), respectively. The short-time linear growth of MSD suggests a short-time diffusive mechanism. Since the short-time linear growth of MSD always occurs before the ballistic regime ($\sim$$t^2$) where RBC-NP collisions start to occur, such short-time diffusive driver should logically be the Brownian effect. Therefore, the initial linear MSD growth at low shear rates (with $\phi$=0$\sim$0.4) or high shear rates (with $\phi$=0.4) can be explained by the Brownian diffusive time scale being much shorter than the RBC-NP collision time scale under such hemorheological conditions. However, under high shear rate (e.g. $\dot{\gamma}$$\ge$2000 $s^{-1}$), the diffusive behavior ($\sim$$t$) turns to ballistic ($\sim$$t^2$) as $\phi$ increase to 0.4; see figure \ref{fig:MSD} (c,d,g,h). This is likely due to the high convective (high shear rate) and low inertial (low haematocrit) effects that reduce the RBC-NP collision time scale to be comparable to the Brownian diffusion time scale. As a result, the Brownian diffusion of NP is overwhelmed by the ballistic behavior caused by insufficient RBC-NP collisions.

\begin{figure}
 \centering
 \includegraphics[width=0.217\columnwidth]{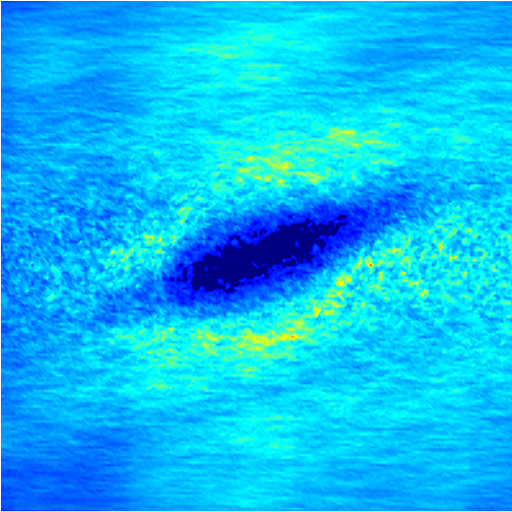}
 \includegraphics[width=0.217\columnwidth]{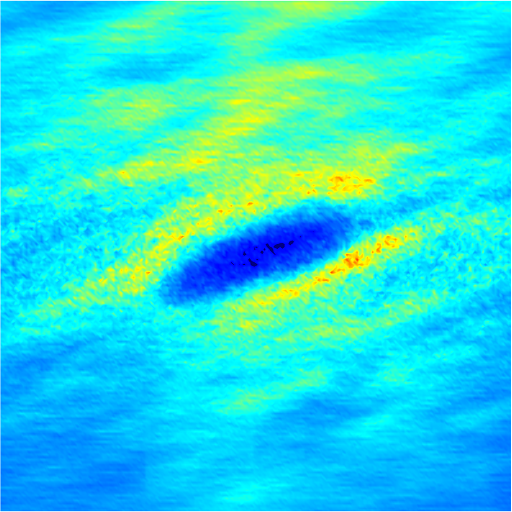}
 \includegraphics[width=0.217\columnwidth]{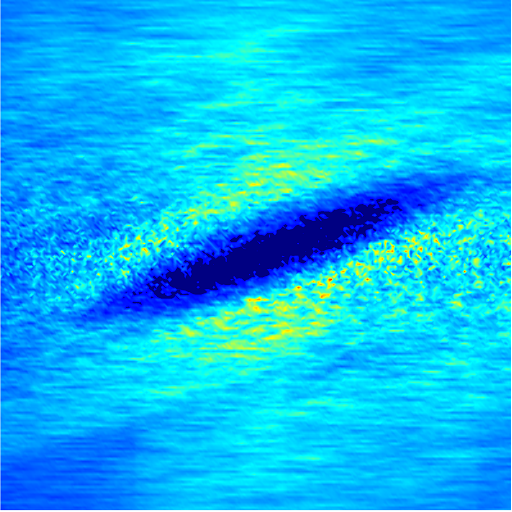}
 \includegraphics[width=0.267\columnwidth]{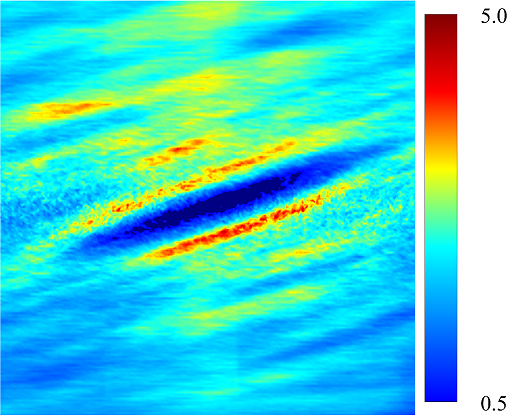}
  \begin{minipage}{0.18\linewidth}
      \centering
  	 (\textit{a}) $\dot{\gamma}$=100 $s^{-1}$
  	 \\
  	  $\phi$=0.1
  \end{minipage} 
  \begin{minipage}{0.22\linewidth}
     \centering
 	(\textit{b}) $\dot{\gamma}$=100 $s^{-1}$
 	\\
 	 $\phi$=0.4
  \end{minipage} 
   \begin{minipage}{0.22\linewidth}
     \centering
 	(\textit{c}) $\dot{\gamma}$=2000 $s^{-1}$
 	\\
 	 $\phi$=0.1
  \end{minipage}
   \begin{minipage}{0.24\linewidth}
     \centering
 	(\textit{d}) $\dot{\gamma}$=2000 $s^{-1}$
 	\\
 	 $\phi$=0.4
  \end{minipage}
 \caption{RBC-NP partial pair distribution function, $g_{21}(\boldsymbol{r})$, projected on the $xy$ plane under various hemorheological conditions and averaged in the strain range of $t\dot{\gamma}$=1$\sim$3.}
\label{fig:mstst}
\end{figure}

In the intermediate-time regime, $t\dot{\gamma}$=$O(1)$$\sim$$O(10)$, anomalous dispersive behavior is observed as RBCs start to elongate and rotate to be aligned with the streamwise direction due to shear. 
For cases at high shear rates ($\dot{\gamma}$$\ge$$2\ 000\ s^{-1}$) and high haematocrit ($\phi=0.4$), we observe a sublinear growth of MSD, representing a temporary hindrance of the NP mobility. This hindrance effect is caused by a string-ordered microstructure of NP in the shearing plane accompanied by the elongation and alignment of RBCs; see the contrast between upper inset of figure \ref{fig:MSD} (c) and inset of figure \ref{fig:MSD} (b) (also see supplementary movie 2). Such string-ordered NP distribution can be better visualized by plotting the RBC-NP partial pair distribution function (PPDF), $g_{21}(\boldsymbol{r})$, projected on the $xy$ plane (the computing procedure of PPDF in the short-time regime is discussed in \S Appendix \ref{appB}).  As shown in figure \ref{fig:mstst}, $g_{21}(\boldsymbol{r})$ in $xy$ plane exhibits streaks showing intensified distribution particularly near the RBC surface under high shear rate and high haematocrit, while in contrast no significant string structure is found in low shear rate ($\dot{\gamma}$$\leq$$100\ s^{-1}$) (see supplementary movie 1) or/and low haematocrit ($\phi=0.1$) cases, as shown in figure \ref{fig:mstst} (a-c). Note that suspension string-like structure often occurs in sheared monodisperse colloidal suspensions at equilibrium as a consequence of the absence of interparticle lubrication interactions at high concentration \citep{Xue1990,FossJFM2000} or the presence of long-range repulsive forces at low concentration \citep{Higdon2010PRE}. However, in the current case, ordering of NP occurs at a non-equilibrium state that involves the change of the RBC suspension structure from a uniformly distributed and randomly oriented configuration to a streamwise-aligned and elongated configuration under high shear and high haematocrit.
The sublinear growth of MSD is followed by a super-ballistic behavior ($\sim$$t^3$) at high shear rate ($\dot{\gamma}$=$10\ 000\ s^{-1}$), 
which might be associated with the large jump of NP between strings when the NP phase gradually evolves from the string-like structure towards the more uniform structure featuring in the long-time diffusive regime.

In the long-time regime ($t\dot{\gamma}$$>$$100$), MSD reaches the second linear-growth stage \textcolor{black}{($\sim$$t$)}, where RESID becomes the dominate diffusive mechanism leading to an uniform distribution of NP (see lower inset of figure \ref{fig:MSD} (c)). Such three-stage (diffusive/super-ballistic/diffusive) anomalous dispersion behavior has previously been reported in the dispersion of Brownian particles subjected to external forces \citep{Siegle2010}. 
Here, we show that such dispersive anomalies also occur in a sheared NP-RBC bidisperse suspension, where the anomalous NP dispersion is, however, driven by an internal shear-induced mechanism, i.e., the elongation and the alignment of the concentrated RBCs along streamwise direction under high shear. Such transient shear-induced morphological adaptation of RBCs contributes to extra mobility of the NP phase, seemingly playing a role of external forces exerted on the NP phase.

Figures \ref{fig:MSD} (e, f) depict the longitudinal and off-diagonal MSD evolution at $\dot{\gamma}$=$10\ 000\ s^{-1}$. Results for lower shear rate show similar MSD behaviors and are not presented for discussion. Because the initial transient regime is neglected, the longitudinal and off-diagonal MSDs yield classical ballistic-diffusive transitions similar to the dispersive behavior of rigid particle suspensions under shear \citep{FossJFM1999,ClausenJFM2011}. The off-diagonal MSD under high shear starts with positive values during the ballistic regime, exhibits a crossover transition involving a change of sign, and eventually maintains negative values in the diffusive regime. 
Absolute values are shown for the off-diagonal MSDs with the sign at certain temporal stage denoted in figure \ref{fig:MSD} (f). The change of sign from positive (+) to negative (-) is a hallmark of the dominant NP migration direction shifted from along the extensional axes (1st and 3rd quadrants) to along the compressive axes (2nd and 4th quadrants) of the flow, as also observed in sheared colloidal suspensions \citep{FossJFM1999}.

\subsection{Hemorheological dependence of NP long-time diffusivity}\label{sec:hemo}
\noindent In this section, we focus on examining the NP long-time diffusive behavior. The NP long-time diffusivities are evaluated in the long time regime (after 100 strain units) and listed in table \ref{tab:srht}. Experimental statistics from various sources \citep{Grabowski1972,Antonini1978,Diller1980,Wang1985} are selected for comparison to gain credibility of the simulation results.
Given the distinct time scales associated with the Brownian ($\tau_B$$\sim$$10^{-4}\ s$) and the long-time RBC-enhanced diffusion (${\tau}_R$$\sim$$100/\dot{\gamma}$) phenomenon, the coupling of BD and RESID follows simple superposition, $\mathsfi{D}_{ij}^{\infty}=\mathsfi{D}_{ij}^R+\mathsfi{D}^B \delta_{ij}$, as confirmed in previous studies, e.g., \citet{Liu2018a}, where $\mathsfi{D}_{ij}^R$ denotes the RESID. Since $\mathsfi{D}^B$ at infinite dilution follows SE relation with negligible dependence on flow conditions, the hemorheological response of $\mathsfi{D}_{ij}^\infty$ is essentially determined by $\mathsfi{D}_{ij}^R$.

\subsubsection{Shear-rate dependence}\label{sec:sr}
\begin{figure}
  \centering
  \includegraphics[width=0.42\columnwidth]{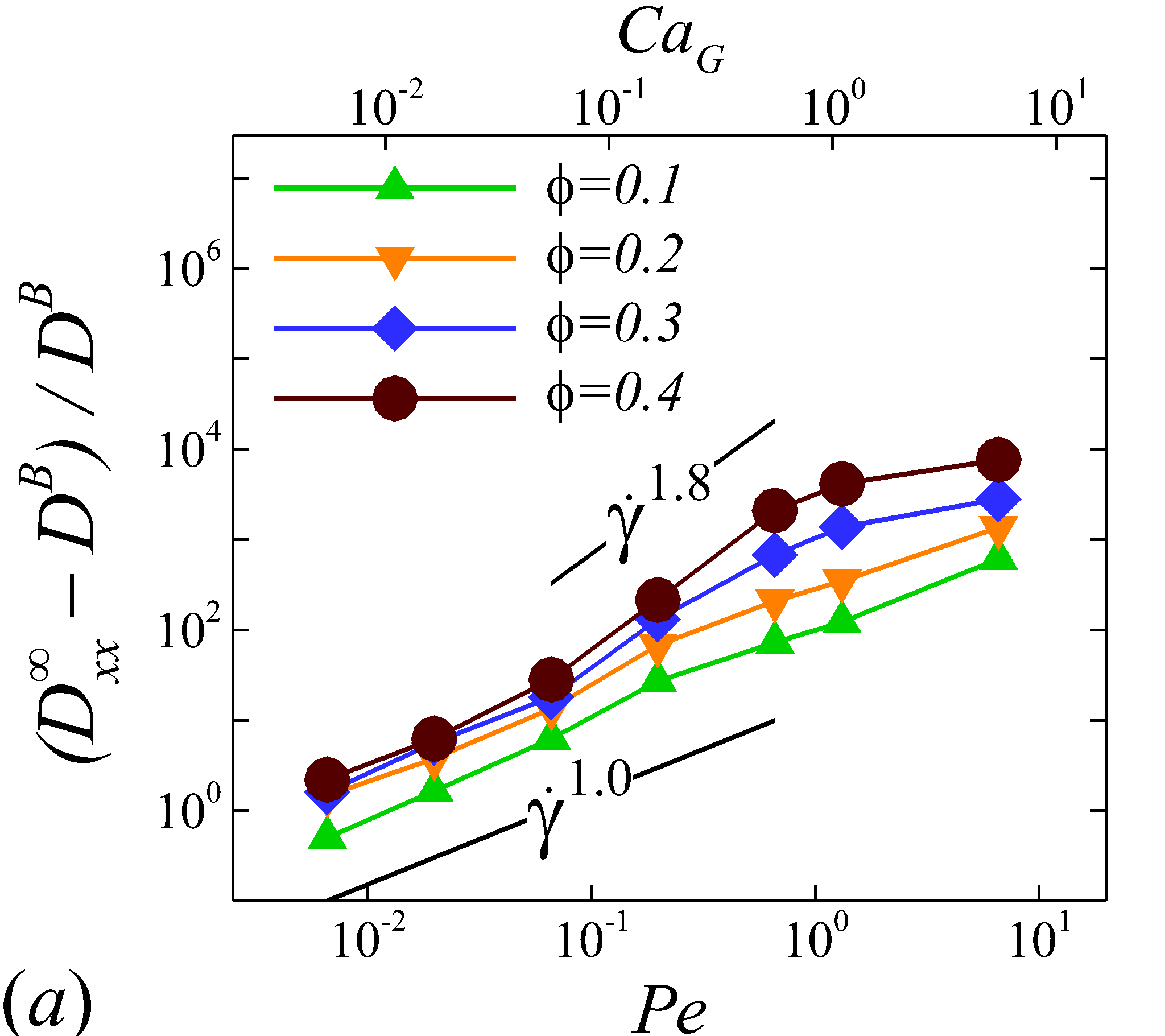}
  \includegraphics[width=0.47\columnwidth]{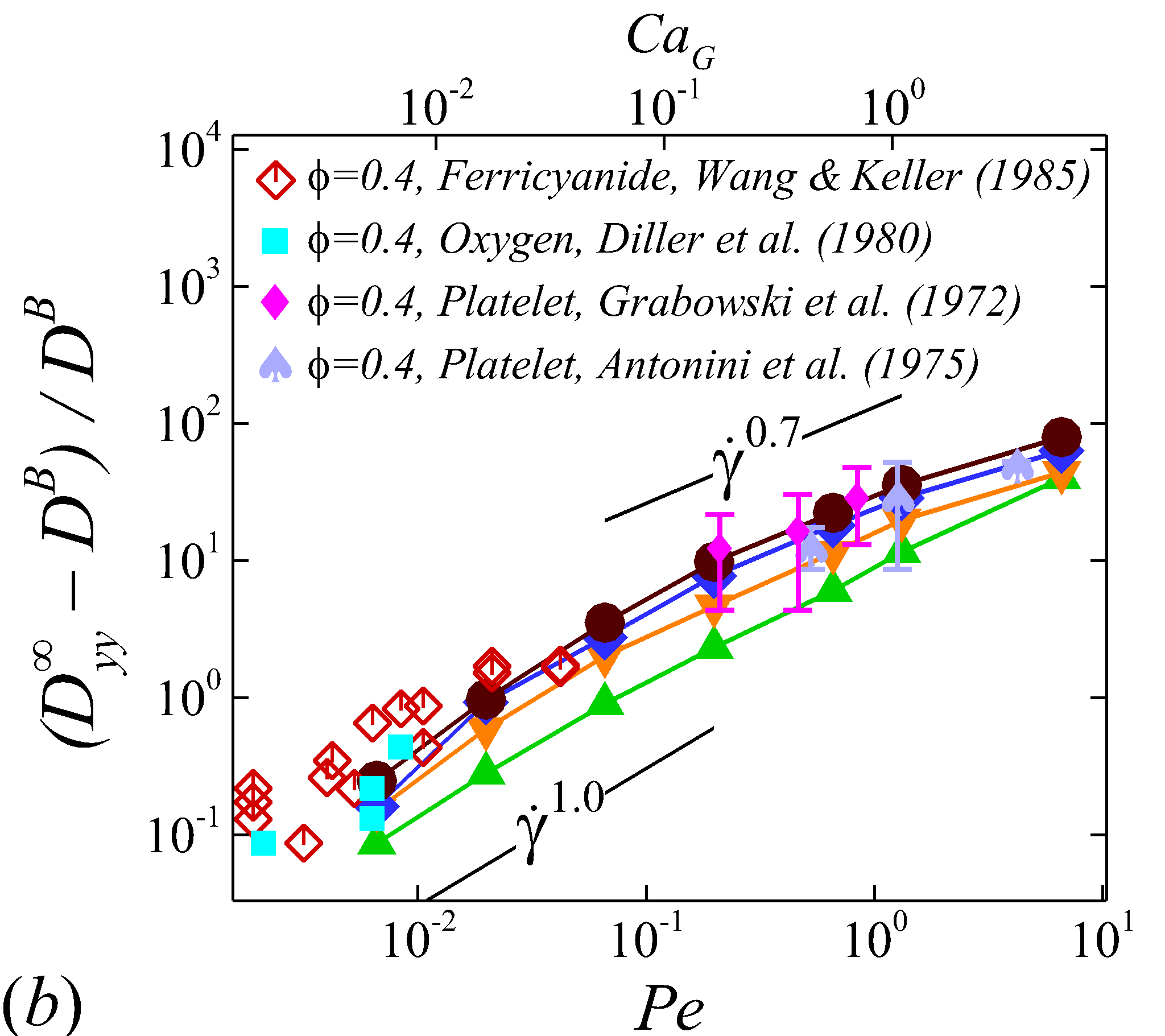}
  \vspace{0.1 cm}
  \includegraphics[width=0.42\columnwidth]{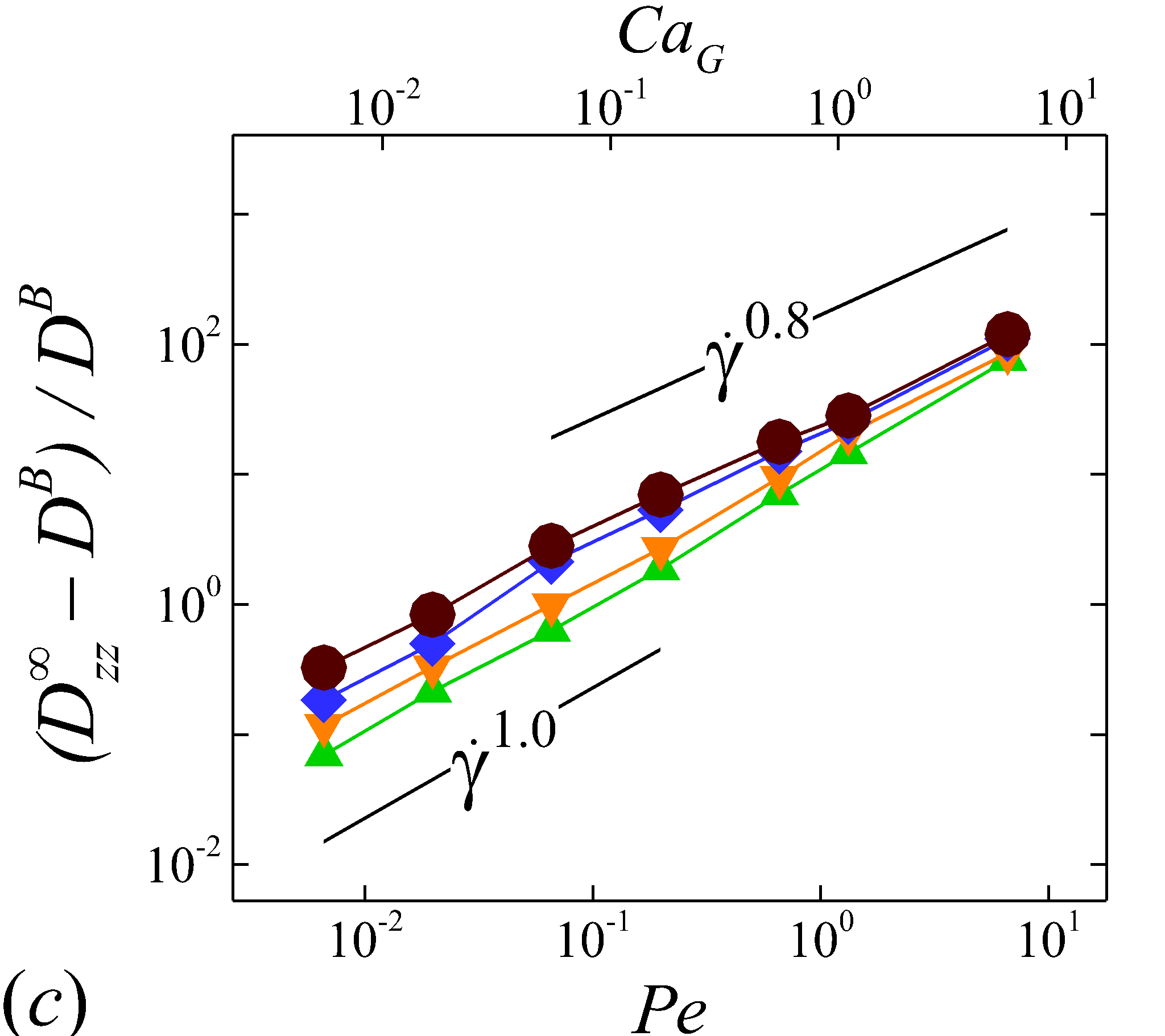}
  \hspace{0.0 cm}
  \includegraphics[width=0.42\columnwidth]{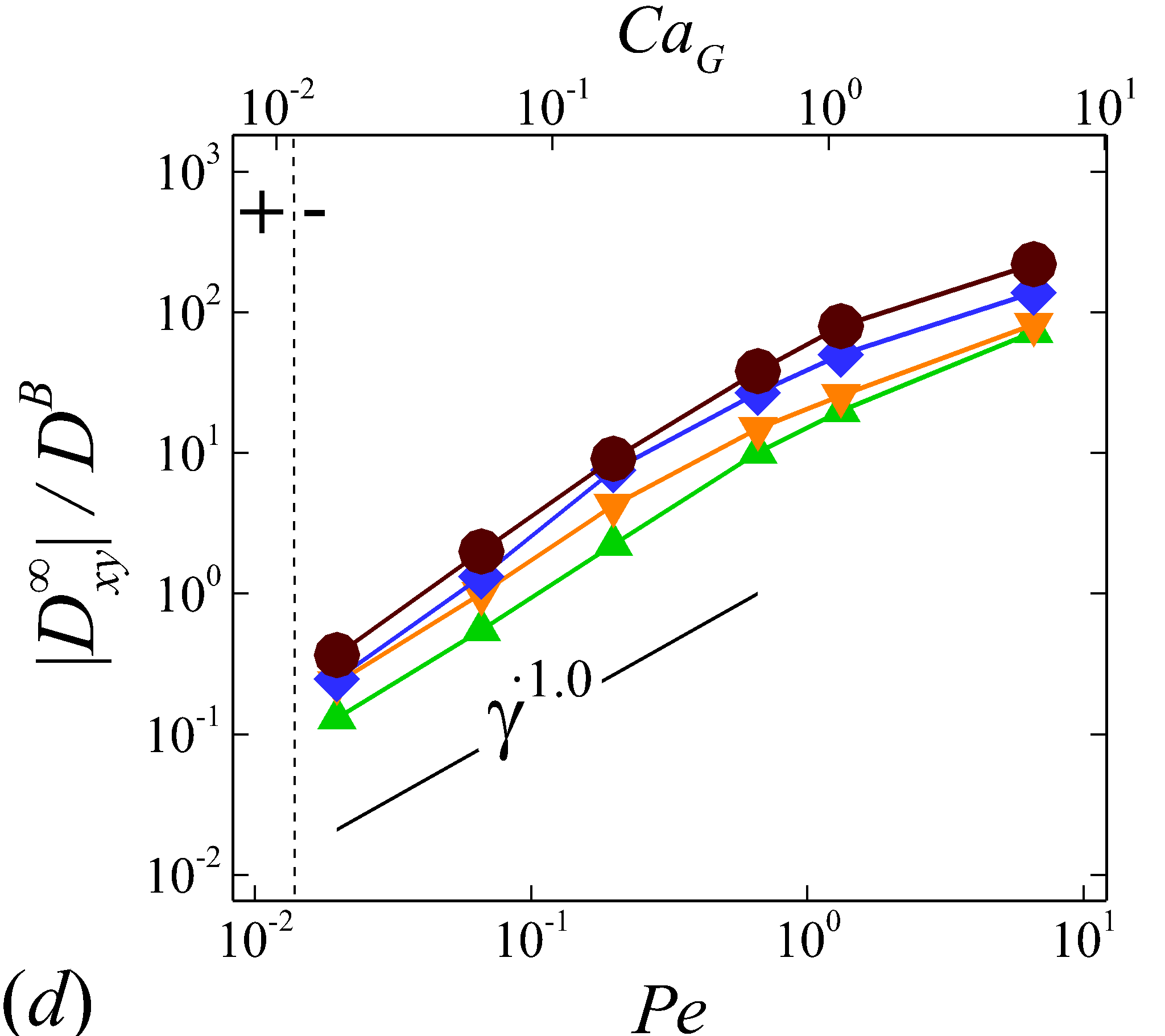}
  \hspace{0.55 cm}
  \caption{Normalized particle long-time RESID in (a) $xx$, (b) $yy$, (c) $zz$ and (d) $xy$ directions against normalized shear rate ($\Pen$ or $Ca_G$) in a log-log scale. Each curve at specific $\phi$ forms by connecting $\hat{\mathsfi{D}}_{ij}^R$ data at shear rate, $\dot{\gamma}=$10, 30, 100, 300, 1000, 2000 and 10\ 000 $s^{-1}$, from left to right; $\hat{\mathsfi{D}}_{xy}^R$ at $\dot{\gamma}$=$10\ s^{-1}$ is not shown for scaling purpose due to its small magnitude. Experimental results for RBC enhanced solute \citep{Diller1980,Wang1985} and platelet \citep{Grabowski1972,Antonini1978} diffusivity in the velocity gradient ($yy$) direction are plotted for comparison. NPs of size $2a_1$=100 $nm$ yield $\mathsfi{D}^B$=3.78 $\mu m^2/s$.}
\label{fig:sr}
\end{figure}

Figure \ref{fig:sr} plots the normalized RESID, $\hat{\mathsfi{D}}_{ij}^R=\mathsfi{D}_{ij}^\infty / \mathsfi{D}^B-\delta_{ij}$ against normalized shear rate, $\Pen$ (or $Ca_G$). To first validate the simulation results, experimental results of solute or cell velocity-gradient ($yy$) diffusivity in sheared human or animal blood are selected for comparison; see figure \ref{fig:sr} (b). \cite{Wang1985} measured the augmentation of ferricyanide solute diffusivity in both bovine and human RBC suspensions using a rotating Couette flow device. \citet{Diller1980} measured the enhanced radial diffusivity of oxygen solute in human blood using a tube oxygenator device. \citet{Grabowski1972} calculated the platelet velocity-gradient diffusivity in cavine blood flowing through a channel by measuring the platelet rate of adhesion/deposition to a foreign surface attached to the flow chamber. Similar experiment was later conducted by \citet{Antonini1978} to measure the platelet radial diffusivity in human blood. These measured diffusivity in 40\% haematocrit normalized based on our notation compares favorably with the calculated $\hat{\mathsfi{D}}_{yy}^R$ at $\phi$=0.4 based on the simulation, as shown in figure \ref{fig:sr} (b).

Depending on the level of shear rate imposed at various haematocrit, the diffusion tensor of NP in sheared blood shows different shear-rate dependence.
At low shear rates ($\dot{\gamma}$$\leq$100), all RESID terms exhibit linear dependence on shear rate ($\sim$$\dot{\gamma}$), matching the linear $\dot{\gamma}$ scaling of shear-induced diffusivity in rigid particle suspensions \citep{FossJFM1999,Sierou2004}. This is also consistent with the insignificant RBC morphological changes at $\dot{\gamma}$$\leq$$100\ s^{-1}$, as shown in figures \ref{fig:RBCNP1} (a-c). 

At intermediate shear rates ($100$$\leq$$\dot{\gamma}$$\leq$$2\ 000$), significant streamwise elongation of RBCs occurs with increasing shear rate; see figure \ref{fig:RBCNP1} (c-e). As a result, nonlinear $\dot{\gamma}$ scaling is observed in all diagonal RESID terms. Specifically, cross-stream diffusivities, $\hat{\mathsfi{D}}_{yy}^R$ and $\hat{\mathsfi{D}}_{zz}^R$, show sublinear scales ($\sim$$\dot{\gamma}^m$, $m$=$0.7$$\sim$$0.8$), while streamwise diffusivity, $\hat{\mathsfi{D}}_{xx}^R$, exhibits superlinear scales ($\sim$$\dot{\gamma}^n$, $n$=$1$$\sim$$1.8$). In contrast to the nonlinear scaling in diagonal diffusivities, $\hat{\mathsfi{D}}_{xy}^R$ maintains largely a linear scale at the intermediate shear-rate regime, as shown in figure \ref{fig:sr} (d). It is also noted that such nonlinear shear-rate scaling in diagonal RESID terms is most prominent at intermediate to high haematocrit ($\phi$$>$$0.1$), which implies the RBC deformability plays less important role in altering the RESID shear-dependence at low haematocrits. This observation is consistent with the results of shear-augmented solute diffusivity in model-RBC suspensions at various RBC deformability reported by \cite{Wang1985}, where they found changing RBC deformability barely affects the augmentation at particle volume concentration of 0.1 or less. \citet{Wang1985} also find a sublinear scaling of $\sim$$\dot{\gamma}^\beta$ ($0.67$$\leq$$\beta$$\leq$0.89) for the solute diffusivity in velocity-gradient directions, while our simulations observe a exponent of $\beta$=0.7.

At high shear rates ($\dot{\gamma}$$\ge$$2\ 000$) and intermediate to high haematocrit ($\phi$$>$0.1), all RESID terms except $\hat{\mathsfi{D}}_{zz}^R$ exhibit reduced shear-rate dependence compared to the intermediate shear-rate regime. In the velocity-gradient ($y$) direction, the hindrance of diffusion is due to the fact that the concentrated RBCs become more aligned and elongated with flow such that RBCs act as obstacles against the NP cross-stream diffusion in $y$-direction; see figure \ref{fig:RBCNP1} (f). Such effect however shows less hindrance on $\hat{\mathsfi{D}}_{zz}^R$, as it does not forbid the NP migration in the vorticity ($z$) direction. 
In the streamwise ($x$) direction, the reduction of the shear dependence of $\hat{\mathsfi{D}}_{xx}^R$ is likely to be associated with the saturation of the RBC elongation.
Owing to the compound effects in both $x$ and $y$ directions, the off-diagonal diffusivty, $\hat{\mathsfi{D}}_{xy}^R$, also exhibit certain reduction of the shear-rate dependence.

\subsubsection{Haematocrit dependence}\label{sec:ht}
\begin{figure}
  \centering
  \includegraphics[width=0.42\columnwidth]{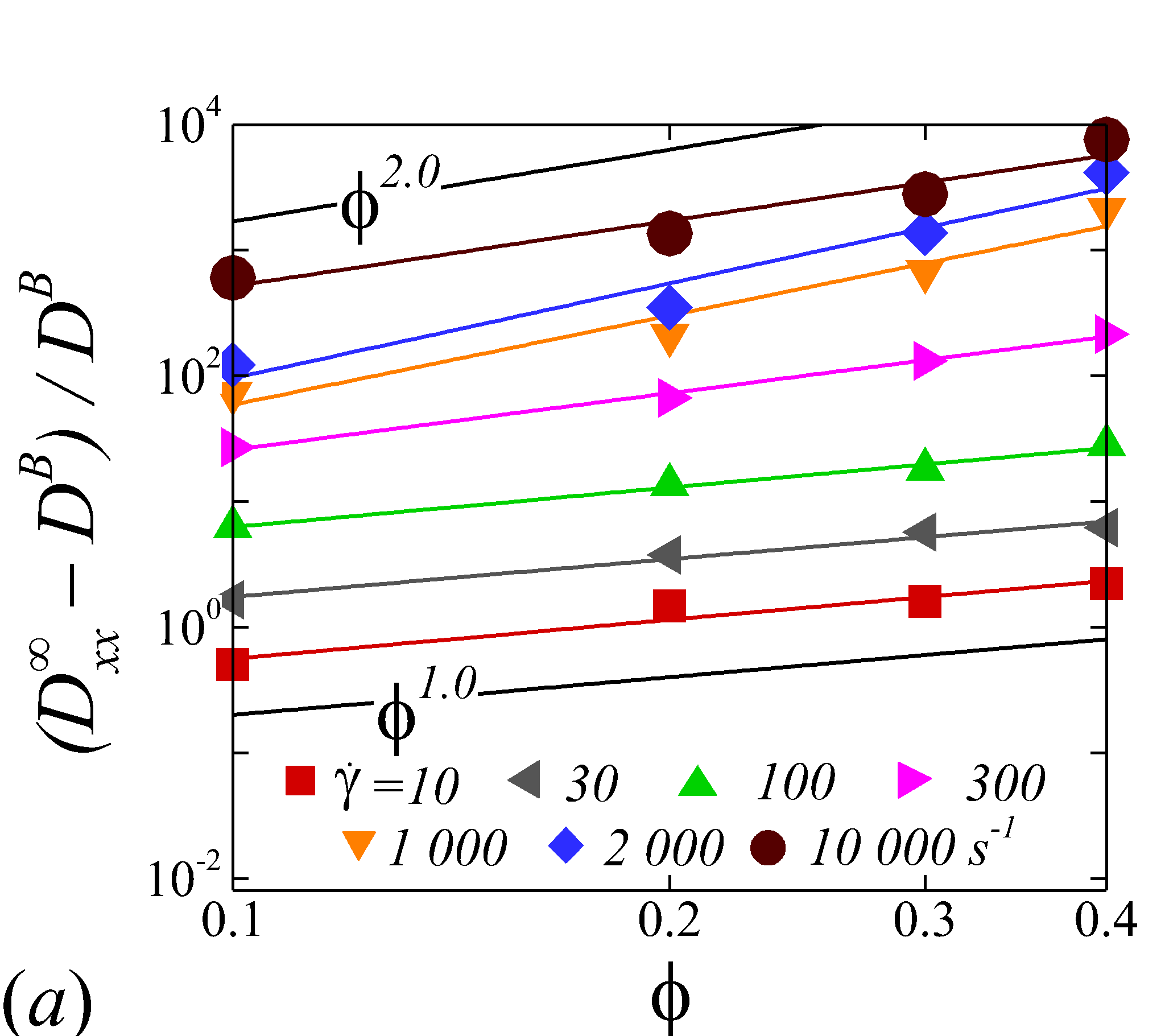}
  \includegraphics[width=0.42\columnwidth]{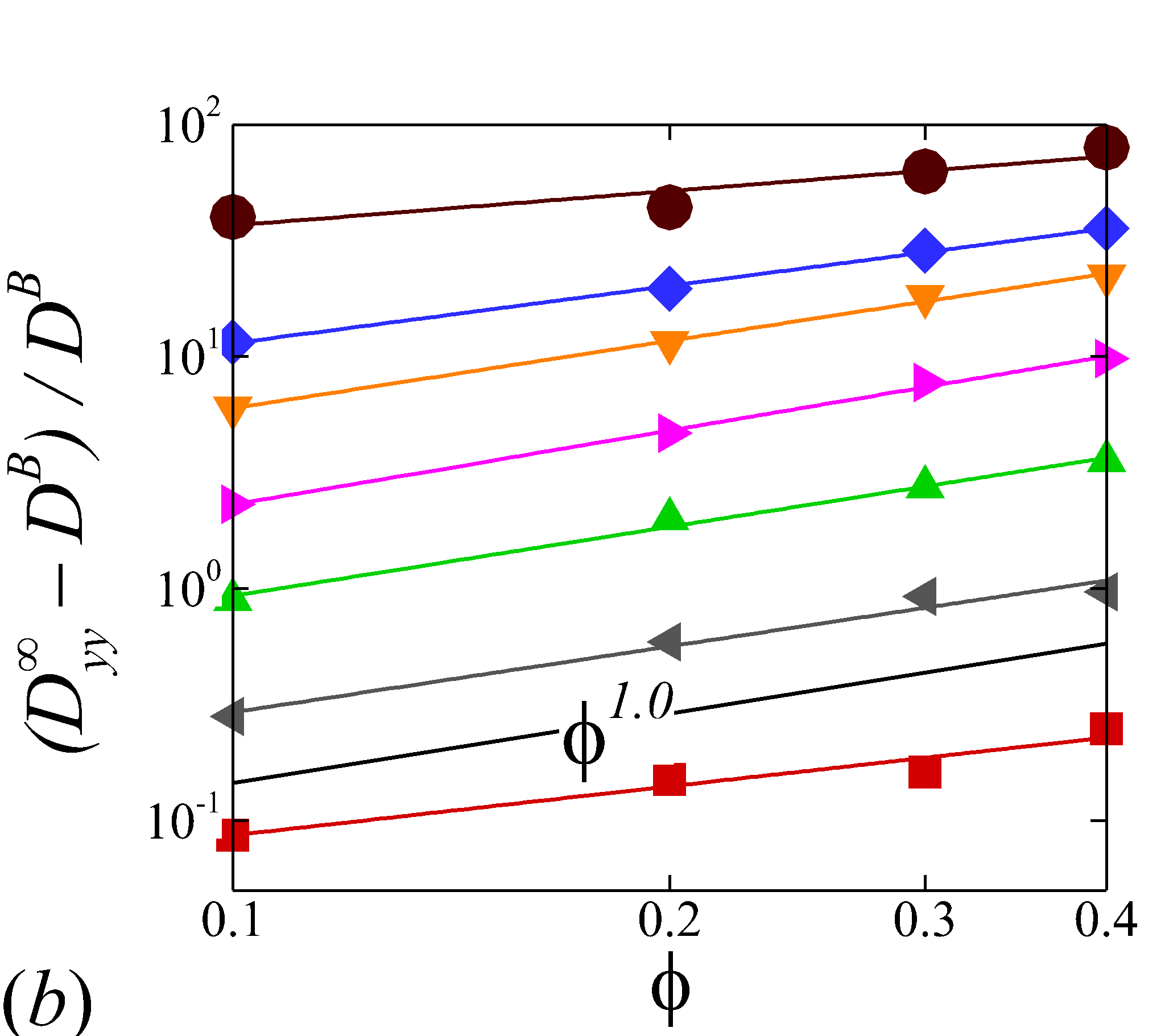}
  \includegraphics[width=0.42\columnwidth]{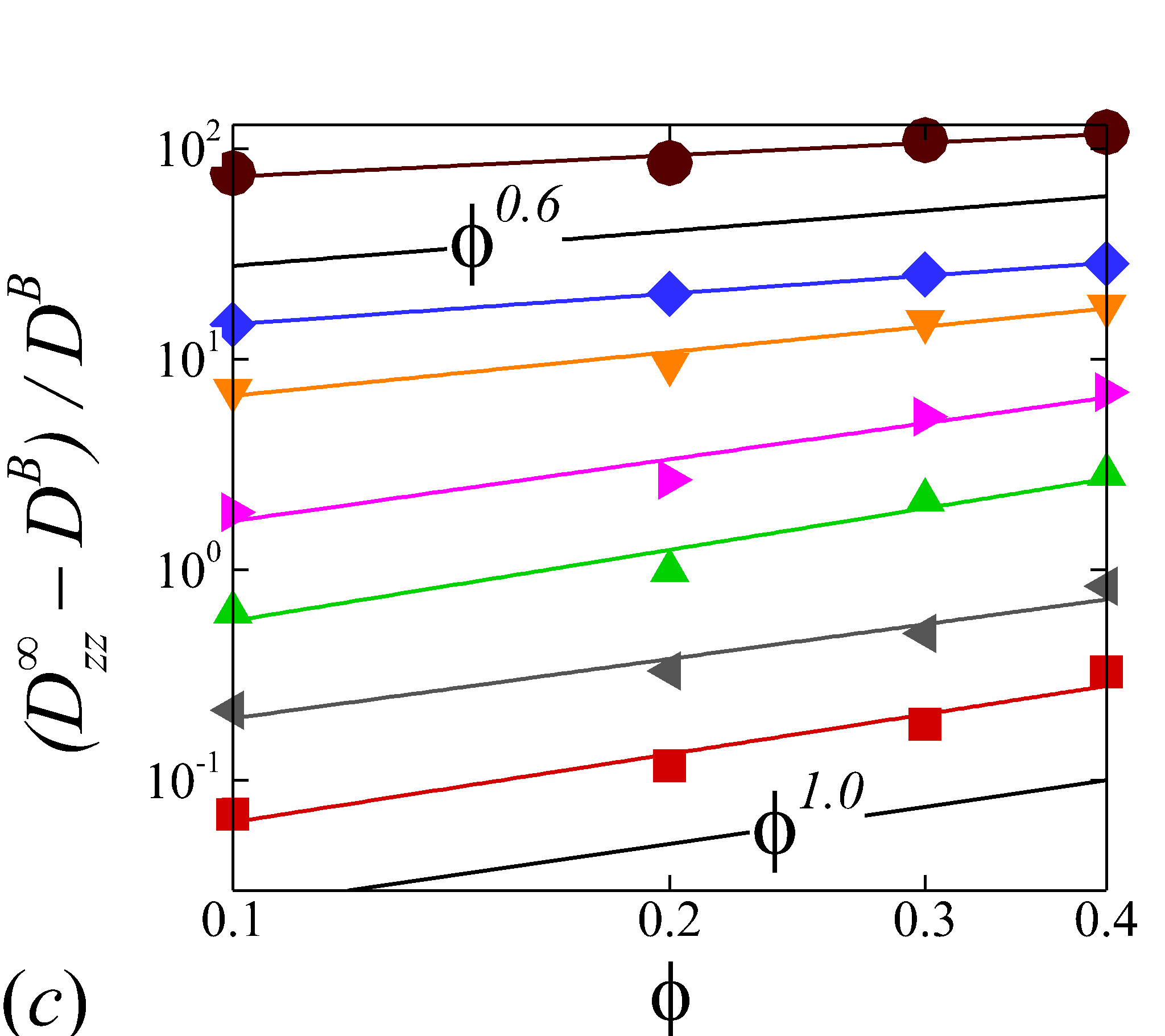}
  \includegraphics[width=0.42\columnwidth]{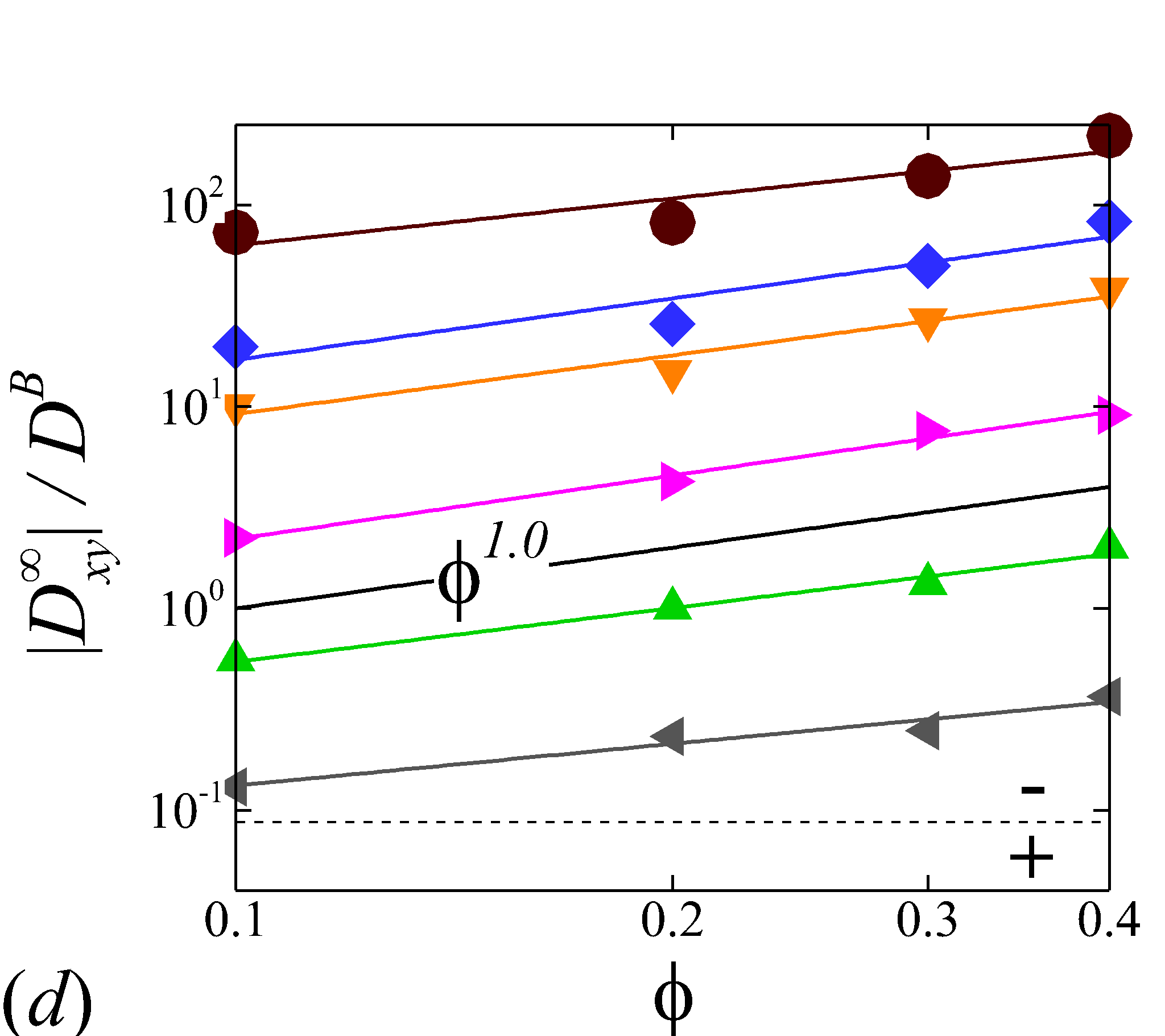}
  \caption{Normalized particle long-time RESID in (a) $xx$, (b) $yy$, (c) $zz$ and (d) $xy$ directions against haematocrit under different shear rate in logarithm scale. Lines are the best power-law fits to the data points at specific $\dot{\gamma}$. $\hat{\mathsfi{D}}_{xy}^R$ at $\dot{\gamma}$=$10\ s^{-1}$ (not shown) is positive with negligible magnitude compared to other diffusion terms. NPs of size $2a_1$=100 $nm$ yield $\mathsfi{D}^B$=3.78 $\mu m^2/s$.}
\label{fig:ht}
\end{figure}
Figure \ref{fig:ht} display the same data as figure \ref{fig:sr} but plotted against $\phi$ to show the haematocrit dependence of RESID at different shear rate. 
At low shear rates ($\dot{\gamma}$$\leq$$100$), all diffusion terms manifest a linear $\phi$ scaling. Note that linear concentration dependence of particle self-diffusivity has been observed in sheared monodisperse suspensions in the presence of surface roughness \citep{DaCunha1996} or residual Brownian motion \citep{BradyMorris1997}, which causes the two-body interaction being irreversible and hence giving rise to a diffusive behavior. Since the shear-induced diffusion of NP is driven by NP-RBC interaction, it is likely that the $\phi$ scaling of NP diffusion tensor results from the irreversible two-body interaction between NP and RBC induced by the RBC membrane roughness/flexibility and the NP Brownian effect.
The off-diagonal diffusivity $\hat{\mathsfi{D}}_{xy}^R$ is found to be positive (+) at very low shear rate ($\dot{\gamma}$$\leq$$10\ s^{-1}$) and otherwise negative (-). 
The change-of-sign behavior of the off-diagonal diffusivity designate the dominant displacement direction of NP changing from along the extensional axes to along the compressive axes of the flow as shear rate increases. Similar observation has been reported in sheared monodisperse colloidal suspensions \citep{FossJFM1999}.

As shear rate grows above 100 $s^{-1}$, various $\phi$ scaling arises in different RESID components. 
$\hat{\mathsfi{D}}_{xx}^R$ exhibits a quadratic $\phi$ scaling ($\sim$$\phi^2$), which suggests that in the $x$-direction additional effects exist to drive the NP diffusive motion besides the irreversible pairwise interactions.
$\hat{\mathsfi{D}}_{yy}^R$ and $\hat{\mathsfi{D}}_{xy}^R$ scale linearly with $\phi$, suggesting the irreversible pairwise interaction remains to be the dominant diffusive mechanism. $\hat{\mathsfi{D}}_{zz}^R$ exhibits a transition from linear to sublinear $\phi$ scale ($\sim$$\phi^{0.6}$). 
The mostly linear $\phi$ scale in $\hat{\mathsfi{D}}_{yy}^R$ observed in the current numerical study is consistent with the experimental observation in \cite{Wang1985}, where they show that the solute diffusivity in the velocity-gradient ($y$) direction is augmented by about three folds when the particle concentration increases from 0.1 to 0.31. Since suspension viscosity typically increases with the particle concentration \citep{FossJFM2000}, the enhancement of solute (e.g. NP) diffusivity at increased $\phi$ indicates the RESID is due to the RBC-NP interaction rather than the secondary flow effect \citep{Wang1985}.

\subsection{Microstructure}\label{sec:mst}
\noindent 
The rheological properties of particle suspensions are often determined by the mechanistic phenomenon occurred on the particle length scales. To elucidate the physical mechanisms that govern the hemorheological scaling behaviors of the NP diffusion tensor observed in \S \ref{sec:hemo}, the RBC-NP PPDF, $g_{21}(\boldsymbol{r})$, in the long-time regime are plotted to visualize the configurational microstructure of the RBC-NP bidisperse suspension under various haematocrit and shear rate. The techniques used to compute and project $g_{21}(\boldsymbol{r})$ are discussed in detail in \S Appendix \ref{appB}.

\begin{figure}
  \centering
  \includegraphics[width=0.26\columnwidth]{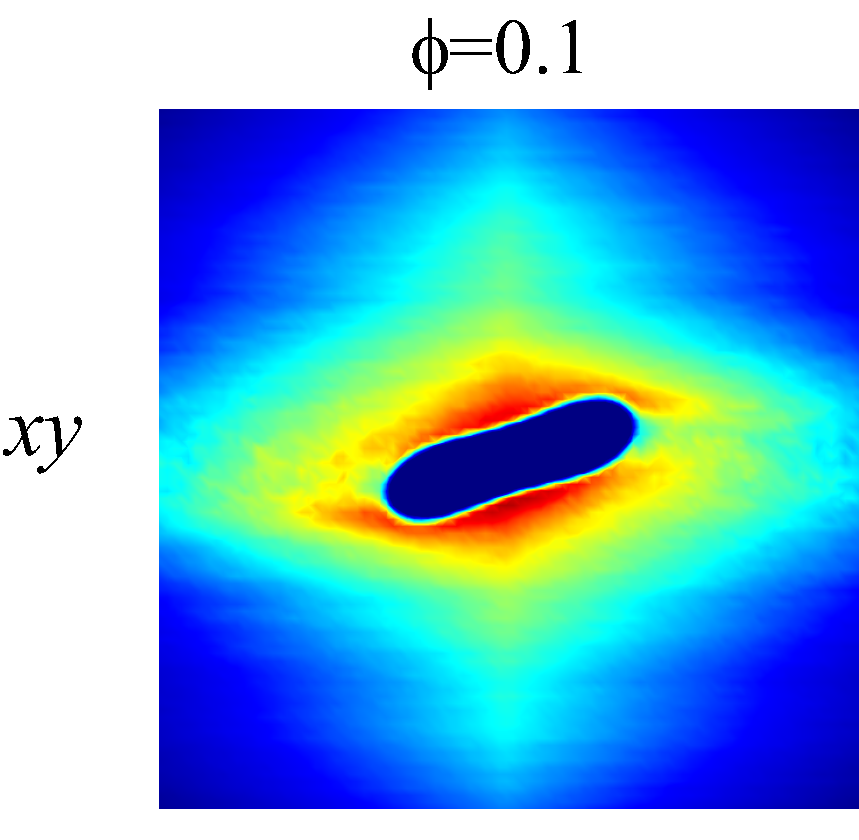}
  \includegraphics[width=0.211\columnwidth]{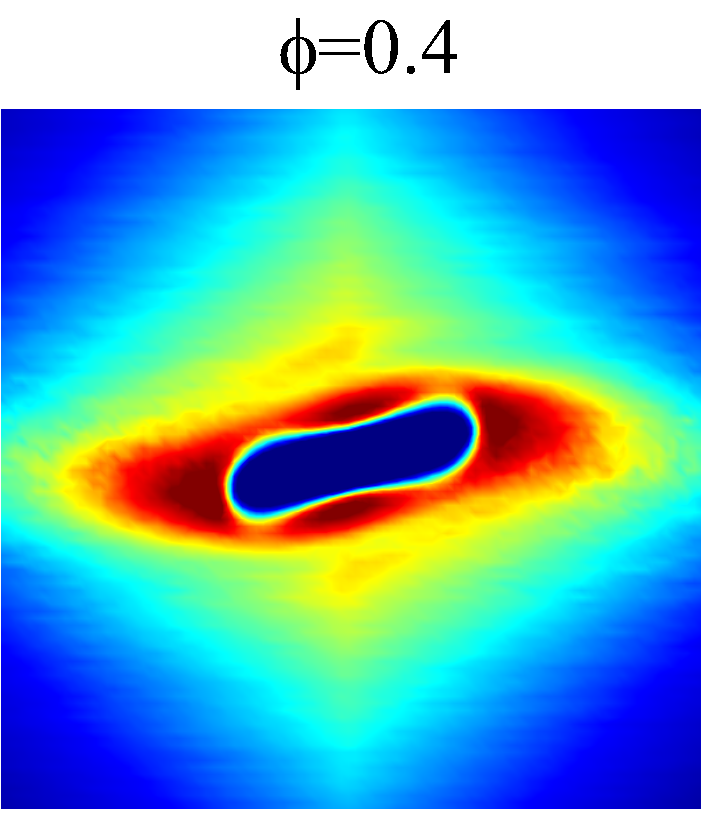}
  \includegraphics[width=0.211\columnwidth]{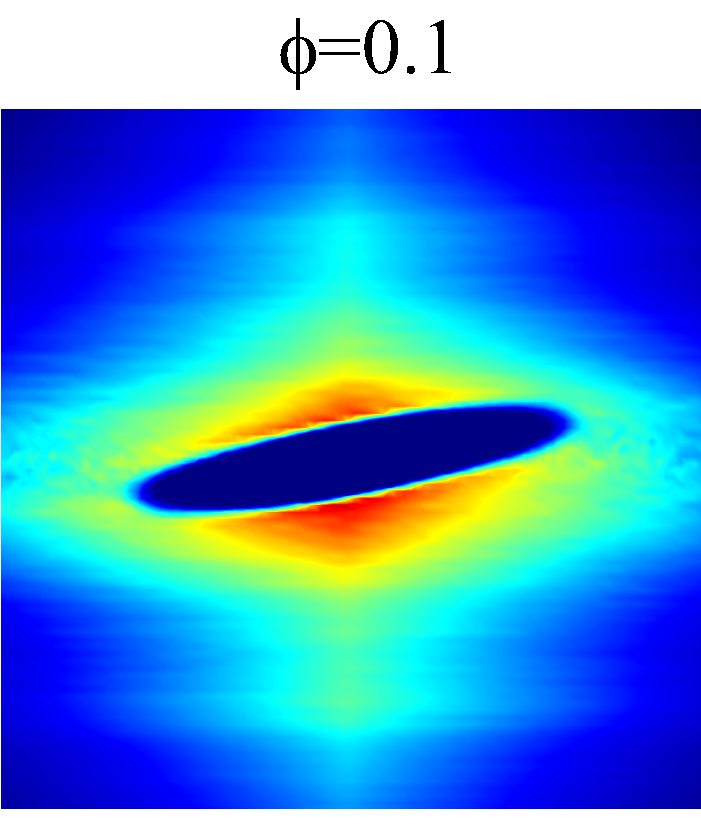}
  \includegraphics[width=0.28\columnwidth]{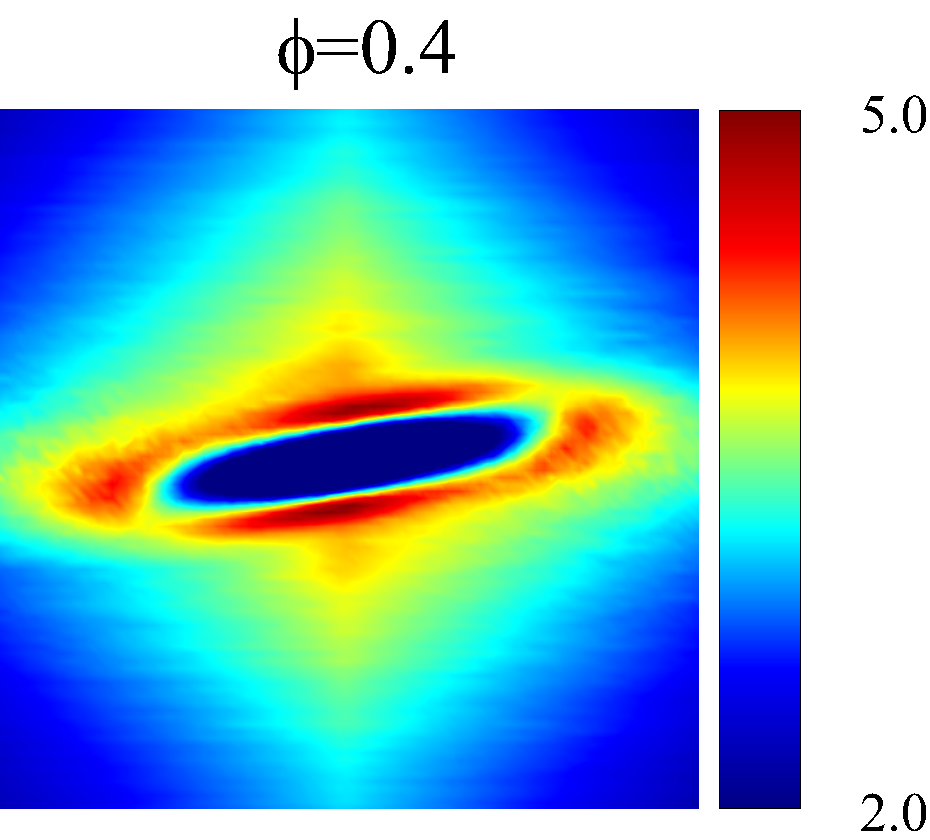}

  \includegraphics[width=0.26\columnwidth]{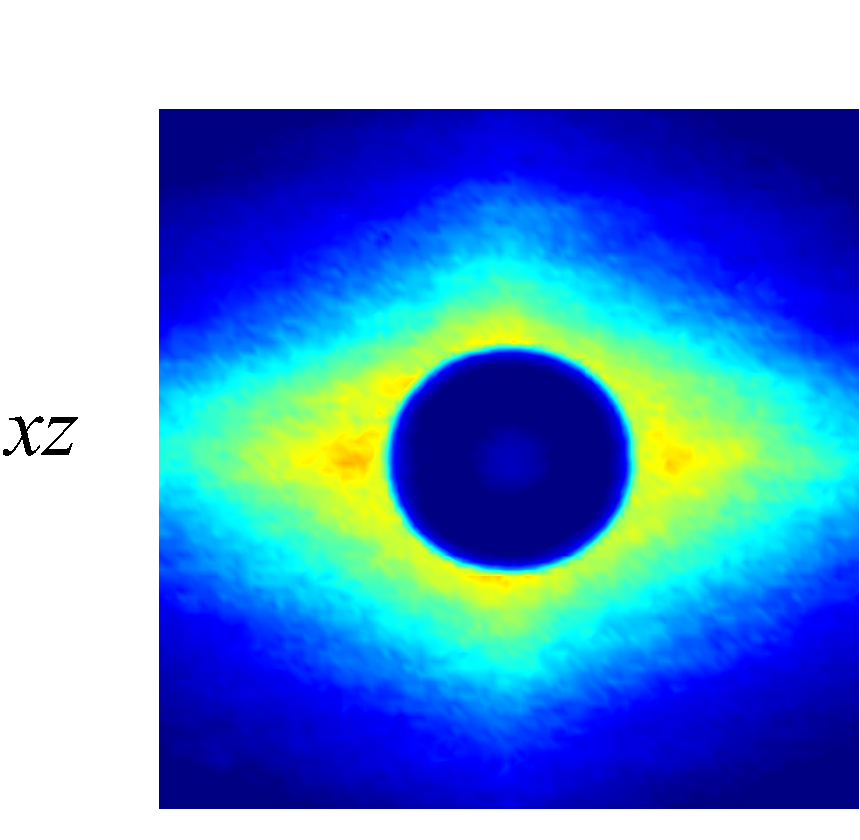}
  \includegraphics[width=0.211\columnwidth]{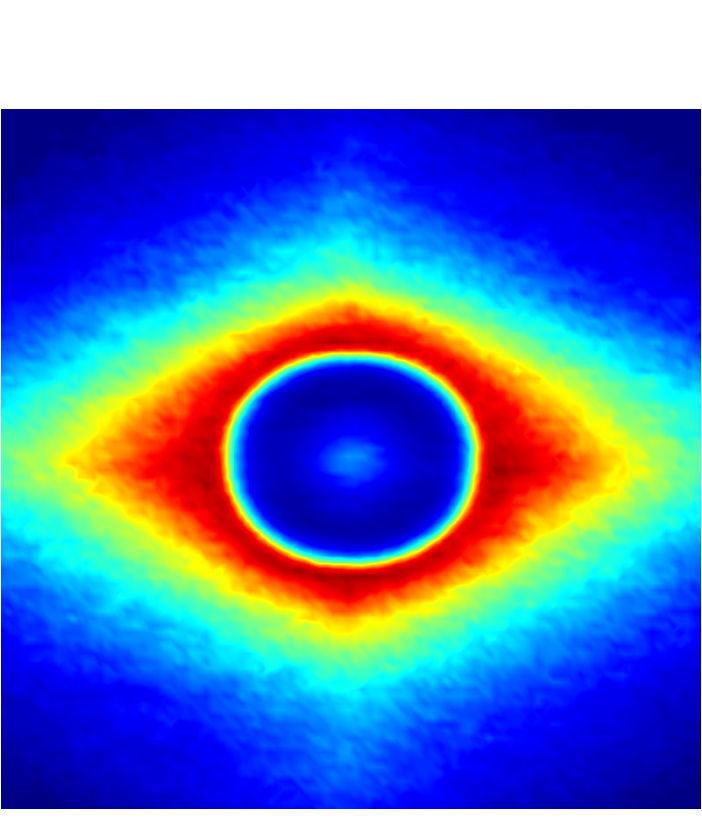}
  \includegraphics[width=0.211\columnwidth]{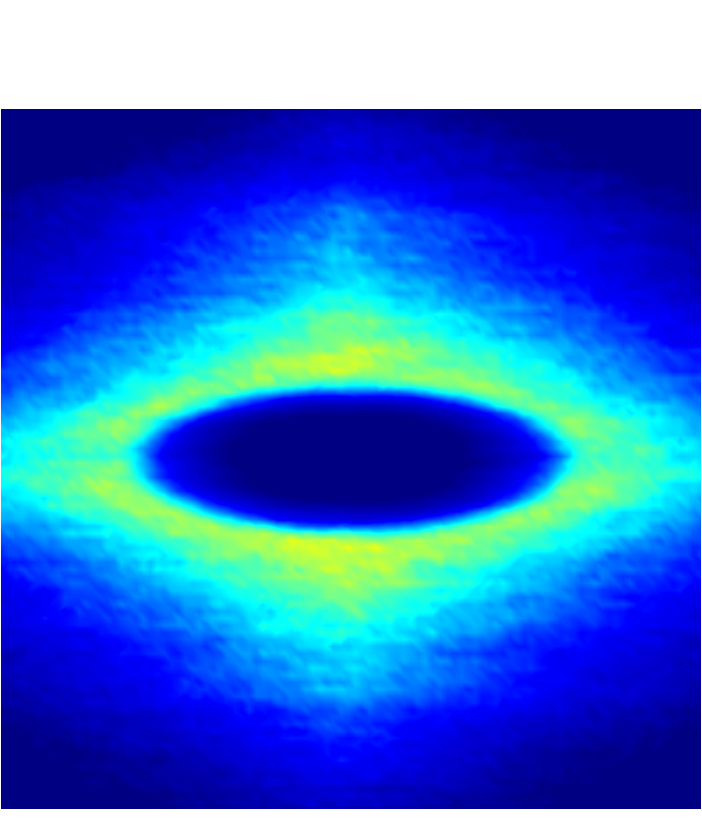}
  \includegraphics[width=0.28\columnwidth]{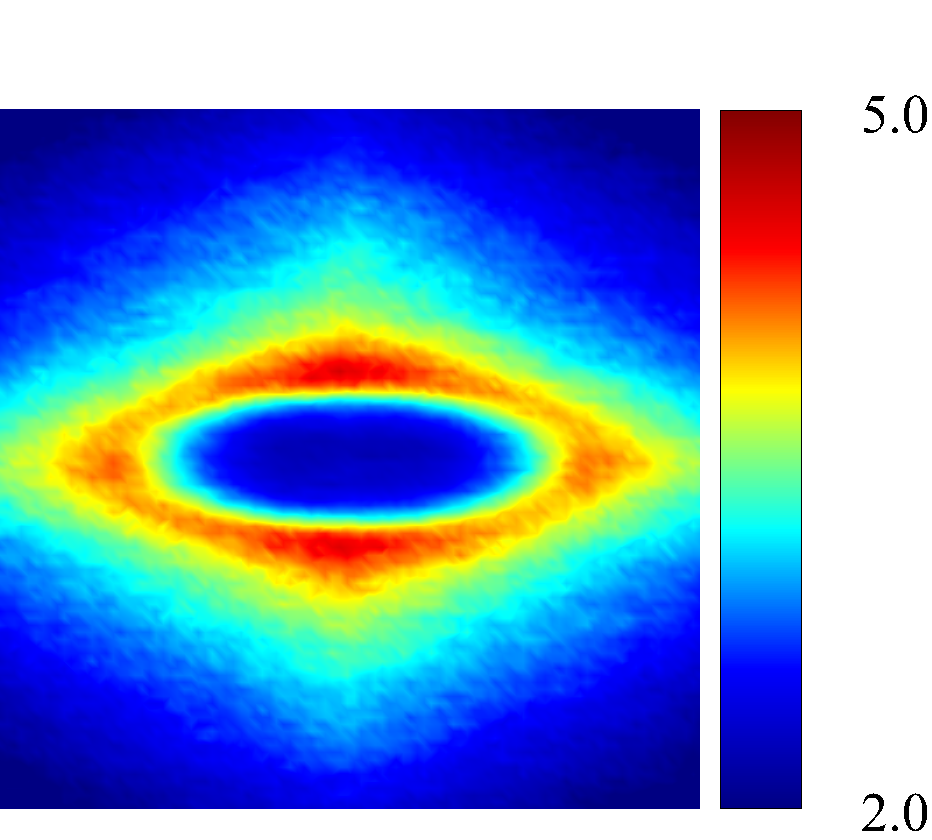}
  
  \includegraphics[width=0.26\columnwidth]{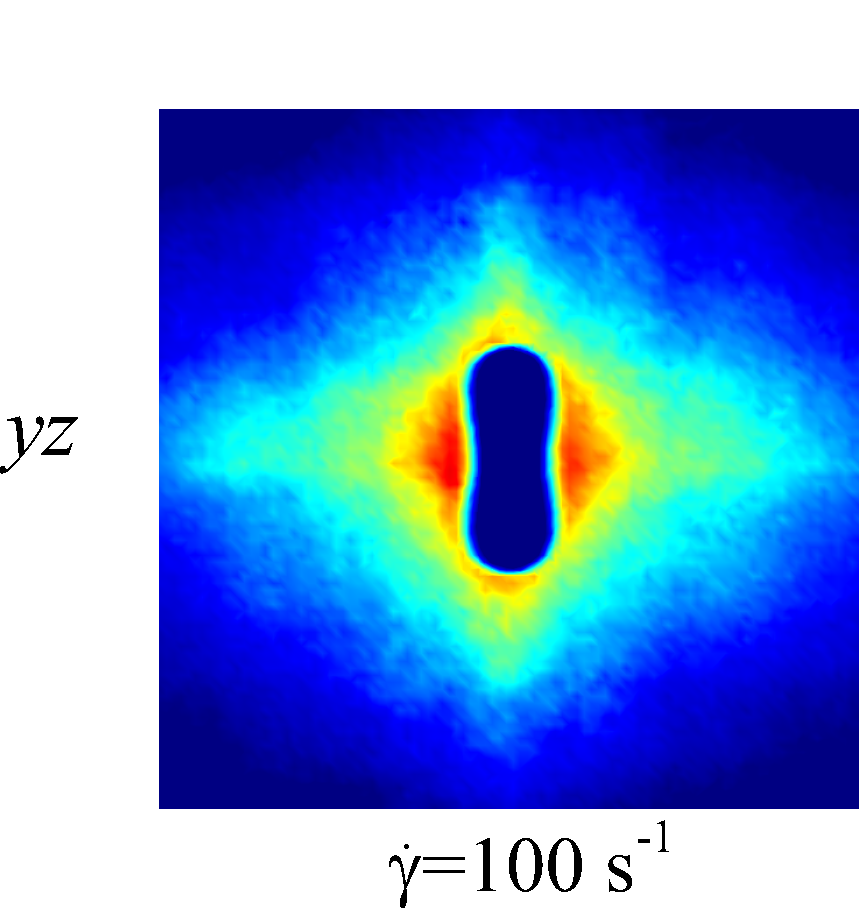}
  \includegraphics[width=0.211\columnwidth]{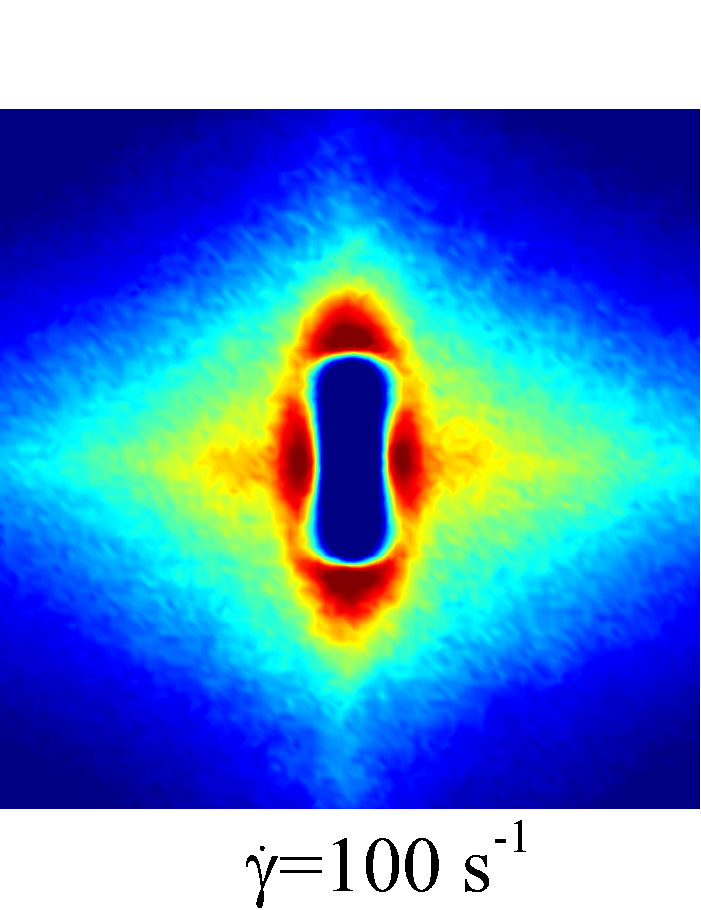}
  \includegraphics[width=0.211\columnwidth]{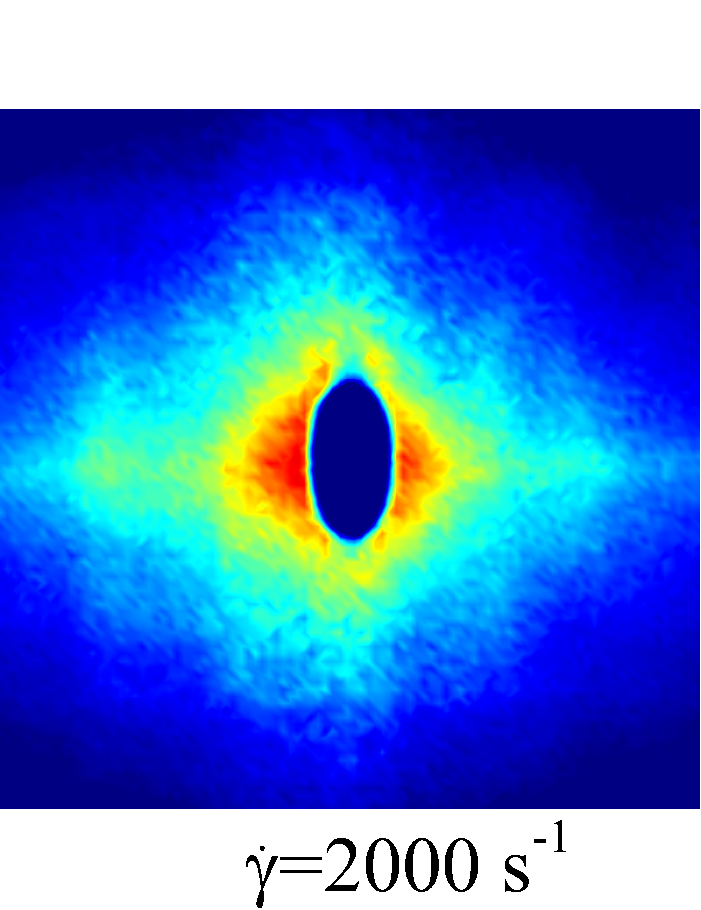}
  \includegraphics[width=0.28\columnwidth]{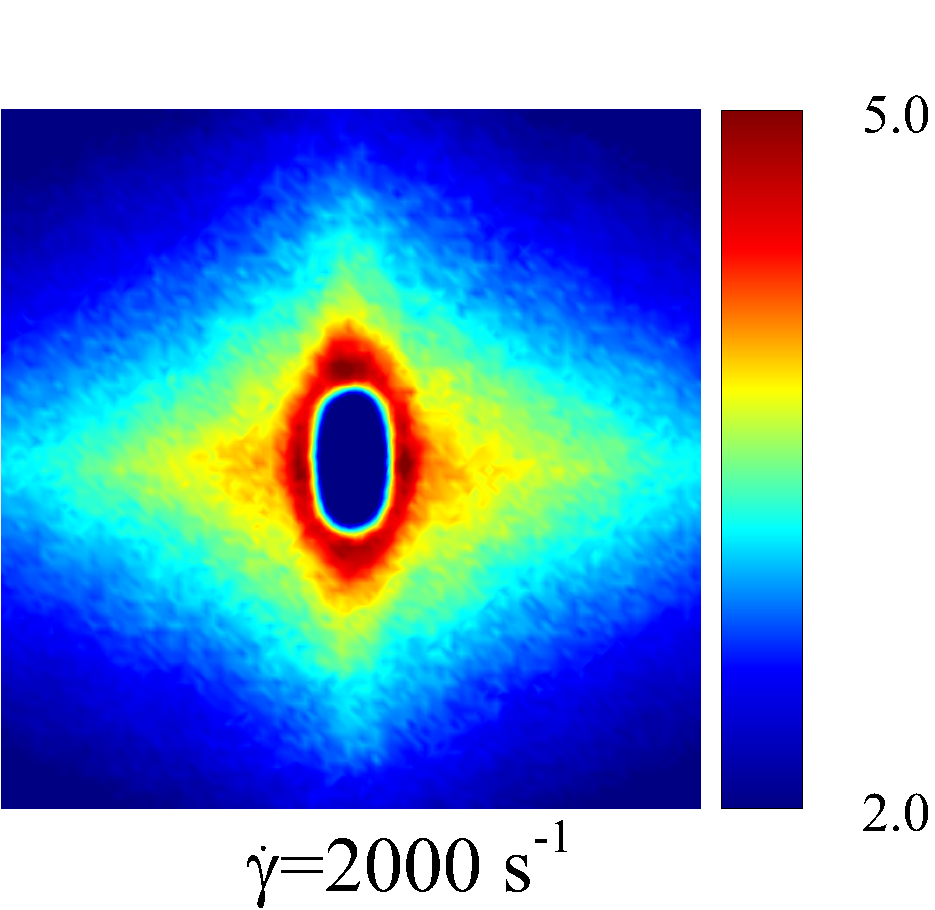}
  \caption{RBC-NP partial pair distribution function, $g_{21}(\boldsymbol{r})$, projected onto the $xy$, $xz$ and $yz$ planes in the long-time regime under various hemorheological conditions, where the PPDF contour has the horizontal edge aligned with the first axis (e.g., $xy$ PPDF contour has the horizontal edge aligned with $x$ axis.). The edge length of the sampling box for computing $g_{21}(\boldsymbol{r})$ is three times of the maximum diameter of the undeformed RBC.}
\label{fig:mstlt}
\end{figure}

Figure \ref{fig:mstlt} presents a matrix of $g_{21}(\boldsymbol{r})$ projections onto the velocity-velocity gradient ($xy$) plane, the velocity-vorticity ($xz$) plane and the velocity gradient-vorticity ($yz$) plane under two shear rates ($\dot{\gamma}$=100 or 2000 $s^{-1}$) and two haematocrits ($\phi$=0.1 or 0.4). In general, all PPDFs feature a large-scale rhombus shape as opposed to typical circular shape commonly observed in rigid-sphere-particle suspensions \citep{FossJR2000,WANG2016JCP,Morris2018JR}. This can be attributed to the disk-shape of RBC that causes geometry-specific anisotropy of the microstructure. The average RBC shapes under various hemorheological conditions, as shown in the central low-intensity region of the PPDF contours, are nicely captured through the PPDF sampling procedure. All PPDFs considered in the $xy$ plane show more intensified distribution (i.e., higher probability of RBC-NP interaction) near the RBC disk surfaces, of which the surface normal directions are more aligned with the compressive axes (in 2nd and 4th quadrants) in accordance with the negative values of $\hat{\mathsfi{D}}^R_{xy}$. In the $xz$ and $yz$ plane, the PPDF distribution tends to be symmetric about the principal axes ($x$, $y$ or $z$), which explains the zero values of the $xz$ and $yz$ diffusivities. 

As the hemorheological condition changes, the detailed configuration of the PPDF within the rhombus structure also varies. At low haematocrit ($\phi$=0.1), the $xy$ PPDF shows fore-aft low intensity similar to rigid particle suspensions \citep{Higdon2011JR}. However, the break of the fore-aft symmetry, being different from the rigid particle suspensions, seems to be related to the RBC orientation algned with the extensional flow axes that is further caused by the tank-treading motion of the RBC membrane \citep{ReasorJFM2013}. As the haematocrit increases to $\phi$=0.4, the overall PPDF around the RBC becomes more intensified, meaning the local NP concentration near RBC surface increases. More interestingly, the fore-aft low PPDF region observed under low haematocrit gets intensified substantially. The change of the NP microstructure with increased haematocrit can be explained by smaller inter-cell separation and hence NP getting squeezed in a smaller inter-cell region. At low shear rate ($\dot{\gamma}$=100 $s^{-1}$), RBC shows a close-to-equilibrium biconcave shape. Increasing shear rate to $\dot{\gamma}$=2\ 000 $s^{-1}$ results in a significant elongation of the average RBC shape in the flow ($x$) direction accompanied by certain contraction in the velocity-gradient ($y$) and vorticity ($z$) directions.

The above configurational changes of $g_{21}(\boldsymbol{r})$ under various shear rates and haematocrits provide possible mechanistic explanations for the nonlinear $\phi$ dependence observed at high shear rates, as discussed in \S \ref{sec:ht}. The quadradic dependence of $\hat{\mathsfi{D}}^R_{xx}$ ($\sim$$\phi^2$), shown in figure \ref{fig:ht} (a), is likely due to the occurrance of three-body RBC-NP interactions in the longitudinal ($x$) direction caused by the compound effect of the elevated NP concentration at the RBC fore-aft surface and the elongation of the RBC in the $x$ direction. The reduced $\phi$ dependence in $\hat{\mathsfi{D}}_{zz}^R$ at high shear rate, as shown in figure \ref{fig:ht} (c), can be attributed to the relatively large contraction of the RBC in the vorticity direction, which reduces the effective $\phi$ in the $z$ direction. The less reduction of the $\phi$ dependence in the $y$ direction, as shown in figure \ref{fig:ht} (c), is owing to the $y$ direction contraction of RBC under shear being less significant than that in the $z$ direction, as clearly indicated in the PPDF contours.

\subsection{Role of RBC deformability}\label{sec:RBCm}
\noindent 
The above PPDF analysis shows prominent RBC morphological change with elevated shear rate, which suggests the RBC deformability may play an important role in causing the nonlinear $\dot{\gamma}$ dependence of the NP diffusion tensor. In this section, we perform numerical experiments to further quantitatively explain the nonlinear $\dot{\gamma}$ scaling of RESID in the intermediate shear-rate regime (100$\leq$$\dot{\gamma}$$\leq$2\ 000 $s^{-1}$), as observed in \S \ref{sec:sr}. Changing shear rate alters both the fluid inertia and RBC deformability, quantified by $\Pen$ and $Ca_G$, respectively. To interrogate the isolated effect of RBC deformability ($Ca_G$), we fix $Ca_G$ by scaling up $G$ while increasing $\Pen$ (through increasing $\dot{\gamma}$).
Two Capillary numbers, $Ca_G$=$0.055$ and $0.55$, are considered corresponding to the $Ca_G$ regime where the nonlinear shear-rate dependence of RESID occurs. For these simulations, we select a fixed haematocrit of $\phi$=$0.4$ and a NP size of $2a_1$=$100\ nm$.
Table \ref{tab:nexp1} lists all parameters and the NP diffusivity values for the cases tested.

\begin{table}
 \begin{center}
  \begin{tabular}{ccccccccccc}
    $\dot{\gamma}\ [s^{-1}]$ & $\phi$ & $a_1/a_2$ & $N^{RBC}$ & $N^{NP}$ & $\Pen$ & $Ca_G$ & $\mathsfi{D}_{xx}^{\infty}/\mathsfi{D}^B$ & $\mathsfi{D}_{yy}^{\infty}/\mathsfi{D}^B$ & $\mathsfi{D}_{zz}^{\infty}/\mathsfi{D}^B$ & $\mathsfi{D}_{xy}^{\infty}/\mathsfi{D}^B$ \\[4pt]
      10 & \textcolor{black}{0.4} & 0.017 & \textcolor{black}{208} & 5\ 000 & 0.0066 & 0.055 & \textcolor{black}{3.4} & \textcolor{black}{1.2} & \textcolor{black}{1.2} & \textcolor{black}{0.06}\\[1.pt]
      100 & \textcolor{black}{0.4} & 0.017 & \textcolor{black}{208} & 5\ 000 & 0.066 & 0.055 & \textcolor{black}{29.2} & \textcolor{black}{4.1} & \textcolor{black}{3.8} & \textcolor{black}{-2.0}\\[1.pt]
       1\ 000 & \textcolor{black}{0.4} & 0.017 & \textcolor{black}{208} & 5\ 000 & 0.66 & 0.055 & \textcolor{black}{317.6} & \textcolor{black}{31.6} & \textcolor{black}{23.8} & \textcolor{black}{-36.1} \\[1.pt]
       10\ 000 & \textcolor{black}{0.4} & 0.017 & \textcolor{black}{208} & 5\ 000 & 6.60 & 0.055 & \textcolor{black}{2999.8} & \textcolor{black}{286.9} & \textcolor{black}{248.6} & \textcolor{black}{-416.0}\\[1.pt]
       
       10 & \textcolor{black}{0.4} & 0.017 & \textcolor{black}{208} & 5\ 000 & 0.0066 & 0.55 & \textcolor{black}{14.6} & \textcolor{black}{1.1} & \textcolor{black}{1.1} & \textcolor{black}{0.35}\\[1.pt]
       100 & \textcolor{black}{0.4} & 0.017 & \textcolor{black}{208} & 5\ 000 & 0.066 & 0.55 & \textcolor{black}{247.6} & \textcolor{black}{3.3} & \textcolor{black}{2.6} & \textcolor{black}{-1.66}\\[1.pt]
       1\ 000 & \textcolor{black}{0.4} & 0.017 & \textcolor{black}{208} & 5\ 000 & 0.66 & 0.55 & \textcolor{black}{2095.9} & \textcolor{black}{23.2} & \textcolor{black}{18.8} & \textcolor{black}{-38.1}\\[1.pt]
       10\ 000 & \textcolor{black}{0.4} & 0.017 & \textcolor{black}{208} & 5\ 000 & 6.60 & 0.55 & \textcolor{black}{31698.3} & \textcolor{black}{209.0} & \textcolor{black}{158.3} & \textcolor{black}{-499.3}\\[1.pt]
  \end{tabular}
  \caption{Simulation data for numerical experiments concerning the isolated RBC deformation effect in NP-RBC suspensions. The RBC membrane shear modulus, $G$, is re-scaled to numerically match specific $Ca_G$ while changing shear rate. Here, the NP size is fixed to $2a_1$=$100\ nm$ RBC has an effective radius of $a_2$=$2.9\ \mu m$. Brownian diffusivity is calculated by $\mathsfi{D}^B$=$k_B T/6\mu \pi a_1$ at a temperature of $T$=$310\ K$.}{\label{tab:nexp1}}
 \end{center}
\end{table}

\begin{figure}
\centering
\includegraphics[width=0.45\columnwidth]{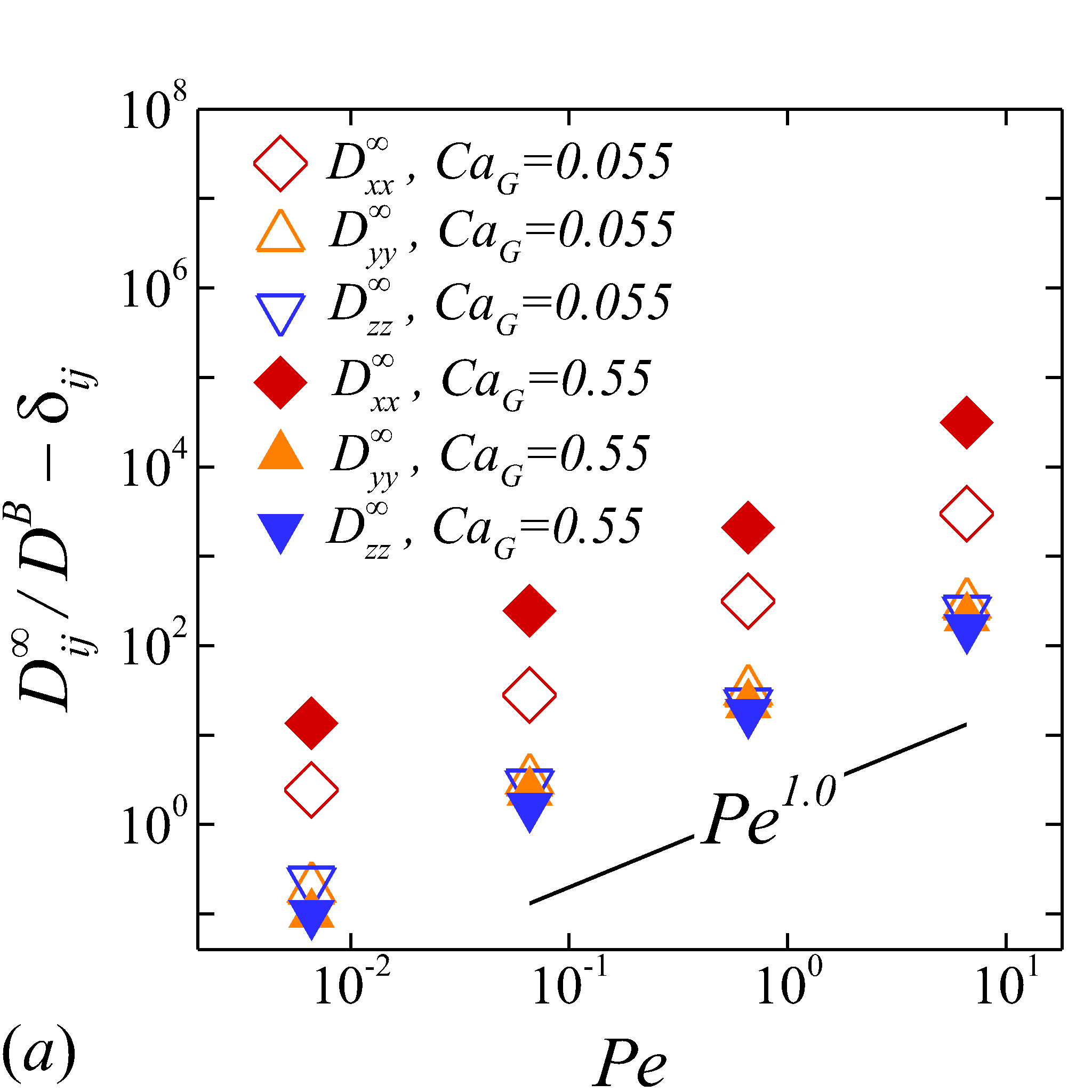}  
\hspace{0.1 cm}
\includegraphics[width=0.45\columnwidth]{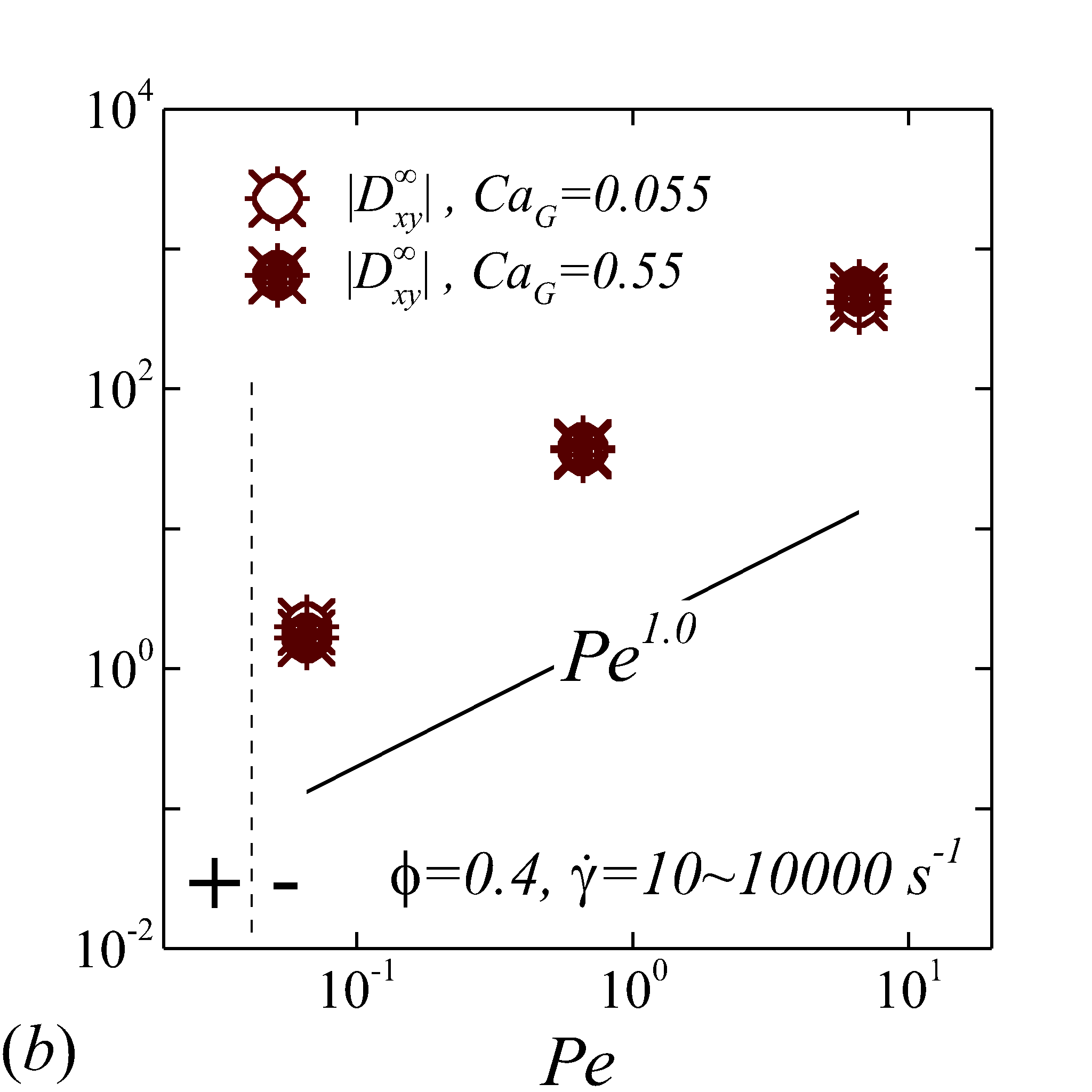}
\caption{\textcolor{black}{The $\Pen$ dependence of the (a) diagonal and (b) off-diagonal RESID terms at fixed capillary number (by rescaling $G$) with $\phi$=$0.4$. Two RBC capillary numbers, $Ca_G$=$0.55$ and $0.055$, are considered to see the isolated effect of changing RBC deformability ($Ca_G$) or adjusting the inertial effect ($\Pen$).}}
\label{fig:Ca}
\end{figure}

As presented in figure \ref{fig:Ca} (a, b), instead of exhibiting nonlinear $\Pen$ scaling as observed in \ref{sec:sr}, all RESID components show linear dependence on $\Pen$ at fixed RBC capilary number of $Ca_G$=$0.55$ or $0.055$ across the wide range of shear rates. This recovers the linear $\Pen$ dependence of solute/self-diffusivity in sheared rigid particle suspensions \citep{Zydney1988,FossJFM1999,Sierou2004} since we intentionally fix the RBC deformation ($Ca_G$) while adjusting $\Pen$.

Moreover, an increase of $Ca_G$ (from 0.055 to 0.55) leads to an increase of $\hat{D}_{xx}^R$ but a decrease of $\hat{\mathsfi{D}}_{yy}^R$ and $\hat{\mathsfi{D}}_{zz}^R$, as shown in figure \ref{fig:Ca} (a). These results are direct evidence indicating that the nonlinear shear-rate scaling of the diagonal RESID terms observed in \S \ref{sec:sr} is due to the variation of RBC deformability ($Ca_G$) induced by changing shear rate at fixed $G$. More specifically, the increase of $Ca_G$ with increasing shear rate at 100$<$$\dot{\gamma}$$<$$2\ 000\ s^{-1}$ under fixed $G$ leads to the superlinear shear-rate dependence of longitudinal diffusivity and meanwhile the sublinear dependence of cross-stream diffusivities. In figure \ref{fig:Ca} (b), the off-diagonal diffusivity shows insignificant change at different $Ca_G$, in consistency with its mostly linear shear-rate dependence shown in figure \ref{fig:sr} (d).

\subsection{Relevant length scale}\label{sec:NPsize}
\begin{table}
 \begin{center}
  \begin{tabular}{ccccccccccc}
    $\dot{\gamma}\ [s^{-1}]$ & $\phi$ & $a_1/a_2$ & $N^{RBC}$ & $N^{NP}$ & $\Pen$ & $Ca_G$ & $\mathsfi{D}_{xx}^{\infty}/\mathsfi{D}^B$ & $\mathsfi{D}_{yy}^{\infty}/\mathsfi{D}^B$ & $\mathsfi{D}_{zz}^{\infty}/\mathsfi{D}^B$ & $\mathsfi{D}_{xy}^{\infty}/\mathsfi{D}^B$ \\[4pt]
       1\ 000 & 0.2 & 0.017 & 104 & 5\ 000 & 0.66 & 0.55 &   211.1 & 12.5 & 10.5 & -14.9\\[1.pt]
       1\ 000 & 0.2 & 0.034 & 104 & 5\ 000 & 5.28 & 0.55 &   801.1 & 26.3 & 21.9 & -34.1\\[1.pt]
       1\ 000 & 0.2 & 0.069 & 104 & 5\ 000 & 42.28 & 0.55 &   1491.9 & 53.3 & 38.1 & -64.7\\[1.pt]
       1\ 000 & 0.2 & 0.14 & 104 & 500 & 338.23 & 0.55 &   3309.1 & 106.9 & 84.9 & -186.8\\[1.pt]
       10 & 0.2 & 0.017 & 104 & 5\ 000 & 0.0066 & 0.55 & 3.1 & 1.0 & 1.1 &  0.2\\[1pt]
       100 & 0.2 & 0.017 & 104 & 5\ 000 & 0.066 & 0.55 & 40.9 & 2.3 & 1.8 & -1.3\\[1.pt]
       10\ 000 & 0.2 & 0.017 & 104 & 5\ 000 & 6.60 & 0.55 &  6109.4 & 113.0 & 102.2 & -270.5\\[1.pt]
  \end{tabular}
  \caption{Simulation data for numerical experiments concerning the relevant length scale in NP-RBC suspensions. NP size is varied in the range of $2a_1$=50$\sim$800 $nm$. RBC has an effective radius of $a_2$=$2.9\ \mu m$. Brownian diffusivity is calculated by $\mathsfi{D}^B$=$k_B T/6\mu \pi a_1$ at temperature $T$=$310\ K$. For the $800\ nm$ case, the number of NP is set to 500 to satisfy the dilute condition for the NP volume fraction. The last three cases are performed with adjusted $G$ to fix $Ca_G$.}{\label{tab:nexp2}}
 \end{center}
\end{table}
\noindent Although the shear-adaption of the RBC deformability is shown to be responsible for the nonlinear shear-rate dependence of the RESID, it is still unclear in what way it alters the RESID. In this section, we further identify the characteristic length scale relevant to the RESID to gain more in-depth understanding of this nonlinear phenomenon.

In a series of experiments measuring the particle diffusivity in a concentrated non-colloidal suspension, \citet{Breedveld1998} show that the ratio of the diffusivity of fluid tracers to the self-diffusivity of non-colloidal particles is close to unity. Through DNS of the cellular blood flow in a micro-channel, \citet{Zhao2012} show that the cross-stream diffusivity of platelets ($\sim$2$\ \mu m$) in blood exhibits similar magnitude to that of passive tracers. These findings imply that the shear-induced diffusivity is insensitive to the size of the scarce (in terms of volume fraction), small particles. Following these results, we hypothesize that RESID is insensitive to the size of NP. Moreover, the characteristic length scale associated with the RESID should be the size of RBC. To confirm this hypothesis, we evaluate the RESID at fixed hemorheological condition with various NP size. The NP size is kept in submicron-scale such that the NP-RBC size ratio satisfies $a_1/a_2$$\ll$$1$. The volume fraction of the NP phase are kept below 0.32$\%$ to satify the dilute condition. The set up and measured diffusivities are listed in table \ref{tab:nexp2}.

\begin{figure}
\centering
\includegraphics[width=0.35\columnwidth]{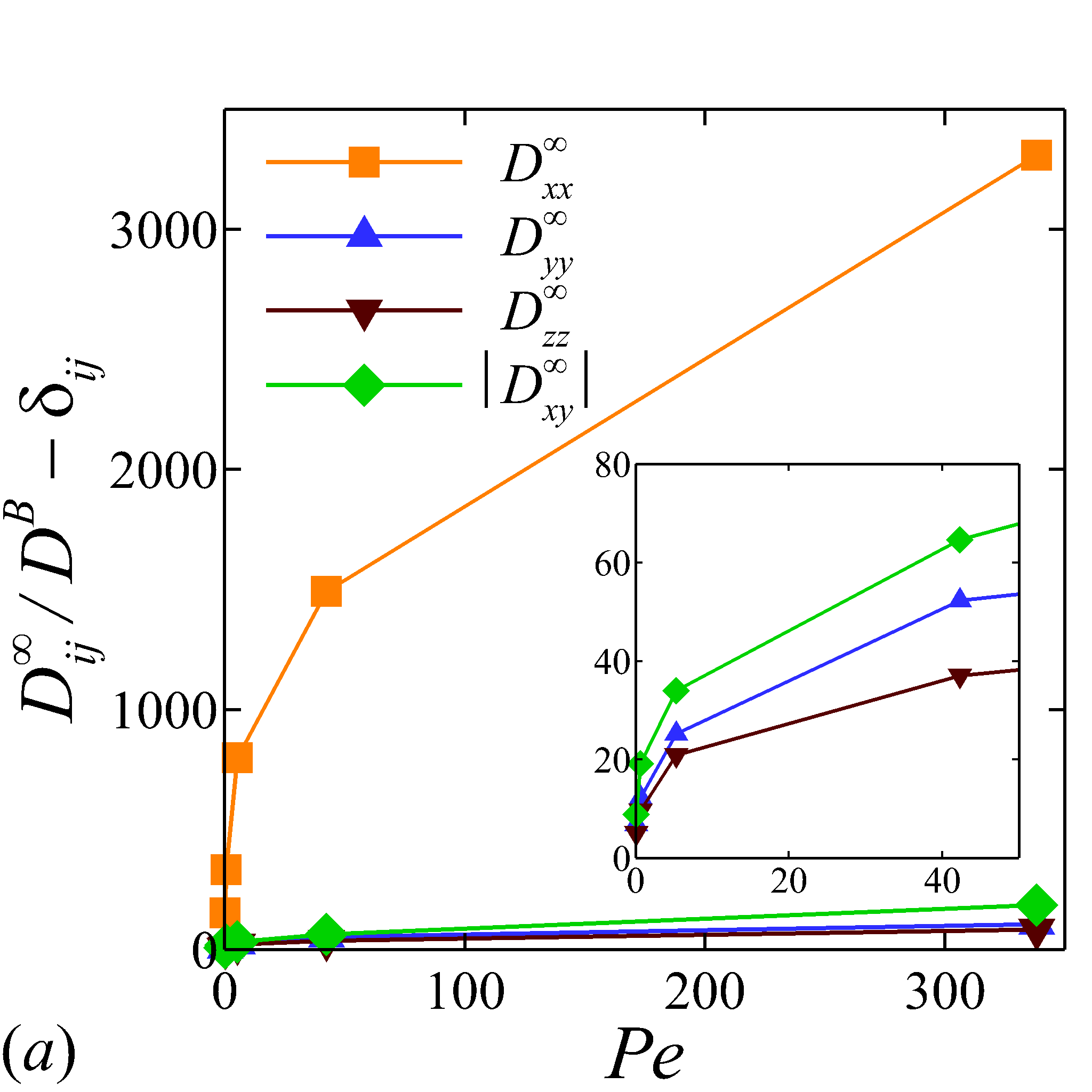}
\includegraphics[width=0.31\columnwidth]{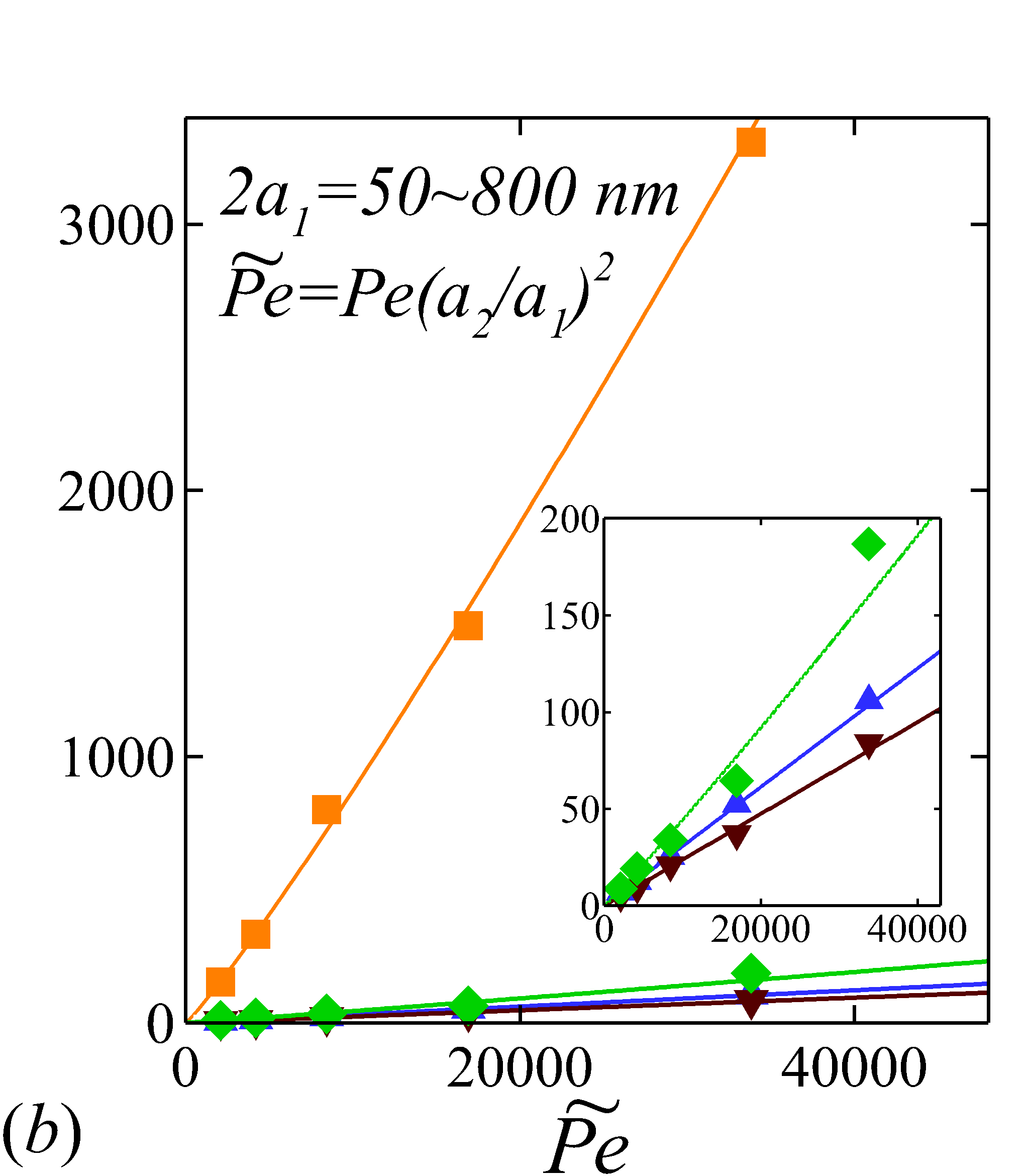}
\includegraphics[width=0.33\columnwidth]{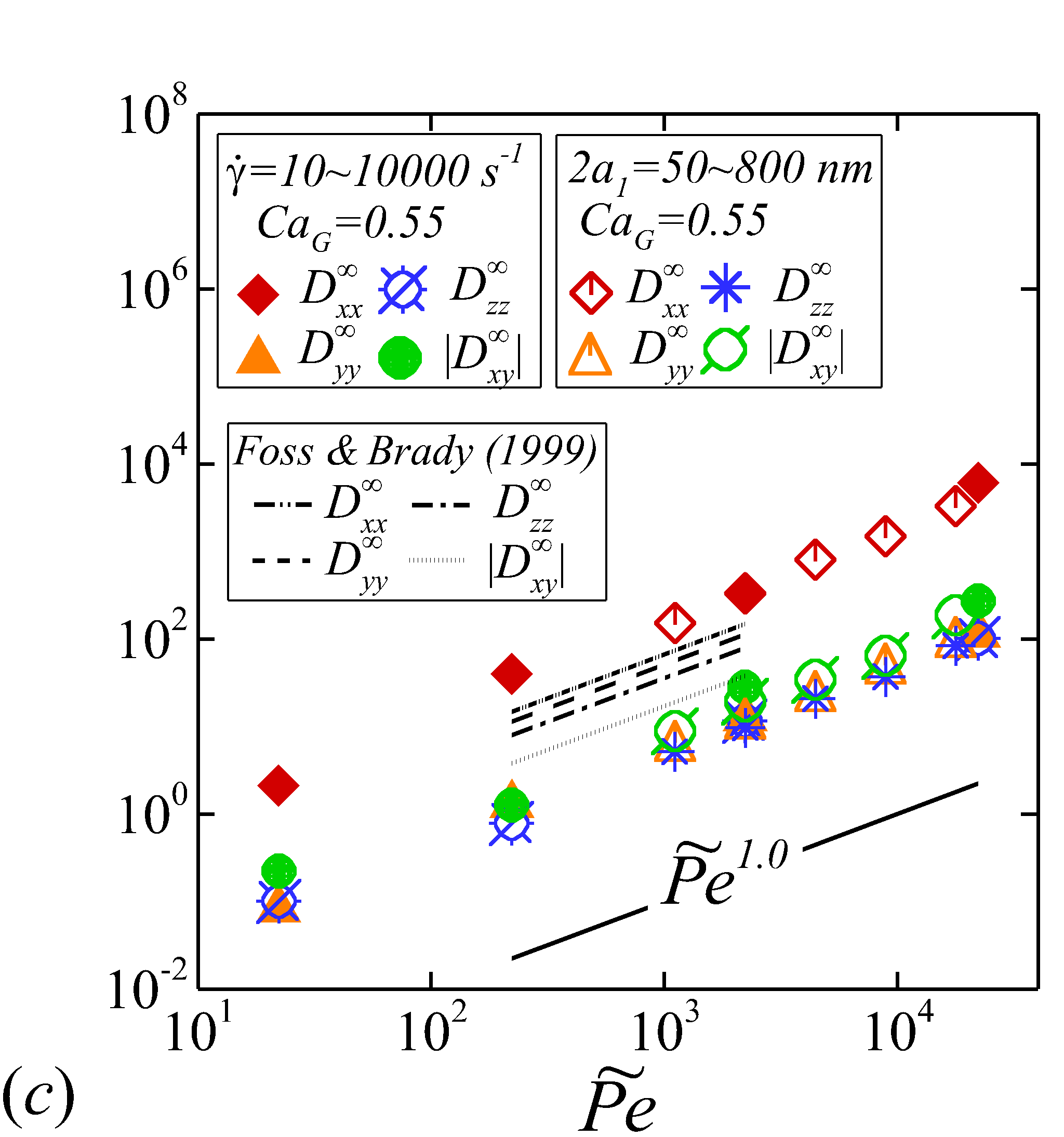}

\caption{RESID plotted against (a) P\'eclet number $\Pen$ and (b) the rescaled P\'eclet number, $\ptilde$, at $\dot{\gamma}$=$1\ 000\ s^{-1}$ and $\phi$=$0.2$ in logarithm scales. Both $\Pen$ and $\ptilde$ are adjusted with NP size in the ranges of $2a_1$=$50$$\sim$$800\ nm$. \textcolor{black}{(c) Linear dependence of RESID on $\ptilde$ at fixed $Ca_G=0.55$ and $\phi$=$0.2$; $\ptilde$ is adjusted by shear rate (with $2a_1$=$100\ nm$) or NP size (with $\dot{\gamma}$=$1\ 000\ s^{-1}$). The diffusion tensor in monodisperse colloidal suspensions \citep{FossJFM1999} are also plotted for comparison, where $\ptilde$ degrades to $\Pen$.}}
\label{fig:size}
\end{figure}

In Figure \ref{fig:size} (a), we plot the RESID versus $\Pen$ at $\dot{\gamma}$=$1\ 000\ s^{-1}$ and $\phi$=$0.2$. Here, $\Pen$ is adjusted by NP size in the range of $50$$\leq$$2a_1$$\leq$$800\ nm$. A sublinear $\hat{\mathsfi{D}}_{ij}^R$-$\Pen$ relationship is observed for all diffusion coefficients. Figure \ref{fig:size} (b) further plots the same data as figure \ref{fig:size} (a), but with $\Pen$ rescaled by the NP-RBC size ratio as
\begin{equation}
  \ptilde=\Pen(\frac{a_2}{a_1})^2=\frac{\dot{\gamma}a_2^2}{\mathsfi{D}^B}. 
  \label{eqn:Pemod}
\end{equation}
This simple rescaling leads to a strong linear relationship between all RESID terms and the rescaled P\'eclet number, $\ptilde$. Moreover, the observed linear relationship, $(\mathsfi{D}_{ij}^\infty/\mathsfi{D}^B-\delta_{ij})$$\sim$$\textit{O}(\ptilde)$, can be deduced to show that $\mathsfi{D}_{ij}^R$$\sim$$\textit{O}(\dot{\gamma}a_2^2)$, which indicates that the dimensional RESID is indeed insensitive to the NP size within the range of NP sizes considered in the current study. This also suggests NP size at submicron plays a secondary role in affecting the NP apparent diffusivity primarily through altering BD.
\begin{figure}
   \centerline{
\includegraphics[width=0.98\columnwidth]{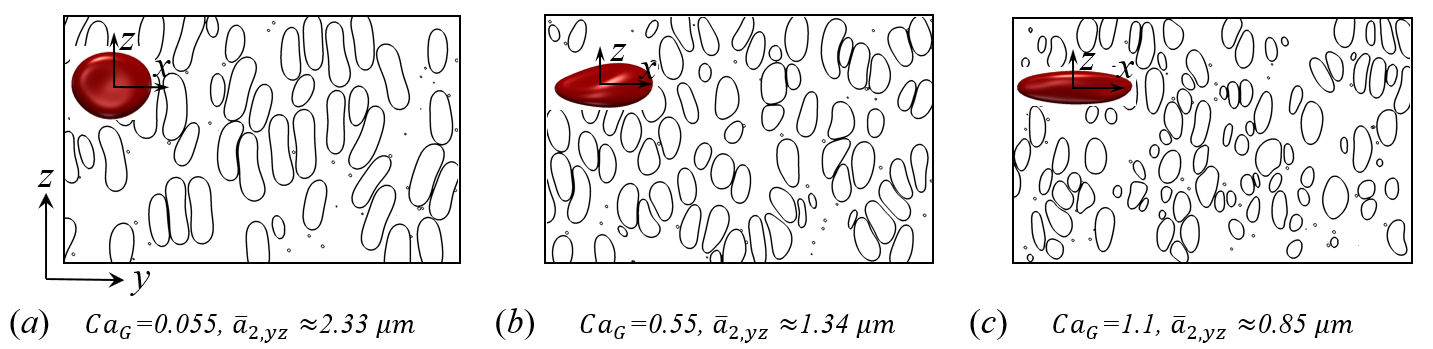}
}
\caption{\textcolor{black}{RBC and NP distributions and morphology in the cross-stream plane ($y$-$z$) with (a) $Ca_G$=$0.055$, (b) $Ca_G$=$0.55$ and (c) $Ca_G$=$1.1$ at $\Pen$=$0.066$ and $\phi$=$0.4$. Large non-circular disks are RBC cross-sectional contours. Small circles are NP cross-sectional contours. $\bar{a}_{2,yz}$ can be estimated from the PPDF analysis.}}
\label{fig:RBCNP2}
\end{figure}

To further justify the universality of the rescaled P\'eclet number, $\ptilde$, i.e., whether the RBC size is the reasonable length scale associated with RBC-enhanced diffusion or not, we plot RESID against $\ptilde$ at $Ca_G$=$0.55$ and $\phi$=$0.2$, with $\ptilde$ adjusted by either shear rate or NP size. As shown in figure \ref{fig:size} (c), the data points for particular RESID coefficient are found to be well aligned on the same linear $\ptilde$ scaling line. \textcolor{black}{This directly confirms that the RBC size is the characteristic length scale governing the RBC-enhanced diffusion, and $\ptilde$ is a more general nondimensional term quantifying the RESID.}

\textcolor{black}{Identifying RBC size being the RESID length scale helps further understanding the nonlinear shear-rate dependence of the RESID.} In figure \ref{fig:RBCNP2}, we present snapshots of RBC and NP distribution and morphology in the mid cross-stream cross-sectional plane. As shown in figure \ref{fig:RBCNP2}, the $x$-direction elongation of RBCs under high shear causes a contraction of RBCs in the $yz$ plane, which causes a reduction of the average effective RBC size (i.e., the length scale $\bar{a}_{2,yz}$) and hence a decrease of the effective $\ptilde$ in cross-stream directions. This explains the sublinear shear-rate scaling of the cross-stream RESID when increasing $Ca_G$. Likewise, the super-linear shear-rate scaling of the longitudinal RESID can be attributed to the elongation of RBC in the streamwise direction, i.e. the increase of the RESID length scale in the longitudinal direction. The change of the RESID length scales in different principal directions due to shear is the root cause of the nonlinear shear-rate dependence of the NP diffusion tensor.

\subsection{Empirical correlations}\label{sec:corr}
\noindent In this section, we construct empirical correlations for the long-time NP diffusion tensor in sheared blood based on the scaling observations in the previous sections. \textcolor{black}{Since all RESID coefficients scales linearly with $\ptilde$ at fixed $Ca_G$, and the nonlinear shear-rate dependence of RESID is primarily due to the variation of $Ca_G$, as demonstrated in \S \ref{sec:RBCm} and \S \ref{sec:NPsize}; there hence exists a scaling relation, $\hat{\mathsfi{D}}_{ij}^R$$\sim$$\textit{O}(\phi^{p_1}\ptilde Ca_G^{p_2})$, such that the exponents $p_1$ and $p_2$ can be estimated through matching the nonlinear $\phi$ and $\dot{\gamma}$ scalings, respectively, as observed in \S \ref{sec:hemo}.} Therefore, empirical correlations of NP long-time diffusivities can be constructed as functions of $\phi$, $\ptilde$ and $Ca_G$ in the hemorheological range of 10$\leq$$\dot{\gamma}$$\leq$2\ 000 $s^{-1}$ and 0.1$\leq$$\phi$$\leq$0.4 as following 
\begin{subequations}
\begin{equation}
\frac{\mathsfi{D}_{xx}^{\infty}}{\mathsfi{D}^B}=
    \begin{cases}
      1+\mathcal{C}_{xx}^l\phi\ptilde, & \ \ \ \ \ \ 0.0055\leq Ca_G\leq 0.055 \\
      1+\mathcal{C}_{xx}^h\phi^{Ca_G+0.945}\ptilde Ca_G^{0.8}, & \ \ \ \ \ \ 0.055< Ca_G\leq 1.10  
    \end{cases}
  \label{eqn:corr1}
\end{equation}
\begin{equation}
  \frac{\mathsfi{D}_{yy}^{\infty}}{\mathsfi{D}^B}=
  \begin{cases}
      1+\mathcal{C}_{yy}^l\phi\ptilde, & \ \ \ \ \ \ \ \ \ 0.0055\leq Ca_G\leq 0.055 \\
      1+\mathcal{C}_{yy}^h\phi\ptilde Ca_G^{-0.3}, & \ \ \ \ \ \ \ \ \ 0.055< Ca_G\leq 1.10
    \end{cases}
  \label{eqn:corr2}
\end{equation}
\begin{equation}
  \frac{\mathsfi{D}_{zz}^{\infty}}{\mathsfi{D}^B}=
  \begin{cases}
      1+\mathcal{C}_{zz}^l\phi\ptilde, & \ \ 0.0055\leq Ca_G\leq 0.055 \\
      1+\mathcal{C}_{zz}^h\phi^{1.022-0.4Ca_G}\ptilde Ca_G^{-0.2}, & \ \ 0.055< Ca_G\leq 1.10 
    \end{cases}
  \label{eqn:corr3}
\end{equation}
\begin{equation}
  \frac{\mathsfi{D}_{xy}^{\infty}}{\mathsfi{D}^B}=\frac{\mathsfi{D}_{yx}^{\infty}}{\mathsfi{D}^B}=
  \begin{cases}
      \mathcal{C}_{xy}^l, & \ \ \ \ \ \ \ \ \ \ \ \ \ \ \ 0.0055\leq Ca_G\leq 0.055 \\
      \mathcal{C}_{xy}^h\phi \ptilde, & \ \ \ \ \ \ \ \ \ \ \ \ \ \ \ 0.055< Ca_G\leq 1.10
    \end{cases} 
  \label{eqn:corr4}
\end{equation}
\end{subequations}
where $\mathcal{C}_{ij}^q$ with $q$$\in$$\{l,h\}$ are constants fitting the correlation values with the simulation measurements at low or high shear rates. The dimensional NP diffusion tensor can then be written in terms of the conventional shear-induced diffusion scaling, $\textit{O}(\dot{\gamma}a_2^2)$, as
\begin{equation}
  \mathsfbi{D}^{\infty}=\mathsfi{D}^B\mathsfbi{I}+\phi\dot{\gamma}a_2^2 \mathsfbi{M}^q, 
  \label{eqn:corr9}
\end{equation}

\noindent where the anisotropic tensor, $\mathsfbi{M}^q$, according to the severity of the RBC deformability ($Ca_G$=$\mu\dot{\gamma}a_2/G$) at different level of shear rates, yields piecewise expressions:
\begin{subequations}
\begin{equation}
  \mathsfbi{M}^{l} = 
\begin{pmatrix}
\mathcal{C}_{xx}^{l} & \mathcal{C}_{xy}^{l} &                        0  \\ 
\mathcal{C}_{yx}^{l} & \mathcal{C}_{yy}^{l} &                        0  \\ 
                       0 &                        0 & \mathcal{C}_{zz}^{l}
\end{pmatrix},\ \ \ \ \ \ \ \ \ \ \ \ \ \ \ \ \ 10\leq\dot{\gamma}\leq 100\ s^{-1}
  \label{eqn:corr10}
\end{equation}
\begin{equation}
\begin{aligned}
  \mathsfbi{M}^{h} = 
\begin{pmatrix}
\mathcal{C}_{xx}^{h}\phi^{Ca_G-0.055} Ca_G^{0.8} & \mathcal{C}_{xy}^{h} &                        0  \\  
\mathcal{C}_{yx}^{h} & \mathcal{C}_{yy}^{h}/Ca_G^{0.3} &                        0  \\ 
                       0 &                        0 & \mathcal{C}_{zz}^{h} /(\phi^{2Ca_G+0.11} Ca_G)^{0.2}
\end{pmatrix},\ \ \ \ \ \ \ \ \\ \ \ \ \ \ \ 100<\dot{\gamma}\leq2\ 000\ s^{-1}
\end{aligned}
  \label{eqn:corr11}
\end{equation}
\end{subequations}
where $\mathsfbi{M}^q$ quantifies the anisotropic behavior of the NP diffusion tensor and characterizes its departure from the conventional shear-induced diffusion scaling, $\textit{O}(\dot{\gamma}a_2^2)$, due to the presence of the deformable RBC phase. Table \ref{tab:corr} lists all the constants, $\mathcal{C}_{ij}^q$. Each constant is obtained by calculating the slope of the best linear fit to the data set, ($\mathsfi{D}_{ij}^{(R,m)}$,$\mathsfi{D}_{ij}^{(R,th)}/\mathcal{C}_{ij}^q$), for specific $i$, $j$ and $q$ at various $\dot{\gamma}$ and $\phi$. Here, $\mathsfi{D}_{ij}^{(R,m)}$=$\mathsfi{D}_{ij}^{\infty}-\mathsfi{D}^B \delta_{ij}$ denotes the measured RESID evaluated by the calculated $\mathsfi{D}_{ij}^{\infty}$ subtracted by the theoretical BD; $\mathsfi{D}_{ij}^{(R,th)}$=$\phi\dot{\gamma}a_2^2 \mathsfi{M}_{ij}^q$ denotes the theoretical diffusivity based on the proposed empirical correlations. $\mathcal{C}_{xy}^l$ is set to zero given its small magnitude. $\mathcal{C}_{xy}^h$ is negative, suggesting the predominant direction of NP migration in the long-time regime is along the contractile flow direction under high shear flow \citep{FossJFM1999}. For cases at $\dot{\gamma}$$<$10 $s^{-1}$ and $\phi$$<$0.1, $\mathsfbi{D}^{\infty}$ can be approximated by the isotropic Brownian diffusivity provided the small magnitude of the RESID. 

\begin{figure}
   \centerline{
\includegraphics[width=0.85\columnwidth]{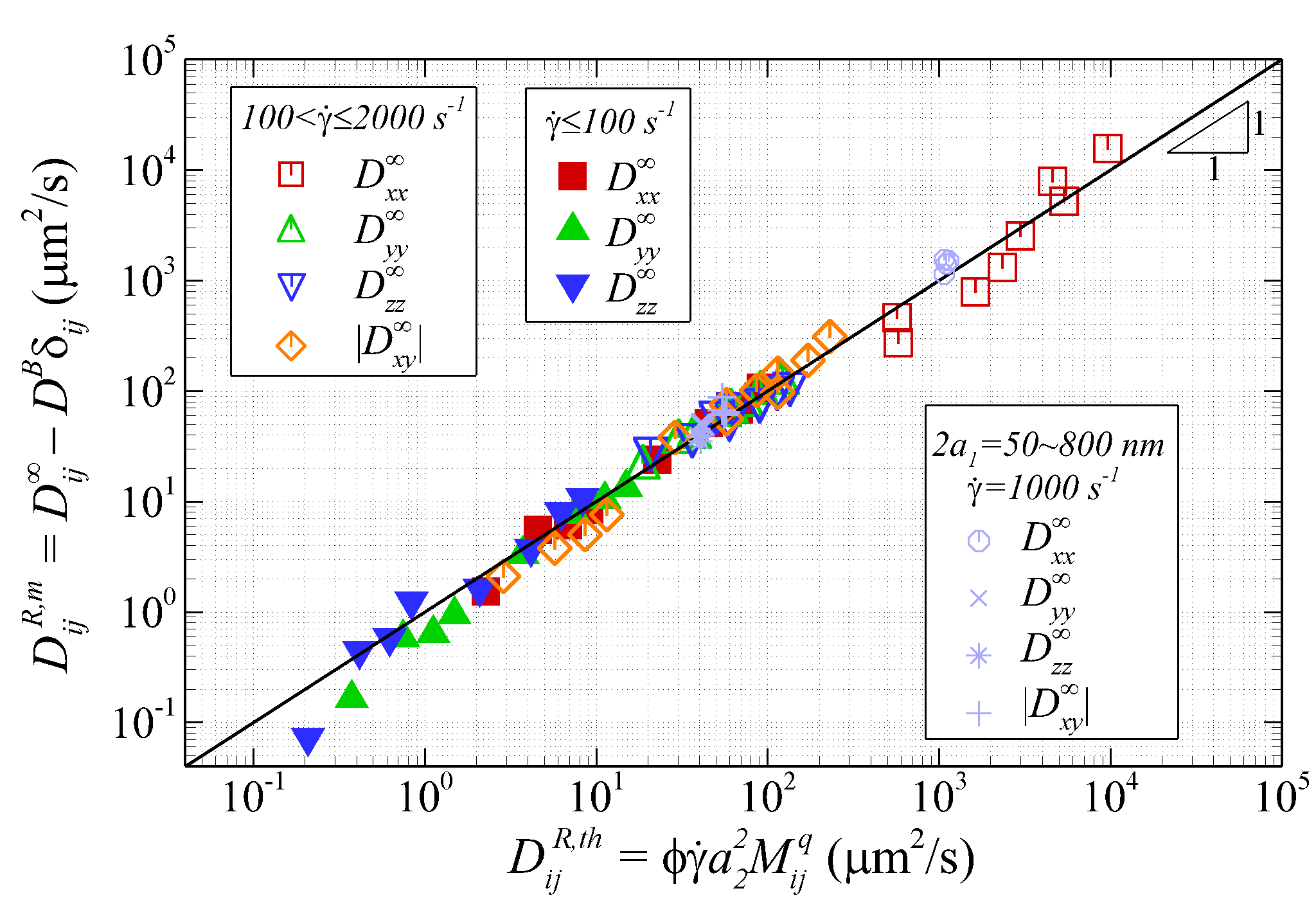}
}
\caption{Measured NP RESID, $\mathsfi{D}_{ij}^{(R,m)}=\mathsfi{D}_{ij}^{\infty}-\mathsfi{D}^B \delta_{ij}$, as a function of the theoretical RESID, $\mathsfi{D}_{ij}^{(R,th)}=\phi\dot{\gamma}a_2^2 \mathsfi{M}_{ij}^q$, based on the empirical correlations presented in equations \ref{eqn:corr10} and \ref{eqn:corr11}. For cases with $\dot{\gamma}$$\leq$$100\ s^{-1}$ (filled symbols), $\mathsfi{M}_{ij}^l$ is applied to calculate $D_{ij}^{(R,th)}$. For cases with $\dot{\gamma}$$>$$100\ s^{-1}$ (open symbols), $\mathsfi{M}_{ij}^h$ is applied to calculate $\mathsfi{D}_{ij}^{(R,th)}$.}
\label{fig:corr}
\end{figure}
\begin{table}
 \begin{center}
  \begin{tabular}{cccccccc}
    $\mathcal{C}_{xx}^{l}$ & $\mathcal{C}_{xx}^{h}$ & $\mathcal{C}_{yy}^{l}$ & $\mathcal{C}_{yy}^{h}$ & $\mathcal{C}_{zz}^{l}$ & $\mathcal{C}_{zz}^{h}$ & $\mathcal{C}_{xy}^{l}$ & $\mathcal{C}_{xy}^{h}$ \\[4pt]
       0.281 & 2.86 & 0.0432 & 0.0181 & 0.0241 & 0.0135 & 0 &  -0.0332 \\[2.0 pt]
  \end{tabular}
  \caption{Correlation constants for the empirical correlations of NP diffusion tensor in sheared cellular blood flow.}{\label{tab:corr}}
 \end{center}
\end{table}

Figure \ref{fig:corr} plots the measured NP diffusivities (subtracted by BD) versus the theoretical estimation based on the empirical correlations. The good collapse of the diffusivity measurements on the $\mathsfi{D}_{ij}^{(R,m)}$=$\mathsfi{D}_{ij}^{(R,th)}$ line demonstrates the empirical correlations can well reproduce the numerically measured diffusivity tensor of NP in sheared blood.

\section{Conclusions}\label{sec:discon}
The dispersion of NP in cellular blood flow under unbounded, homogeneous shear has been investigated over a wide range of shear rate and haematocrit using a LB-LD-SL multiscale complex blood flow solver. In the short-time regimes, NP dispersive anomalies are observed and attributed to the transient morphology and orientation change of RBCs under high shear and high haematocrit. In the long-time regimes, results for the long-time diffusivity in the velocity gradient direction agree well with existing experimental data. The long-time NP diffusion tensor has been described as a function of shear-rate and haematocrit with various power-law scalings. 

\textcolor{black}{By plotting the RBC-NP PPDF, the NP microstructure in sheared blood has been visualized for the first time that features a rhombus configuration with the detailed inner structure changes according to specific hemorheological conditions. The RBC-NP PPDF analysis also suggests a novel approach to visualizing the average RBC morphology in concentrated RBC suspensions subject to different hemorheological conditions. Under high shear rate, the $\phi^2$ dependence in $\hat{\mathsfi{D}}_{xx}^R$ is proposed to be related to the $x$-elongation of RBC and elevated RBC-NP PPDF near the fore-aft region, which together suggests possible more-than-two-body interaction occurred particularly in the streamwise direction. The sublinear $\phi$ dependence is suggested to be related to the reduced effective $\phi$ owing to the substantial contraction of RBC in the vorticity direction subject to high shear. The RBC-NP bidisperse suspension presents an example of highly anisotropic microstructure of particle suspensions caused by the compound effect of particle-shape/orientation anisotropy and shear-flow anisotropy.}

It is also found that there exists a critical shear rate ($\sim$100 $s^{-1}$) around which the RESID shear-rate dependence changes from linear to nonlinear scale. Through numerical experiments, the transition to nonlinear shear-rate scaling of RESID has been related to the prominent change of average RBC morphological state between different shear rate. Specifically, the superlinear scalings ($\dot{\gamma}^{1\sim1.8}$) of $\hat{\mathsfi{D}}_{xx}^R$ are due to the streamwise elongation of RBC, while the sublinar scalings ($\dot{\gamma}^{0.7\sim0.8}$) of $\hat{\mathsfi{D}}_{yy}^R$ and $\hat{\mathsfi{D}}_{zz}^R$ are associated with the cross-stream contraction of RBC (in response to the streamwise elongation). The morphological changes under shear alter the RESID length scale in different principal directions, which has been demonstrated to be the fundamental cause of the nonlinear shear-rate dependence of the RESID. This mechanism is also worth to be distinguished from the causes of the nonlinear shear-rate scaling of the self-diffusivity of RBCs \citep{Gross2014,Mountrakis2016} or deformable capsules \citep{ClausenJFM2011}, where latter has been attributed to the heterogeneous interparticle `collision' due to the cell deformability \citep{Kumar2012}. However, the nonlinear shear-rate dependence of NP diffusion in sheared blood, based on our interrogation, is more associated with a `one-way' mechanism, i.e., the RBC morphological adaptation to shear flow changes the RESID length scale which further alters the NP diffusion rate.

The determination of the rescaled P\'elect number being a more general nondimensional term to describe the severity of RESID enables the comparison between the bidisperse NP-RBC suspension and the monodisperse colloidal suspensions. In the latter scenario where the particle size ratio is one, $\ptilde$ drops to $\Pen$ and `RESID' drops to the particle self-diffusivity. The self-diffusion tensor reported by \citet{FossJFM1999} in a sheared monodisperse colloidal suspensions are plotted in figure \ref{fig:size} (c) for comparison. In both monodisperse and bidisperse scenario, $\hat{D}_{xx}^R$ shows the greatest magnitude among all diffusivity terms; $\hat{D}_{yy}^R$ is slightly greater than $\hat{D}_{zz}^R$. In general, $\hat{D}_{ij}^R$ in RBC-NP suspension shows higher anisotropy than the monodisperse case owing to the geometric asymmetry of RBCs ($x$ being greater than $y$ and $z$ dimensions). Such geometric asymmetry effect can be further increased with $Ca_G$, leading to higher anisotropy of the diffusivity tensor.
In the monodisperse scenario, $\hat{D}_{xy}^R$ is the smallest; while in the NP-RBC bidisperse scenario, $\hat{D}_{xy}^R$ is greater than $\hat{D}_{yy}^R$ and $\hat{D}_{zz}^R$ due to the severe diffusive effect in $x$ direction.

This work, to the authors' knowledge, offers the first detailed study of the complete 3D NP diffusion tensor in cellular blood flow over a wide range of shear rate and haematocrit. The proposed empirical correlations for the NP diffusion tensor offers a constitutive relation that can be adopted by effective continuum models to pursue large-scale NP biotransport applications (e.g. $in\ vivo$ NP drug delivery) with better accuracy.

\vspace{0.3cm}
\section*{Acknowledgments}
The authors acknowledge the support from Sandia National Laboratories under grant 2506X36 and the compuatational resource provided by National Science Foundation under grant TG-CT100012. The authors appreciate the suggestion by one of the reviewers to consider the pair distribution function of NP around RBC in all three planes. Z. Liu acknowledges the constructive discussions with Dr. Jeremy B. Lechman, Prof. Eugene C. Eckstein and Prof. Kurt B. Wiesenfeld. Sandia National Laboratories is a multimission laboratory managed and operated by National Technology and Engineering Solutions of Sandia LLC, a wholly owned subsidiary of Honeywell International Inc. for the U.S. Department of Energy’s National Nuclear Security Administration under contract DE-NA0003525. This paper describes objective technical results and analysis. Any subjective views or opinions that might be expressed in the paper do not necessarily represent the views of the U.S. Department of Energy or the United States Government.

\appendix
\textcolor{black}{
\section{Sensitivity to NP-RBC contact model}\label{appA}
The NP-RBC short-distance interaction is through Morse potential that forbids NP from penetrating the RBC membrane. The potential parameters are adjusted to match the measured inter-cell potential energy, as discussed in \citet{Liu2004}. The Morse potential function is given as
\begin{equation}
  U_{M}(r)=D_e[e^{-2\beta(r-r_0)}-2e^{-\beta(r-r_0)}],\ \ (r\leq r_0) 
  \label{eqn:Morse}
\end{equation}
where $r$ is the normal distance between the particle center to the RBC surface, $r_0$ is a cut-off distance in which no interaction forces are present, $D_e$ is the potential well depth and $\beta$ is a scaling factor. The surface energy is set to $D_e=1\times10^7k_BT$ and the equilibrium distance is set to $r_0=a_1+10\ nm$. 
In figure \ref{fig:appA}, we explore the sensitivity of the NP diffusivity calculation to the adjustment of the Morse model parameters. Figure \ref{fig:appA} (a) and (b) show the $D^{\infty}_{yy}$ and $D^{\infty}_{zz}$ calculation exhibits less than 5\% variation when changing the magnitude of the energy and equilibrium distance, respectively, by up to $\pm 60\%$. This indicates the NP diffusion is largely driven by the hydrodynamic interaction rather than the direct contact between NP and RBC membrane.
}

\textcolor{black}{
\begin{figure}
\centerline{
\includegraphics[width=0.48\columnwidth]{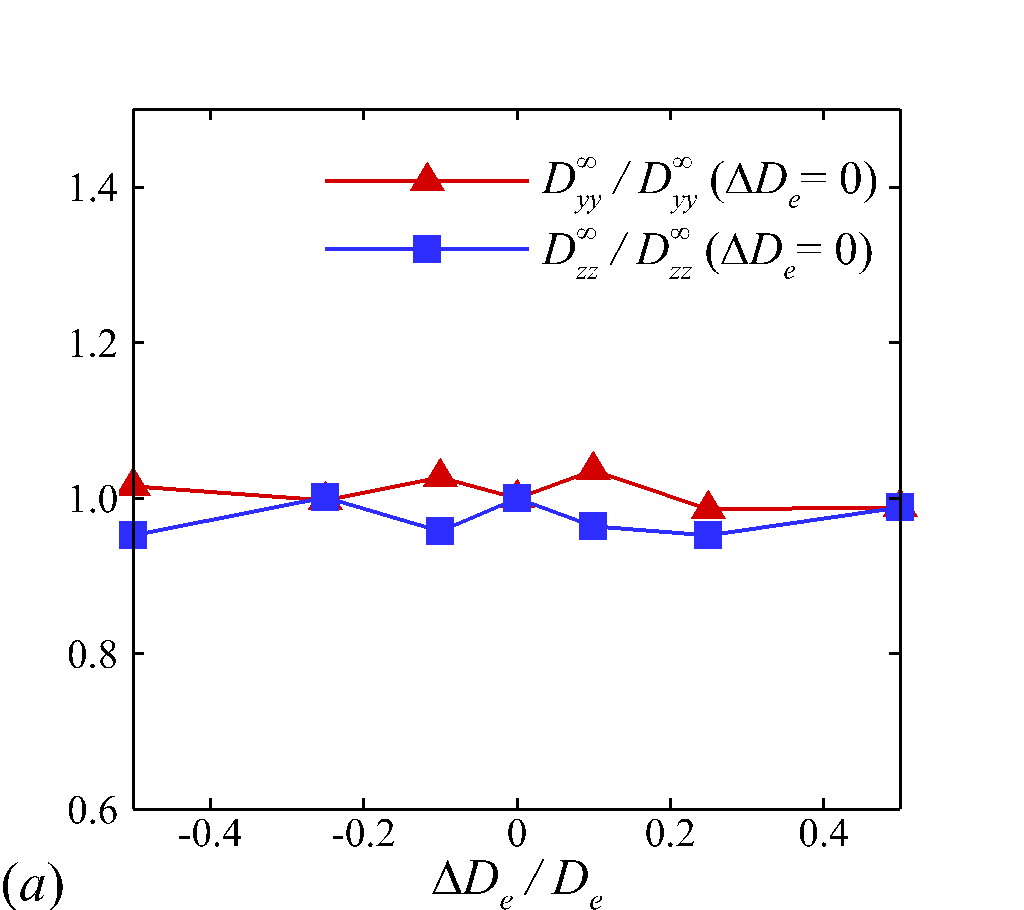}
\includegraphics[width=0.48\columnwidth]{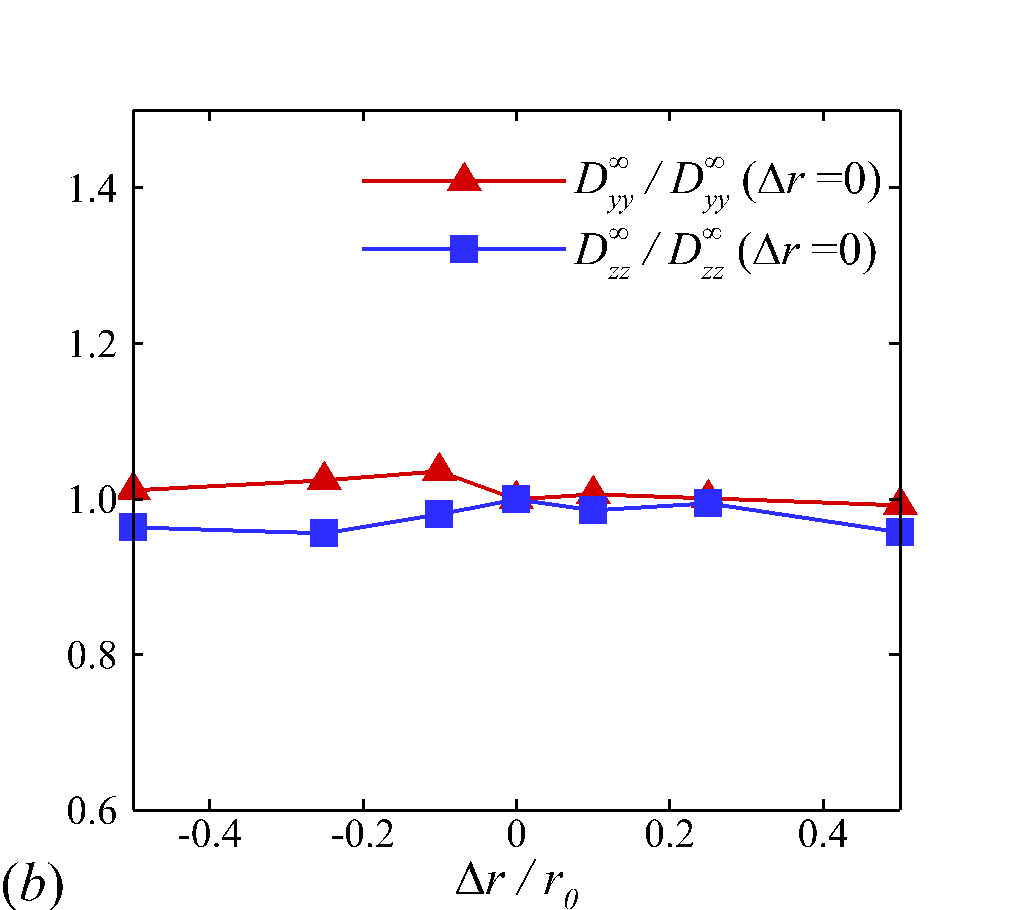}
}
\caption{\textcolor{black}{(a) NP long-time diffusivities in the cross-stream directions plotted against the relative variation of the Morse potential (a) well depth and (b) equilibrium distance.}}
\label{fig:appA}
\end{figure}
\section{Calculation of the partial pair distribution function}\label{appB}
In the bidisperse RBC-NP suspension system, the NP-RBC partial pair distribution function (PPDF), $g_{21}(\boldsymbol{r})$, quantifies the conditional probability of finding a NP (species 1) at a position of $\boldsymbol{r}$ with respect to the geometric center of a single RBC (species 2). This quantity can be calculated as
\begin{equation}
  g_{21}(\boldsymbol{r})=\frac{1}{n_2}\bigg\langle \sum_{i\in LB}\sum_{j\in b} \frac{\delta(\boldsymbol{r}-\boldsymbol{r}_2^i+\boldsymbol{r}_1^j)}{n_1^i L^3_b}\bigg\rangle, 
  \label{eqn:ppdf}
\end{equation}
where $L_b$ defines the cubic box size for PPDF sampling around one RBC, index $j$ goes through all NPs within the local sampling box, and $i$ goes through all RBCs in the computational LB domain. $r_2^i$ and $r_1^j$ denotes the position of RBC and NP, respectively. $n_1^i$ shows the number concentration of NP within the sampling box, while $n_2$ denotes the number concentration of RBC in the entire domain. The angle bracket represents the ensemble average among independent realizations, which in the current case is through time averaging given the ergodic hypothesis. Similar techniques have recently been used in calculating PPDFs in bidisperse and polydisperse rigid particle suspensions \citep{WANG2016JCP,Morris2018JR}. Projection of $g_{21}(\boldsymbol{r})$ to principal planes follows the integration procedure discussed in \citet{Higdon2011JR} but with a smaller integration interval, [$-a_2/4$, $a_2/4$], to better capture the RBC morphological change.
}

\textcolor{black}{
The PPDF sampling box size for all cases is selected to be three times of the RBC maximum diameter at equalibrium in biconcave shape, i.e., $L_b$=24 $\mu m$. To capture the PPDF in the long-time regime ($t\dot{\gamma}$$>$$100$) with detailed suspension microstructure, a total number of 300 strain units are employed for time averaging. In the short-regime ($t\dot{\gamma}$$\sim$1), owing to limited time steps associated with specific suspension configuration (e.g. the string ordered configuration), about 3 strain units are adopted for averaging.
}


\bibliography{jfmbib}
\bibliographystyle{jfm}

\end{document}